\documentclass[12pt, draftclsnofoot, onecolumn]{IEEEtran}

\usepackage{tabls}
\usepackage{amsmath,amssymb,amscd,latexsym,dsfont,amsthm}
\usepackage{cite}
\usepackage{psfrag}
\usepackage{multicol}
\usepackage{color,comment}
\usepackage{balance}
\usepackage{color}
\usepackage{float}
\usepackage{graphicx}
\usepackage{caption}
\usepackage{subfigure}

\newtheorem{theorem}{Theorem}
\newtheorem{corollary}{Corollary}
\newtheorem{remark}{Remark}

\begin{document}
%-- Paper title
\title{Securing Multi-User Broadcast Wiretap Channels with Finite CSI Feedback}

%------------------
\author{% 
Amal Hyadi,~\IEEEmembership{Member,~IEEE}, Zouheir Rezki,~\IEEEmembership{Senior~Member,~IEEE}, and~Mohamed-Slim~Alouini,~\IEEEmembership{Fellow,~IEEE}\vspace{-0.8cm}
\thanks{%
\mbox{}}
\thanks{
A. Hyadi is with the Electrical and Computer Engineering Department, McGill University, Montreal, Canada.[e-mail: amal.hyadi@mcgill.ca].%}
%\thanks{
~\! Z. Rezki is with the Electrical and Computer Engineering Department, University of Idaho, Moscow, ID, US. [e-mail: zrezki@uidaho.edu].%}
%\thanks{
~\! M.-S. Alouini is with the Division of Computer, Electrical, and Mathematical Sciences \& Engineering (CEMSE), King Abdullah University of Science and Technology (KAUST), Thuwal, Makkah Province, Saudi Arabia. [e-mail: slim.alouini@kaust.edu.sa].
}
\thanks{
%The research reported in this publication was supported by CRG 2 grant from the Office of Sponsored Research at King Abdullah University of Science and Technology (KAUST).
~\! This work was presented in part at the 2016 IEEE Information Theory Workshop (ITW'2016) and the 2016 IEEE Global Communications Conference (GLOBECOM'2016), Washington, DC, USA.
}
}%
%-----------------

%-- Make the title area
\maketitle 
\enlargethispage{0.8cm}
\thispagestyle{empty}
%-- Abstract
\begin{abstract}\vspace{-0.4cm}
In this work, we investigate the problem of secure broadcasting over block-fading wiretap channels with limited channel knowledge at the transmitter. More particularly, we analyze the effect of having a finite rate feedback on the throughput of multi-user broadcast wiretap channels. We consider that the transmitter is only provided by a $b$-bits feedback of the main channel state information (CSI) sent by each legitimate receiver, at the beginning of each fading block, over error-free public links with limited capacity. Also, we assume that the transmitter is aware of the statistics of the eavesdropper's CSI but not of its channel's realizations. Under these assumptions of CSI uncertainty, we characterize the ergodic secrecy capacity of the system when a common message is broadcasted to all legitimate receivers, the ergodic secrecy sum-capacity when multiple independent messages are transmitted, and the ergodic secrecy capacity region for the broadcast channel with confidential messages (BCCM). In all three scenarios, we show that as long as the transmitter has some knowledge of the main CSI, obtained even through a 1-bit CSI feedback, a non-zero secrecy rate can still be achieved. The impact of having the feedback sent over a binary erasure channel (BEC) is also investigated for the BCCM case. Here again, and even with the possibility of having the feedback bits erased, a positive secrecy rate can still be achieved as long as the erasure event is not a probability-one event. An asymptotic analysis of the obtained results is provided for the high SNR regime, and the scaling law of the system, when the number of legitimate receivers is large, is also presented. \vspace{-0.2cm}
\end{abstract}
%-- Keywords
\begin{IEEEkeywords}\vspace{-0.2cm}
Ergodic secrecy capacity, channel state information, broadcast channel, block-fading channel, binary erasure channel, limited feedback.
\end{IEEEkeywords}

\newpage

%-------------------------------------------------------------------------------------------------------------
\section{Introduction}
%-------------------------------------------------------------------------------------------------------------
For many years, the security challenge has been mainly addressed at the application layer using cryptographic techniques. However, with the emergence of ad-hoc and decentralized networks and the next wave of innovative wireless systems, the need for less complex securing methods had become a necessity. It is mainly for this reason that wireless physical layer security has gained a lot of attention from the research community in the last few years. The core idea behind information theoretic security is to exploit the characteristics of the wireless channel, such as noise and fading, to limit the amount of information that can be extracted at the\linebreak physical layer. 
%-------------------------------------------------------------------------------------------------------------
\subsection{Literature Review}
%-------------------------------------------------------------------------------------------------------------
The first research steps on information theoretic security were taken by Shannon in his pioneering work on cipher systems~\cite{Shannon_J1}. Shannon showed that a perfectly secure transmission could be achieved at the bit level using a shared secret key. This key should be, however, at least as long as the secret message itself, and should only be known by the legitimate entities. The physical layer security paradigm was extended later on, by Wyner, to a more promising setup that doesn't require the use of a secret key~\cite{Wyner_J1}. Wyner's model considers a degraded wiretap channel where the source exploits the structure of the communication link to transmit a message reliably to the intended receiver while asymptotically leaking no information to the eavesdropper. Ulterior works generalized Wyner's work to the case of non-degraded channels~\cite{Csiszar_J1}, Gaussian channels~\cite{Leung_J1}, and fading channels \cite{Gopala_J1,Liang_J1}, to cite only a few. 

The secrecy performance of multi-user systems has been of interest in a number of recent research works.  For the broadcast multi-user scenario, the secrecy capacity of parallel and fading channels, assuming perfect main channel state information (CSI) at the transmitter (CSIT), was considered in~\cite{Ashish_J3} and\cite{Ekrem_J2} while the case of imperfect main channel estimation was elaborated in~\cite{Hyadi_J1}. For the multiple access scenario, the authors in~\cite{Tekin_C1} and \cite{Ekrem_C1} investigated the secrecy capacity of degraded wiretap channels. The problem of analyzing the secrecy capacity of multiple antenna systems has also been of great interest. The secrecy capacity for the multiple-input single-output (MISO) wiretap Gaussian channel was proven in \cite{Ashish_C1} and \cite{LiJ_J2}. Another work \cite{Ashish_J1} characterized the secrecy capacity for the MISO case, with a multiple-antenna eavesdropper, when the main and the eavesdropping channels are known to all terminals. The secrecy capacity of the multiple-input multiple-output (MIMO) transmission with a multiple-antenna eavesdropper was considered in \cite{Oggier_J1} and~\cite{Ashish_J2} when the channel matrices are fixed and known to all terminals. The secrecy capacity region of the Gaussian MIMO broadcast wiretap channel was derived~in~\cite{Ekrem_J1}.

The channel model, presented by Csisz\'{a}r and K\"{o}rner in~\cite{Csiszar_J1}, is regarded as the broadcast channel with confidential messages (BCCM), in which the source also has a common message to transmit in addition to the confidential message. The secrecy capacity region of the parallel BCCM and the fading BCCM with perfect CSIT were characterized in~\cite{Liang_J1}. Further results on the BCCM can be found in~\cite{LiuR_J1,Xu_J1,Zou_C1}.

Taking full advantage of the ability of the physical layer to secure wireless transmissions, requires a complete knowledge of the channel state information (CSI) at the transmitter (CSIT); which is difficult to have in practical scenarios. One way to overcome this challenge is by using feedback. With no secrecy constraints, the finite feedback problem has been extensively studied in the literature~\cite{Heath_C1,Blum_C1,Love_J2,Lau_J1,Murthy_J1,Love_J1}. For the case of a single user transmission, the secrecy throughput when the feedback is finite has been investigated in~\cite{Rezki_J1} for single antenna wiretap channels, in~\cite{Hyadi_J3}~and~\cite{LiuS_J1} for MIMO block-fading channels, and in~\cite{LiN_J1} for the fast fading MIMO case. The noise leakage problem when transmitting artificial noise with limited CSI feedback was analyzed in~\cite{Zhang_J2}~and~\cite{LinSC_J1}. 
Recently, the work in~\cite{Bassi_C1,Bassi_J1,Bassi_C2} investigated the problem of secure transmission over the wiretap channel with generalized feedback. The considered model assumes that the transmitter observes general feedback that is correlated to the channel outputs of the receiver and the eavesdropper. This feedback link is mainly utilized to generate a secret key that can encrypt the message either wholly or partially. The obtained results show that the transmitter and the legitimate receiver can agree on a secret key simultaneously with the transmission of a message. Also, these results represent a generalization of multiple ulterior works.  

The uncertainty of the CSIT can also be the result of an error of estimation at the transmitter or of a delayed feedback information. The authors in the following works, \cite{ZhengTX_J1,Rezki_J2,Hyadi_J1,ChuZ_J1,Zhou_J1}, examined the case when the uncertainty is the result of an error of estimation, and \cite{Ferdinand_J1,HuangY_J1,Hu_J1} considered the wiretap channel with outdated CSI. A synopsis of how different levels of CSIT impact the system's security is provided in \cite{LiuTY_J2} and a detailed state-of-the-art review of physical layer security with CSIT uncertainty is presented in \cite{Hyadi_J2}. 

The particular case where the transmitter does not have any knowledge about the eavesdropper's channel, not even the statistical knowledge or the distribution of the wiretapper gain, is found in the framework of arbitrary varying eavesdropper channel~\cite{HeX_J1}. Under such an assumption on the eavesdropper's channel state and assuming that the number of antennas of the eavesdropper is limited, the authors in~\cite{HeX_J1} derived the secrecy degrees of freedom (SDoF) of the MIMO wiretap channel. This work was later on extended to the multi-user MIMO setup~in~\cite{HeX_C1},~\cite{HeX_J5}, and~\cite{HeX_J6}. 
%-------------------------------------------------------------------------------------------------------------
\subsection{Contributions Summary}
%-------------------------------------------------------------------------------------------------------------
In this work, we aim to analyze and understand the impact of having a limited knowledge of the CSIT on the ergodic secrecy throughput of multi-user broadcast wiretap channels. In particular, we assume that the transmitter is unaware of the instantaneous channel gains to the legitimate receivers and to the eavesdropper and is only provided by a finite CSI feedback. This feedback is sent by the legitimate receivers through feedback links with limited capacity. At the difference of~\cite{Bassi_C1,Bassi_J1,Bassi_C2}, the feedback link is non-secure and is only used to inform the transmitter about the CSI. Also, we assume that the eavesdropper is aware of the CSI of the legitimate receivers and of the feedback information. Consequently, the CSI and the feedback are not a source of secrecy and cannot be used to generate a secret key for the transmitter. 
Besides, we consider three different types of broadcast channels, namely the common message broadcast channel, the independent messages broadcast channel, and the broadcast channel with confidential messages. The main contributions of this work can be summarized as follows:

\begin{itemize}
\item Under the assumption of finite CSI feedback, we provide lower and upper bounds on the ergodic secrecy capacity and the ergodic secrecy sum-capacity for the common and the independent messages broadcast channels, respectively. In both transmission scenarios, we show that even with a 1-bit CSI feedback, sent at the beginning of each fading block, a positive secrecy rate can still be achieved. For the particular case of infinite feedback, we prove that our bounds coincide, hence, fully characterizing the secrecy capacity and the secrecy sum-capacity in this case. 
\item We investigate the ergodic secrecy capacity region of the multi-user BCCM with finite CSI feedback. In particular, we consider that the transmitter has a common message intended for all system users and a confidential information that needs to be kept secret from one of them. First, we examine the case when the feedback is sent over an error-free link. Interestingly, we show that the by using the 1-bit feedback as an indication bit that compares the users' channels, not only can we achieve a positive secrecy rate, but we can also establish the converse. Any extra bit that could be fed back should be used to adapt the transmission power. Then, we extend the obtained result to the case when the feedback is sent over a binary erasure channel (BEC). Here again, we show that one bit should be used as an indication bit while the extra bits could be used either as redundant indication bits or to adapt the power. 
\item We present an asymptotic study of the obtained results in the high SNR regime. The derived expressions show that the secrecy throughput is bounded and does not depend on the transmission power. In the case of independent messages transmission, we characterize the scaling law of the system and show that even with finite CSI feedback, the secrecy performance scales with $\log\log K$ as the number of legitimate receivers $K$ becomes large. 
\end{itemize}

In all cases, we show that as long as the transmitter has some knowledge of the main CSI, a positive secrecy rate can still be achieved. 
%-------------------------------------------------------------------------------------------------------------
\subsection{Outline of the Paper}
%-------------------------------------------------------------------------------------------------------------
The remaining of this paper is organized as follows. Section~\ref{SM} describes the system model. The main results are summarized in Section~\ref{MR}; the secrecy capacity region of the common message transmission is characterized in subsection~\ref{BCM}, the secrecy sum-capacity of the independent messages case is analyzed in subsection~\ref{BIM}, and the secrecy capacity region of the BCCM is considered in subsection~\ref{BCCM}. The asymptotic analyses in the high SNR regime are presented in Section~\ref{AA}. Finally, selected simulation results are illustrated in Section~\ref{NE}, and Section~\ref{conclusion} concludes the paper. 

\vspace{0.2cm}
\textit{Notations:}
Throughout the paper, we use the following notational conventions. The expectation operation is denoted by $\mathbb{E}[.]$, the conditional expectation, given event $A$, is represented by $\underset{.|A}{\mathbb{E}}[.]$, $\log$ denotes the base two logarithm unless otherwise indicated, and we define $\{\nu\}^+{=}\max (0,\nu)$. The entropy of a discrete random variable $\mathbf{X}$ is denoted by $H(\mathbf{X})$, and the mutual information between random variables $\mathbf{X}$ and $\mathbf{Y}$ is denoted by $I(\mathbf{X};\mathbf{Y})$. We also use the notation $X{\sim}\mathcal{CN}(\mu,\sigma^2)$ to indicate that $X$ is a circularly symmetric complex Gaussian variable with mean $\mu$ and variance~$\sigma^2$. A sequence of length $n$ is denoted by $X^n$, i.e., $X^n{=}\{X(1),X(2),{\cdots},X(n)\}$, a sequence of elements between $i$ and $j$, $i{<}j$, is denoted by $X^{[i,j]}$, i.e., $X^{[i,j]}{=}\{X(i),X(i{+}1),{\cdots},X(j)\}$, $X(k)$ represents the $k$-th element of $X^n$, and $X(l,k)$ denotes the $k$-th element of $X$ in the $l$-th fading block.

%-------------------------------------------------------------------------------------------------------------
\section{System Model}\label{SM}
\begin{figure}[t]
\psfrag{t}[l][l][1.5]{$\text{T}$}
\psfrag{e}[l][l][1.5]{$\text{E}$}
\psfrag{r1}[l][l][1.5]{$\text{R}_1$}
\psfrag{rk}[l][l][1.5]{$\text{R}_k$}
\psfrag{r3}[l][l][1.5]{$\text{R}_K$}
\psfrag{h1}[l][l][1.5]{$h_1$}
\psfrag{hk}[l][l][1.5]{$h_k$}
\psfrag{h3}[l][l][1.5]{$h_K$}
\psfrag{g}[l][l][1.5]{\hspace{-0.1cm}$h_\text{e}$}
\psfrag{feedback}[l][l][1.5]{\hspace{0.3cm}$b$-bit Feedback Link}
\psfrag{l}[l][l][2]{$\left. \rule{0pt}{2.2cm} \right\}$}
\psfrag{transmitter}[l][l][1.5]{Transmitter}
\psfrag{eavesdropper}[l][l][1.5]{Eavesdropper}
\psfrag{receivers}[l][l][1.5]{$\hspace{-0.3cm}K$ Legitimate Receivers}
\begin{center}%\vspace{-0.2cm}
\scalebox{0.4}{\includegraphics{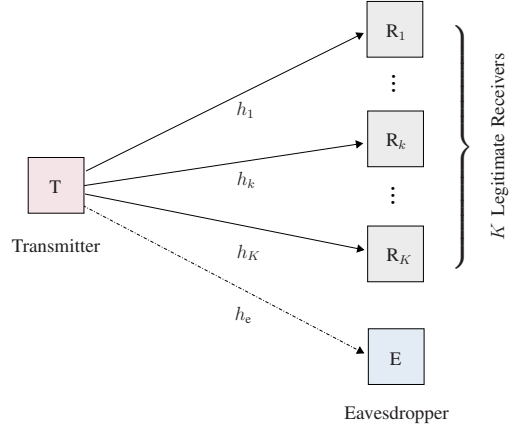}}
\end{center}%\vspace{-0.2cm}
\caption{Multi-User broadcast wiretap channel.}%\vspace{-0.3cm}
\label{fig:Model}
\end{figure}

%-------------------------------------------------------------------------------------------------------------
We consider the problem of secure broadcasting over block-fading wiretap channels when the transmitter has limited knowledge about the main users' CSI. The CSI knowledge is obtained through finite-rate feedback links used by the legitimate receivers to inform the transmitter about their channel prior to data transmission. The feedback links are public, which implies that the CSI information cannot be used as a source of secrecy. Our model of interest consists of a block-fading broadcast wiretap channel where a transmitter (T) communicates with $K$ legitimate receivers ($\text{R}_k, k\in\{1,{\cdots},K\}$) in the presence of an eavesdropper (E), as depicted in~Fig.~\ref{fig:Model}. We consider three different transmission scenarios: 
\begin{itemize}
\item\textbf{$\mathbf{1^\text{st}}$ scenario:} a common message $W$ is broadcasted to all $\text{R}_k$ receivers and has to be kept secret from the eavesdropper. We refer to this scenario as the common message case (CM~case).
\item\textbf{$\mathbf{2^\text{nd}}$ scenario:} multiple independent messages $(W_1,\cdots,W_K)$ are transmitted. Message $W_k$, $k\in\{1,{\cdots},K\},$ is intended for the $k^\text{th}$ legitimate receiver $\text{R}_k$, and all $K$ messages has to be kept secret from the eavesdropper. We refer to this scenario as the independent messages case (IMs case).
\item\textbf{$\mathbf{3^\text{rd}}$ scenario:} a common information $W_0$ is transmitted to all system users, including the eavesdropping node in~Fig.~\ref{fig:Model}, while a confidential information $W_1$ is transmitted to the legitimate receivers $\text{R}_k$ only. Message $W_1$ has to be kept secret from the eavesdropper. In this scenario, the eavesdropper is both a system receiver and an eavesdropper. We denote it in this case as R/E. We refer to this scenario as the common and confidential messages case (CCMs case). 
\end{itemize}

%%%%%%%%%%%%%%%%%%%%%%%%%%%%%%%%%%%%%%%%%%%%%%%%%%%%%%%%%%%%%%%%%%
%%%%%%%%%%%%%%%%%%%%%%%%%%%%%%%%%%%%%%%%%%%%%%%%%%%%%%%%%%%%%%%%%%
\subsection{Received Signals and Channel Assumptions}
Regardless of the considered transmission scenario, the received signals at each legitimate receiver $\text{R}_k, k\in\{1,{\cdots},K\},$ and the eavesdropper are respectively given by
\begin{equation}\label{Sys_Out}
\begin{aligned}
&Y_k(l,j)=h_k(l)X(l,j)+v_k(l,j)\\
&Y_\text{e}(l,j)\hspace{0.06cm}=h_\text{e}(l)X(l,j)\hspace{0.05cm}+w_\text{e}(l,j),
\end{aligned}
\end{equation}
where $l\in\{1,{\cdots},L\}$, with $L$ being the number of fading blocks, $j\in\{1,{\cdots ,}\kappa\}$, with $\kappa$ representing the length of each fading block, $X(l,j)$ is the $j$-th transmitted symbol in the $l$-th fading block, $h_k(l)\!\in\!\mathbb{C}$, $h_\text{e}(l)\!\in\!\mathbb{C}$ are the complex Gaussian channel gains corresponding to each legitimate channel and the eavesdropper's channel, respectively, and $v_k(l,j)\!\in\!\mathbb{C}$, $w_\text{e}(l,j)\!\in\!\mathbb{C}$ represent zero-mean, unit-variance circularly symmetric white Gaussian noises at $\text{R}_k$ and~$\text{E}$, respectively. We consider a block-fading channel where the channel gains remain constant within a fading block. We assume that the channel encoding and decoding frames span a large number of fading blocks, i.e., $L$ is large, and that the blocks change independently from a fading block to another. 
An average transmit power constraint is imposed at the transmitter such that 
\begin{equation}\label{PC}
\frac{1}{n}\sum_{t=1}^n\mathbb{E}\left[|X(t)|^2\right]\leq P_{\text{avg}},
\end{equation}
with $n{=}\kappa L$, and where the expectation is over the input distribution. The transmitted signal $X$ can either corresponds to a common message in the CM case, to a combination of independent messages in the IMs case, or to a combination of common and confidential messages in the CCMs case. 

The channel gains $h_k$ and $h_\text{e}$ are independent, ergodic and stationary with bounded and continuous probability density functions (PDFs). In the rest of this paper, we denote $|h_k|^2$ and $|h_\text{e}|^2$ by $\gamma_k$ and $\gamma_\text{e}$, respectively. We assume that each legitimate receiver is instantaneously aware of its channel gain $h_k(l)$, and the eavesdropper knows $h_\text{e}(l)$. The statistics of the main and the eavesdropping channels are available to all nodes. Further, we assume that the transmitter is not aware of the instantaneous channel realizations of neither channel, and relies on the CSI feedback links to acquire knowledge about the legitimate channels. 

\subsection{Feedback Channel Model}
%%%%%%%%%%%%%%%%%%%%%%%%%%%%%%%%%%%%%%%%%%%%%%%%%%%%%%%%%%%%%
\begin{figure}[h!]
\psfrag{t}[l][l][1.5]{$\text{T}$}
\psfrag{e}[l][l][1.5]{$\text{E}$}
\psfrag{re}[l][l][1.5]{\hspace{-0.1cm}$\text{R/E}$}
\psfrag{r1}[l][l][1.5]{$\text{R}_1$}
\psfrag{rk}[l][l][1.5]{$\text{R}_k$}
\psfrag{r3}[l][l][1.5]{$\text{R}_K$}
\psfrag{feedback}[l][l][1.5]{\hspace{-0.55cm}$b$-bit CSI Feedback Link}
\psfrag{CM case: W0}[l][l][1.5]{\hspace{-0.5cm} CM case: $W$}
\psfrag{IMs case: W1...WK}[l][l][1.5]{\hspace{-0.5cm} IMs case: $(W_1,\cdots,W_K)$}
\psfrag{CM case: Wh01}[l][l][1.5]{CM case: $\hat{W}^{(1)}$}
\psfrag{IMs case: Wh1}[l][l][1.5]{IMs case: $\hat{W}_1$}
\psfrag{CM case: Wh0k}[l][l][1.5]{CM case: $\hat{W}^{(k)}$}
\psfrag{IMs case: Whk}[l][l][1.5]{IMs case: $\hat{W}_k$}
\psfrag{CM case: Wh0K}[l][l][1.5]{CM case: $\hat{W}^{(K)}$}
\psfrag{IMs case: WhK}[l][l][1.5]{IMs case: $\hat{W}_K$}
\psfrag{CM case:  secrecyconstraintC}[l][l][1.5]{CM case: $\frac{1}{n}I(W;Y_\text{e}^n){\rightarrow}0$}
\psfrag{IMs case:  secrecyconstraintI}[l][l][1.5]{IMs case: $\frac{1}{n}I(W_1,\cdots,W_K;Y_\text{e}^n){\rightarrow}0$}
\psfrag{W0 W1}[l][l][1.5]{\hspace{-0.3cm}$(W_0,W_1)$}
\psfrag{Wh01 Wh11}[l][l][1.5]{\hspace{-0.3cm}$(\hat{W}_0^{(1)},\hat{W}_1^{(1)})$}
\psfrag{Wh0k Wh1k}[l][l][1.5]{\hspace{-0.3cm}$(\hat{W}_0^{(k)},\hat{W}_1^{(k)})$}
\psfrag{Wh0K Wh1K}[l][l][1.5]{\hspace{-0.3cm}$(\hat{W}_0^{(K)},\hat{W}_1^{(K)})$}
\psfrag{Wh0K+1}[l][l][1.5]{\hspace{-0.3cm}$\hat{W}_0^{(K+1)}$}
\psfrag{secrecyconstraint}[l][l][1.5]{\hspace{-0.3cm}$\frac{1}{n}I(W_1;Y_\text{e}^n){\rightarrow}0$}
\centering
\vspace{-0.5cm}
\subfigure[Each legitimate receiver has its own CSI feedback link.]{\scalebox{0.38}{\hspace{-1cm}\includegraphics{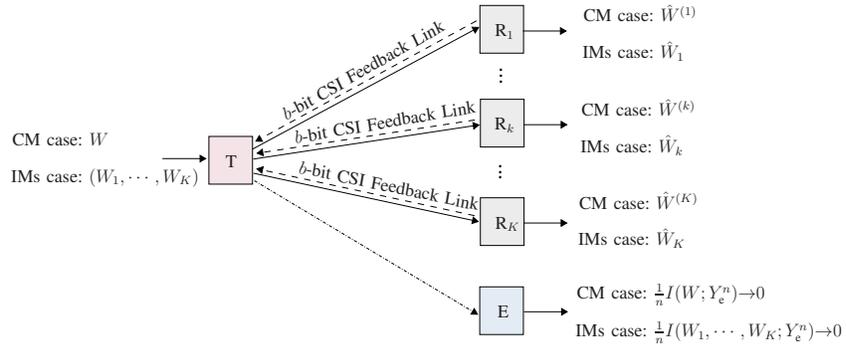}}\label{F_model_a}}
\par\bigskip
\subfigure[The legitimate receivers share a unique CSI feedback link.]{\scalebox{0.38}{\hspace{-1cm}\includegraphics{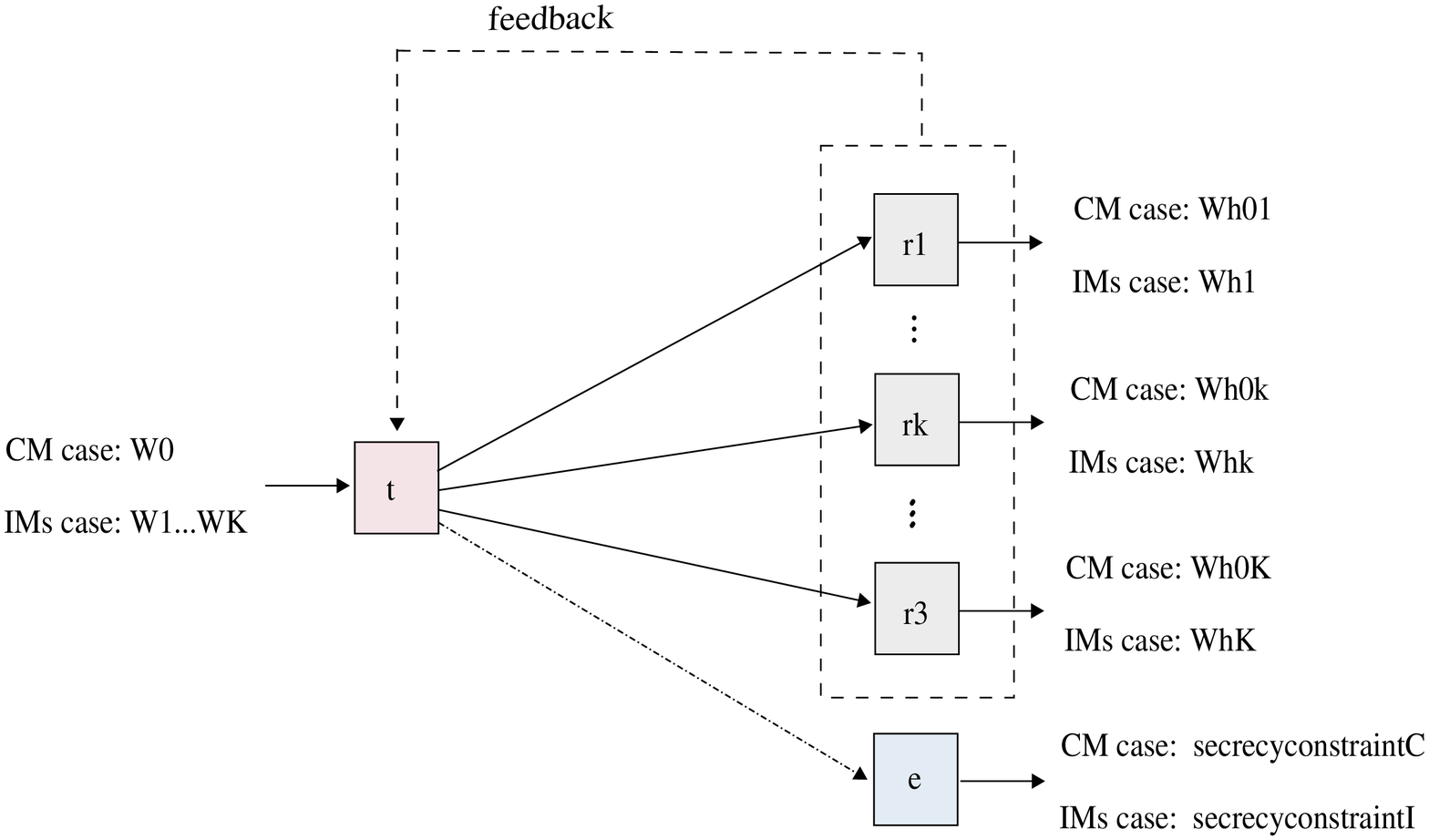}}\label{F_model_b}}
\par\bigskip
\subfigure[All system users share the same CSI feedback link (Applicable in the CCMs case).]{\scalebox{0.38}{\hspace{-1cm}\includegraphics{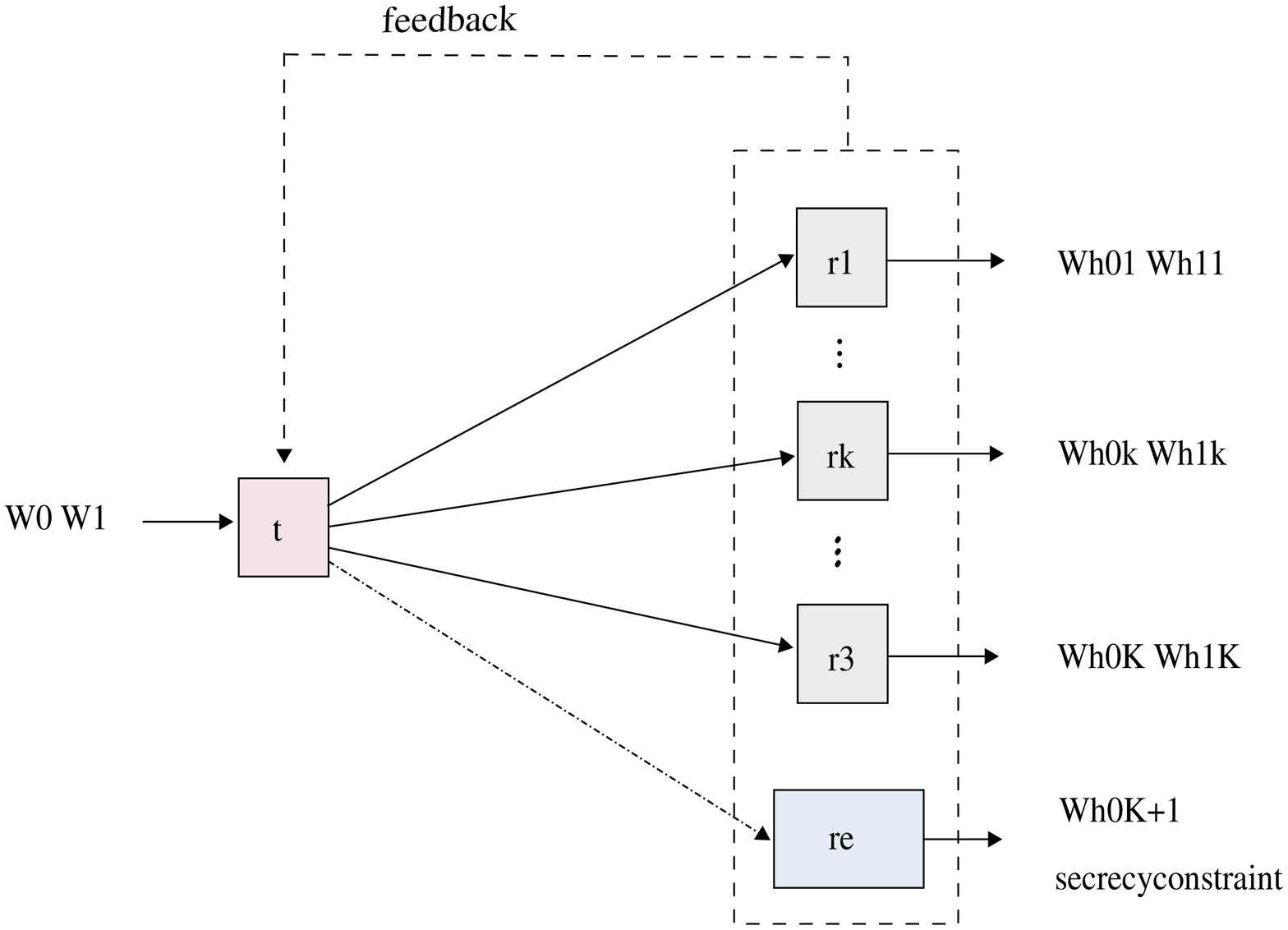}}\label{F_model_c}}
%\vspace{-0.8cm}
\caption{CSI Feedback Model.}%\vspace{-0.2cm}
\label{fig:F_model}
\end{figure}
%%%%%%%%%%%%%%%%%%%%%%%%%%%%%%%%%%%%%%%%%%%%%%%%%%%%%%%%%%%%%
We consider two possible CSI feedback situations, either each legitimate receiver has its own CSI feedback link to the transmitter, cf.~Fig.~\ref{F_model_a}, or all receivers have a unique CSI feedback link that they need to share, cf.~Fig.~\ref{F_model_b} and~cf.~Fig.~\ref{F_model_c}. In the latter case, the receivers cooperate to send the CSI feedback. 

In the case of secure common and multiple messages transmissions, we focus on the feedback model where each receiver has its own CSI feedback link, cf.~Fig.~\ref{F_model_a}. The adopted strategy consists, then, on partitioning the main channel gain support into $Q$ intervals $[\tau_1, \tau_2), \cdots,$ $[\tau_q, \tau_{q+1}), \cdots , [\tau_Q, \infty)$, where $Q{=}2^b$. That is, during each fading block, each legitimate receiver $\text{R}_k$ determines in which interval, $[\tau_q, \tau_{q+1})$ with $q{=}1,\cdots ,Q$, its  channel gain $\gamma_k$ lies and feeds back the associated index $q$ to the transmitter. At the transmitter side, each feedback index $q$ corresponds to a power transmission strategy $P_q$ satisfying the average power constraint. We assume that all nodes are aware of the main channel gain partition intervals $[\tau_1, \tau_2), \cdots , [\tau_q, \tau_{q+1}), \cdots , [\tau_Q, \infty)$, and of the corresponding power transmission strategies\linebreak $\{P_1, \cdots , P_Q\}$. Later on in the paper, we will show that the obtained results for the CM case and the IMs case are also valid when a unique feedback link is available.

When transmitting common and confidential information, we particularly focus on the second feedback model where a unique CSI feedback link is used, cf.~Fig.~\ref{F_model_c}. Here again, every system user, including R/E in this case, has to determine in which partition interval its channel gain lies. The users should, then, cooperate to decide which channel information to transmit over the feedback link. We provide more details on this setup in Section~\ref{BCCM}, where we also show how the results change when each receiver has its own CSI feedback link. We consider both the case when the CSI feedback is sent over an error-free channel, and the case when the feedback information is sent over a BEC with erasure probability~$\epsilon$.

In all scenarios, we consider that the eavesdropping node knows all channel gains and also tracks all CSI feedback links. Therefore, the feedback information can not be used by the transmitter as a source of secrecy. 

\subsection{Coding for Secrecy}
To make the rest of this paper easy to follow, we introduce, here, the adopted coding setup for each transmission model and define the associated achievable secrecy rates.
 
\textbf{CM case:}
In the case of common message transmission, a $(2^{n\mathcal{R}_\text{s}}, n)$ code consists of the following elements:
\begin{itemize}
\item A message set $\mathcal{W}=\left\lbrace 1,2,{\cdots},2^{n\mathcal{R}_\text{s}}\right\rbrace$ with the messages $W\in\mathcal{W}$ independent and uniformly distributed over $\mathcal{W}$;
\item A stochastic encoder $f: \mathcal{W}\rightarrow\mathcal{X}^n$ that maps each message $w$ to a codeword $x^n\in\mathcal{X}^n$;
\item A decoder at each legitimate receiver $g_k: \mathcal{Y}_k^n\rightarrow\mathcal{W}$ that maps a received sequence $y_k^n\in\mathcal{Y}_k^n$ to a message $\hat{w}^{(k)}\in\mathcal{W}$.
\end{itemize}

A rate $\mathcal{R}_\text{s}$ is an \textit{achievable secrecy rate} if there exists a sequence of $(2^{n\mathcal{R}_\text{s}}, n)$ code such that both the average error probability at each legitimate receiver,
%\vspace{-0.2cm}
\begin{equation}
P_{\text{e}_k}=\frac{1}{2^{n\mathcal{R}_\text{s}}}\sum_{w=1}^{2^{n\mathcal{R}_\text{s}}}\text{Pr}\left[\hat{W}^{(k)}\neq W\big|W=w\right],
\end{equation} 
and the leakage rate at the eavesdropper 
\begin{equation}
\frac{1}{n} I(W;Y_e^n,h_e^L,h_1^L,\cdots,h_K^L,F_1^L,\cdots,F_K^L),
\end{equation}
where $F_k^L$ is the sequence of feedback information sent by the $k$-th receiver during $L$ fading blocks, go to zero as $n$ goes to infinity. Note that, we give to the eavesdropper all channels to strengthen the secrecy results. The \textit{secrecy capacity} $\mathcal{C}_\text{s}$ is defined as the maximum achievable secrecy rate, i.e., $\displaystyle{\mathcal{C}_\text{s}\triangleq\sup\mathcal{R}_\text{s},}$ where the supremum is over all achievable secrecy rates. 

\textbf{IMs case:}
When transmitting $K$ independent messages to the legitimate receivers, each intended for a particular user, a $(2^{n\mathcal{R}_1},\cdots, 2^{n\mathcal{R}_K}, n)$ code consists of the following elements:
\begin{itemize}
\item $K$ message sets $\mathcal{W}_k=\left\lbrace 1,2,{\cdots},2^{n\mathcal{R}_k}\right\rbrace$, $k\in\{1,\cdots,K\}$, with the messages $W_k\in\mathcal{W}_k$ independent and uniformly distributed;
\item A stochastic encoder at the transmitter $f: \mathcal{W}_1\times\cdots\times\mathcal{W}_K\rightarrow\mathcal{X}^n$ that maps each message tuple $(w_1,\cdots,w_K)$ to a codeword $x^n\in\mathcal{X}^n$;
\item $K$ decoders, one at each legitimate receiver, $g_k: \mathcal{Y}_k^n\rightarrow\mathcal{W}_1\times\cdots\times\mathcal{W}_K$,\linebreak $k\in\{1,\cdots,K\}$, that maps a received sequence $y_k^n\in\mathcal{Y}_k^n$ to $(\hat{w}_1,\cdots,\hat{w}_K)\in\mathcal{W}_1\times\cdots\times\mathcal{W}_K$.
\end{itemize}%\vspace{0.3cm}

A rate tuple $\left(\mathcal{R}_1,\mathcal{R}_2,\cdots,\mathcal{R}_K\right)$ is said to be achievable if there exists a code such that the average error probability at each legitimate receiver,
%\vspace{-0.2cm}
\begin{equation}
P_{\text{e}_k}=\frac{1}{2^{n\mathcal{R}_k}}\sum_{w_k=1}^{2^{n\mathcal{R}_k}}\text{Pr}\left[\hat{W}_k\neq W_k\big|W_k=w_k\right],
\end{equation} 
and the leakage rate at the eavesdropper %\vspace{-0.3cm}
\begin{equation}
\frac{1}{n} I(W_1,\cdots,W_K;Y_e^n,h_e^L,h_1^L,\cdots,h_K^L,F_1^L,\cdots,F_K^L),
\end{equation}\mbox{}\vspace{-1.2cm}\\
where $F_k^L$ is the sequence of feedback information sent by the $k$-th receiver during $L$ fading blocks, go to zero as $n$ goes to infinity. The secrecy sum-rate is, then, given by $\displaystyle{\widetilde{\mathcal{R}}_\text{s}=\sum_{k=1}^K\mathcal{R}_k},$ and the secrecy sum-capacity is defined, in this case, as $\displaystyle{\widetilde{\mathcal{C}}_\text{s}\triangleq\sup\widetilde{\mathcal{R}}_\text{s}.}$

\textbf{CCMs case:}
A $(2^{n\mathcal{R}_0}, 2^{n\mathcal{R}_1}, n)$ code for the broadcast channel with a common and a confidential messages transmissions consists of the following elements:
\begin{itemize}
\item Two message sets: $\mathcal{W}_0=\left\lbrace 1,2,{\cdots},2^{n\mathcal{R}_0}\right\rbrace$ and $\mathcal{W}_1=\left\lbrace 1,2,{\cdots},2^{n\mathcal{R}_1}\right\rbrace$ with the messages $W_0\in\mathcal{W}_0$ and $W_1\in\mathcal{W}_1$ independent and uniformly distributed over the corresponding sets;
\item A stochastic encoder $f: (\mathcal{W}_0,\mathcal{W}_1)\rightarrow\mathcal{X}^n$ that maps each message pair $(w_0,w_1)$ to a codeword $x^n\in\mathcal{X}^n$;
\item A decoder at each $\text{R}_k$ receiver, $k\in\{1,\cdots,K\}$, $g_k: \mathcal{Y}_k^n\rightarrow (\mathcal{W}_0,\mathcal{W}_1)$ that maps a received sequence $y_k^n\in\mathcal{Y}_k^n$ to a message pair $(\hat{w}_0^{(k)},\hat{w}_1^{(k)})\in (\mathcal{W}_0,\mathcal{W}_1)$;
\item A decoder at the R/E receiver $g: \mathcal{Y}_\text{e}^n\rightarrow\mathcal{W}_0$ that maps a received sequence $y_\text{e}^n\in\mathcal{Y}_\text{e}^n$ to a message $\hat{w}_0^{(K+1)}{\in}\mathcal{W}_0$.
\end{itemize}

A rate pair $(\mathcal{R}_0,\mathcal{R}_1)$ is \textit{achievable} if there exists a sequence of $(2^{n\mathcal{R}_0}, 2^{n\mathcal{R}_1}, n)$ code such that the average error probability at each receiver $\text{R}_k$,
\begin{equation}
P_{\text{e}_{k}}=\frac{1}{2^{n\mathcal{R}_0+n\mathcal{R}_1}}\sum_{w_0=1}^{2^{n\mathcal{R}_0}}\sum_{w_1=1}^{2^{n\mathcal{R}_1}}\text{Pr}\left[(\hat{W}_0^{(k)},\hat{W}_1^{(k)})\neq (W_0,W_1)\big|W_0=w_0 \hspace{0.2cm}\text{and}\hspace{0.2cm} W_1=w_1\right],
\end{equation} 
the average error probability at receiver R/E, 
\begin{equation}
P_{\text{e}_{\text{R/E}}}=\frac{1}{2^{n\mathcal{R}_0}}\sum_{w_0=1}^{2^{n\mathcal{R}_0}}\text{Pr}\left[\hat{W}_0^{(K+1)}\neq W_0\big|W_0=w_0\right],
\end{equation} 
and the leakage rate at receiver R/E, $\frac{1}{n}I(W_1;Y_e^n,h_e^L,h_1^L,\cdots,h_K^L,F^L),$ go to zero as $n$ goes to infinity.
%-------------------------------------------------------------------------------------------------------------
\section{Main Results}\label{MR}
%-------------------------------------------------------------------------------------------------------------
The aim of this work is to study and understand the effect of having a limited knowledge of the CSIT on the ergodic secrecy throughput of multi-user broadcast wiretap channels. This limitation in the knowledge of the CSIT is the downside of the realistic assumption that the feedback links, used by the legitimate receivers to inform the transmitter about their CSI, have finite capacity. In this section, we formulate the obtained results for the CM, the IMs, and the CCMs' transmission scenarios. In all cases, we show that as long as the transmitter has some knowledge of the main CSI, a positive secrecy rate can still be achieved.    
%-------------------------------------------------------------------------------------------------------------
\subsection{Broadcasting a Common Message}\label{BCM}
%-------------------------------------------------------------------------------------------------------------
In this subsection, we examine the case when a unique confidential information is broadcasted to all the legitimate receivers in the presence of an eavesdropper. 

\begin{theorem}\label{TH1}
The ergodic common message secrecy capacity of the block-fading multi-user broadcast wiretap channel with an error free $b$-bit CSI feedback sent by each legitimate receiver, at the beginning of each fading block, is characterized as
\begin{align}
\mathcal{C}_\text{s}^-\leq\mathcal{C}_\text{s}\leq\mathcal{C}_\text{s}^+,
\end{align}
where $\mathcal{C}_\text{s}^-$ and $\mathcal{C}_\text{s}^+$ are given by
\begin{align}
&\mathcal{C}_{\text{s}}^-{=}\hspace{-0.2cm}\min_{1\leq k\leq K}\max_{\{\tau_q;P_q\}_{q=1}^Q}\sum_{q=1}^Q\text{\emph{Pr}}\left[\tau_q{\leq}\gamma_k{<}\tau_{q+1}\right]\underset{\gamma_\text{e}}{\mathbb{E}}\!\left[\left\lbrace\log\!\left(\frac{1{+}\tau_qP_q}{1{+}\gamma_\text{e}P_q}\right)\right\rbrace^{\hspace{-0.1cm}+}\right],\label{th1_a} \\
&\mathcal{C}_{\text{s}}^+{=}\hspace{-0.1cm}\min_{1\leq k\leq K}\max_{\{\tau_q;P_q\}_{q=0}^Q}\sum_{q=0}^Q\text{\emph{Pr}}\left[\tau_q{\leq}\gamma_k{<}\tau_{q+1}\right]\underset{\gamma_\text{e},\gamma_k}{\mathbb{E}}\!\left[\left\lbrace\!\log\!\left(\frac{1{+}\gamma_kP_q}{1{+}\gamma_\text{e}P_q}\right)\!\right\rbrace^+\!\Bigg|\tau_q\!\leq\!\gamma_k\!<\!\tau_{q+1}\right],\label{th1_b}
\end{align}
with $Q{=}2^b$, $\{\tau_q~\!|~\!0{=}\tau_0{<}\tau_1{<}\cdots{<}\tau_Q\}_{q=1}^Q$ are the reconstruction points describing the support of $\gamma_k$ with $\tau_{Q+1}{=}\infty$ for convenience, and $\{P_q\}_{q=1}^Q$ are the power transmission strategies satisfying the average power constraint.%, and $\displaystyle{\Theta_{\tau_q}^{\gamma_k}{=}\text{\emph{Pr}}\left[\tau_q{\leq}\gamma_k{<}\tau_{q+1}\right]}$ for all $q\in\{1,{\cdots},Q\}$.
\end{theorem}

\textit{Proof:} A detailed proof of Theorem \ref{TH1} is provided in Appendix~\ref{App_TH1}.\vspace{0.4cm}

The main difference between the lower and the upper bounds in Theorem~\ref{TH1} is that the feedback information is used to adapt both the transmission rate and the power for the achievable secrecy rate while it is only used to adjust the transmission power for the upper bound. As a matter of fact, the key point in the proof of achievability of~\eqref{th1_a} is that the feedback information is exploited to fix the transmission rate during each coherence block. That is, if the legitimate receiver with the weakest average SNR informs the transmitter that its channel gain falls within the interval $[\tau_q,\tau_{q+1})$, $q\in\{1,{\cdots},Q\}$, the transmitter conveys the codewords at rate $\mathcal{R}_q=\log\left(1{+}\tau_qP_q\right).$ Rate $\mathcal{R}_q$ changes only periodically and is held constant over the duration interval of a fading block. It may seem optimal to let the transmission rate vary with the actual value of the weakest channel gain instead of fixing it with regards to the lower bound of the interval in which it lies. However, in this case, we will lose the $\{.\}^+$ inside the expectation, i.e., the eavesdropper can have a better rate than the legitimate receivers in some fading blocks. The considered setup guarantees that when $\gamma_\text{e}{>}\tau_q$, the mutual information between the transmitter and the eavesdropper is upper bounded by $\mathcal{R}_q$. Otherwise, this mutual information is equal to $\log\left(1{+}\gamma_\text{e}P_q\right)$. 

It is also worth mentioning that, similarly to the case of multi-user common message transmission with no secrecy constraints, the obtained secrecy bounds are limited by the legitimate receiver with the lowest average SNR. It goes without saying that this limitation ensures that all legitimate receivers are able to recover the transmitted message reliably. We can also see from Theorem~\ref{TH1} that even with a 1-bit CSI feedback, sent by each legitimate receiver at the beginning of each fading block, a positive secrecy rate can still be achieved. Of course, as the number of feedback bits increases, the secrecy throughput ameliorates, and when $Q\rightarrow\infty$, our bounds coincide, yielding the result presented in the following corollary. 

\begin{corollary}\label{CR1}
The ergodic common message secrecy capacity of the block fading multi-user broadcast wiretap channel with perfect main CSIT is given by
\begin{align}\label{cr1}
\mathcal{C}_{\text{s}}=\min_{1\leq k\leq K}\max_{P(\gamma_k)}\underset{\gamma_k,\gamma_\text{e}}{\mathbb{E}}\!\left[\left\lbrace\log\!\left(\frac{1{+}\gamma_kP(\gamma_k)}{1{+}\gamma_\text{e}P(\gamma_k)}\right)\right\rbrace^+\right],
\end{align}
with $\mathbb{E} [P(\gamma_k)]\leq P_\text{avg}$. 
\end{corollary}

\textit{Proof:} Corollary \ref{CR1} results directly from the expressions of the achievable rate in \eqref{th1_a} and the upper bound in \eqref{th1_b}, by letting $\text{Pr}\left[\tau_q{\leq}\gamma_k{<}\tau_{q+1}\right]=1/Q$ and taking into consideration that as $Q{\rightarrow}\infty$, the set of reconstruction points, $\{\tau_1,{\cdots},\tau_Q\}$, becomes infinite and each legitimate receiver $\text{R}_k$ is basically forwarding~$\gamma_k$ to the transmitter. $\hfill \square$

To the best of our knowledge, this result has not been reported in earlier works. For the special case of single user transmission, the secrecy capacity in corollary 1 coincides with the result in~\cite[Theorem~2]{Gopala_J1}.

\begin{remark}\label{Rmk1}
The results presented in~Theorem~\ref{TH1} and~Corollary~\ref{CR1} are obtained for the case when each legitimate receiver has its own CSI feedback link. The same results are valid when the legitimate receivers share a unique feedback channel. In this case, the shared link should only be used to send the feedback information associated with the legitimate receiver having the worst average channel gain.
\end{remark}
\begin{remark}\label{Rmk2}
Even though the considered system model only assumes the presence of one eavesdropper, the extension to the multiple eavesdroppers' case is straightforward. In fact, since the distribution of the variable~$\gamma_\text{e}$ is not fixed, the results in~Theorem~\ref{TH1} and~Corollary~\ref{CR1} can also cover the case with multiple wiretappers. That is, in the event when multiple non-colluding eavesdroppers conduct the attack, the distribution of~$\gamma_\text{e}$ should be that of the maximum, i.e.,~$\gamma_\text{e}=\max_{1\leq m\leq M}\gamma_{\text{e}_m}$, where $M$ is the total number of eavesdroppers, and $\gamma_{\text{e}_m}=|h_{\text{e}_m}|^2$, with $h_{\text{e}_m}$ being the channel gain between the transmitter and the $m^\text{th}$ eavesdropper. If the information is secured against the eavesdropper with the best channel, it would also be secured against all the other eavesdroppers. In the case when the wiretappers collude, the distribution of~$\gamma_\text{e}$ should be substituted by that of $\sum_{m=1}^M\gamma_{\text{e}_m}$. The eavesdroppers could be seen, then, as a powerful wiretapper with $M$ antennas. 
\end{remark}
%-------------------------------------------------------------------------------------------------------------
\subsection{Broadcasting Independent Messages}\label{BIM}
%-------------------------------------------------------------------------------------------------------------
Now, we consider the case when multiple independent messages are broadcasted to the legitimate receivers in the presence of an eavesdropper.

\begin{theorem}\label{TH2}
The ergodic secrecy sum-capacity of the block-fading multi-user broadcast wiretap channel with an error free $b$-bit CSI feedback sent by each legitimate receiver, at the beginning of each fading block, is characterized as
\begin{align}
\mathcal{C}_\text{s}^-\leq\mathcal{C}_\text{s}\leq\mathcal{C}_\text{s}^+,
\end{align}
where $\mathcal{C}_\text{s}^-$ and $\mathcal{C}_\text{s}^+$ are given by 
\begin{align}
&\mathcal{C}_{\text{s}}^-{=}\max_{\{\tau_q;P_q\}_{q=1}^Q}\sum_{q=1}^Q\text{\emph{Pr}}\left[\tau_q{\leq}\gamma_\text{max}{<}\tau_{q+1}\right]\underset{\gamma_\text{e}}{\mathbb{E}}\left[\left\lbrace\log\left(\frac{1{+}\tau_qP_q}{1{+}\gamma_\text{e}P_q}\right)\right\rbrace^+\right],\label{th2_a} \\
&\mathcal{C}_{\text{s}}^+{=}\max_{\{\tau_q;P_q\}_{q=0}^Q}\sum_{q=0}^Q\text{\emph{Pr}}\left[\tau_q{\leq}\gamma_\text{max}{<}\tau_{q+1}\right]\underset{\gamma_\text{e},\gamma_\text{max}}{\mathbb{E}}\!\left[\left\lbrace\!\log\!\left(\frac{1{+}\gamma_\text{max}P_q}{1{+}\gamma_\text{e}P_q}\right)\!\right\rbrace^+\Bigg|\tau_q\!\leq\!\gamma_\text{max}\!<\!\tau_{q+1}\right],\label{th2_b}
\end{align}
with $\displaystyle{\gamma_\text{max}{=}\max_{1\leq k\leq K}\gamma_k}$, $Q{=}2^b$, $\{\tau_q~\!|~\!0{=}\tau_0{<}\tau_1{<}\cdots{<}\tau_Q\}_{q=1}^Q$ are the reconstruction points describing the support of $\gamma_\text{max}$ with $\tau_{Q+1}{=}\infty$ for convenience, and $\{P_q\}_{q=1}^Q$ are the power transmission strategies satisfying the average power constraint.%, and $\Theta_{\tau_q}^{\gamma_\text{max}}{=}\text{\emph{Pr}}\left[\tau_q{\leq}\gamma_\text{max}{<}\tau_{q+1}\right]$ for all $q\in\{1,{\cdots},Q\}$.
\end{theorem} \vspace{0.2cm}

\textit{Proof:} A detailed proof of Theorem \ref{TH2} is provided in Appendix~\ref{App_TH2}.\vspace{0.4cm}  

As for the common message case, the main difference between the bounds in Theorem~\ref{TH2} is that the feedback information is used to adapt both the transmission rate and the power for the achievable secrecy sum-rate and only the power in the upper bound. The secrecy sum-rate is achieved by transmitting only to the legitimate user with the best quantized CSI, in a given fading block. Under this strategy, the multi-user broadcast channel reduces to a point-to-point communication with the channel gain distributed as $\max_{1\leq k\leq K}\gamma_k$. One can think that encoding only for the strongest receiver is not valid to establish the secrecy sum-capacity. However, if we look for instance at the two users case, we can easily show that $I(X;Y_1|U){+}I(U;Y_2){=}I(X;Y_1)$, and hence that $\mathcal{R}_1{+}\mathcal{R}_2{\leq}I(X;Y_1)$, with the first receiver being always the strongest one and $U{\rightarrow}X{\rightarrow}Y_1{\rightarrow}Y_2$ forming a Markov chain. The proposed achievability scheme has then a time sharing interpretation to it and even if the result is given in terms of the secrecy sum-rate, the secrecy rate $\mathcal{R}_k$ of each legitimate receiver, $k\in\{1,{\cdots},K\}$, can also be characterized. Indeed, we can write $\mathcal{R}_k\leq\mathcal{C}_{\text{s}}^-\times\text{Pr}[\text{user}~\!k~\!\text{is the strongest receiver}].$

The result in Theorem~\ref{TH2} shows also that even with a 1-bit CSI feedback, sent by each legitimate receiver at the beginning of each fading block, a non-zero secrecy sum-rate can still be achieved. Of course, as the number of feedback bits increases, the secrecy sum-throughput ameliorates, and when $Q\rightarrow\infty$, the bounds on the secrecy sum-capacity coincide, yielding the expression presented in the following corollary.

\begin{corollary}\label{CR2}
The ergodic secrecy sum-capacity of a block fading multi-user broadcast wiretap channel, with perfect main CSIT, is given by
\begin{align}\label{cr2}
\mathcal{C}_{\text{s}}=\max_{P(\gamma_\text{max})}\underset{\gamma_\text{max},\gamma_\text{e}}{\mathbb{E}}\!\left[\left\lbrace\log\!\left(\frac{1{+}\gamma_\text{max}P(\gamma_\text{max})}{1{+}\gamma_\text{e}P(\gamma_\text{max})}\right)\right\rbrace^+\right],
\end{align}
with $\gamma_\text{max}{=}\max_{1\leq k\leq K}\gamma_k$, and $\mathbb{E}[P(\gamma_\text{max})]\leq P_\text{avg}$. 
\end{corollary}

\textit{Proof:} Corollary \ref{CR2} results directly from Theorem~\ref{TH2} by letting $Q\rightarrow\infty$ and following a similar reasoning as for the proof of Corollary~\ref{CR1}. $\hfill \square$

\vspace{0.7cm}
%The presented results for common message and independent messages transmissions, are also valid when multiple non-colluding eavesdroppers conduct the attack. In such a scenario, the transmitter has to limit its transmission with regards to the eavesdropper with the strongest wiretapping channel. Whereas, in the case of colluding eavesdroppers, the results can be extended by replacing the term $\gamma_\text{e}$ with the squared norm of the vector of channel gains of the colluding eavesdroppers, i.e, this case could be seen as if the wiretapping attack is fulfilled by one eavesdropper equipped with multiple antennas and deploying maximum ratio combining (MRC). It is not hard to guess that the strongest the eavesdropper gets, the little is the secrecy we can achieve. Besides, in the analyzed system, we assumed unit variance Gaussian noises at all receiving nodes. The results can be easily extended to a general setup where the noise variances are different. 

\begin{remark}\label{Rmk3}
The results presented in~Theorem~\ref{TH2} and~Corollary~\ref{CR2} consider the case when each legitimate receiver has its own CSI feedback link. The same results are valid when the legitimate receivers share a unique feedback channel. In this case, the shared link should be used to send the feedback information associated with the legitimate receiver having the best channel gain at a given fading block.
\end{remark}
\begin{remark}\label{Rmk4}
As for the CM case, the results in~Theorem~\ref{TH2} and~Corollary~\ref{CR2} can be easily extended to the multiple eavesdroppers' case by adopting the same changes outlined in Remark~\ref{Rmk2}. 
\end{remark}
\begin{remark}\label{Rmk5}
In both the CM case and the IMs case, we assume that the CSI feedback link is error-free. In the event when the feedback information is erased or erroneous, the obtained secrecy rates should be multiplied by the probability of no erasure or the probability of no error, respectively. Note, however, that in this case, the transmitter can opt to omit the CSI feedback when the impact of erasure or error is high. The secret transmission is, then, based on the knowledge of the channel distributions only. This means that we lose the $\{.\}^+$ inside the expectations. 

To better understand this scenario, let us give an example for the MIs case. Assuming that $Pr_\text{e}$ is the probability of erasure or error, the achievable secrecy rate in this case is $\max\left(\overline{R}_\text{s}; Pr_\text{e}\times\mathcal{C}_{\text{s}}^-\right),$ where $\mathcal{C}_{\text{s}}^-$ is given in Theorem~\ref{TH2}, and $$\overline{R}_\text{s}=\max_{1\leq k\leq K}\left\lbrace\underset{\gamma_\text{e},\gamma_k}{\mathbb{E}}\left[\log\left(\frac{1{+}\gamma_kP_t}{1{+}\gamma_\text{e}P_t}\right)\right]\right\rbrace^+,$$ with $P_t$ being a fixed transmission power.
\end{remark}
%-------------------------------------------------------------------------------------------------------------
\subsection{Broadcasting Common and Confidential Messages}\label{BCCM}
%-------------------------------------------------------------------------------------------------------------
In this subsection, we present the ergodic secrecy capacity region of the block-fading BCCM with finite CSI feedback. First, we consider the case of the error-free feedback link. Then, we characterize the achievable secrecy rate region when the feedback link is a BEC.

We should note that, for this transmission scenario, we focus on the case when the system users are cooperating, and the feedback information is sent over a shared CSI feedback link, cf.~Fig.~\ref{F_model_c}. This particular model allows better use of the feedback information to achieve the secrecy capacity. We will also explain how the obtained results change when each user has its own CSI feedback link. 
%-------------------------------------------------------------------------------------------------------------
\subsubsection{Feedback Sent Over an Error-Free Link}\label{BCCM_EF}
%-------------------------------------------------------------------------------------------------------------
Let us first start with the case when only one bit of CSI feedback could be sent. This one bit of feedback will be exploited to indicate to the transmitter whether the channel to R/E is better than that to the legitimate receivers, i.e., the feedback is equal to one when $\min_{1\leq k\leq K}\mathbb{E}\left[\gamma_k\right]>\gamma_\text{e}$, and equal to zero otherwise. This particular use of the feedback allows us to achieve the secrecy capacity of the system, presented in the following theorem.
\begin{theorem}\label{TH3}
The ergodic secrecy capacity region of the block-fading BCCM with a 1-bit CSI feedback, sent at the beginning of each coherence block over an error-free link, is given by
\begin{align}
&\mathcal{C}_\text{s}=\underset{(p_{01},p_{02},p_1)\in\mathcal{P}}{\bigcup}\nonumber\\
&\begin{cases}
(\mathcal{R}_0,\mathcal{R}_1):\\
\mathcal{R}_0\leq\displaystyle{\min_{1\leq k\leq K{+}1}}\!\left\lbrace\displaystyle{\underset{\gamma_k}{\mathbb{E}}\!\left[\log\!\left(\!1{+}\frac{p_{01}\gamma_k}{1{+}p_1\gamma_k}\!\right)\Big|\underline{\gamma}{\in}\mathcal{A}\right]\text{\emph{Pr}}[\underline{\gamma}{\in}\mathcal{A}]}+\underset{\gamma_k}{\mathbb{E}}\!\left[\log\left(1{+}p_{02}\gamma_k\right)\big|\underline{\gamma}{\in}\mathcal{A}^c\right]\text{\emph{Pr}}[\underline{\gamma}{\in}\mathcal{A}^c]\right\rbrace \\
\mathcal{R}_1\leq\displaystyle{\min_{1\leq k\leq K}\underset{\gamma_k,\gamma_\text{e}}{\mathbb{E}}\left[\log\left(1{+}p_1\gamma_k\right){-}\log\left(1{+}p_1\gamma_\text{e}\right)\big|\underline{\gamma}{\in}\mathcal{A}\right]\text{\emph{Pr}}[\underline{\gamma}{\in}\mathcal{A}],}
\end{cases}
\end{align}
where $\gamma_{K+1}=\gamma_\text{e}$, $\underline{\gamma}=[\gamma_1 \cdots \hspace{0.1cm}\gamma_K \hspace{0.2cm}\gamma_e]$, $\mathcal{A}=\left\lbrace\underline{\gamma}:\min_{1\leq k\leq K}\mathbb{E}\left[\gamma_k\right]>\gamma_\text{e}\right\rbrace$ and $$\displaystyle{\mathcal{P}=\left\lbrace(p_{01},p_{02},p_1)\!:(p_{01}{+}p_1)~\!\text{\emph{Pr}}[\underline{\gamma}{\in}\mathcal{A}]{+}p_{02}~\!\text{\emph{Pr}}[\underline{\gamma}{\in}\mathcal{A}^c]{\leq} P_\text{avg}\right\rbrace .}$$
\end{theorem}\vspace{0.2cm}

\textit{Proof:} A detailed proof of Theorem \ref{TH3} is provided in Appendix~\ref{App_TH3}.\vspace{0.4cm}

We can see, from Theorem~\ref{TH3}, that the common message $W_0$ is sent over all coherence blocks while the confidential message $W_1$ is transmitted only over the fading blocks where the instantaneous channel gain of R/E is worst than the average channel gains of the other receivers, i.e., $\underline{\gamma}\in\mathcal{A}{=}\left\lbrace\underline{\gamma}:\min_{1\leq k\leq K}\mathbb{E}\left[\gamma_k\right]>\gamma_\text{e}\right\rbrace$. That is, when $\underline{\gamma}\in\mathcal{A}$, all receivers decode the common message considering the secure message as noise, whereas when $\underline{\gamma}\in\mathcal{A}^c$, since the confidential message is not sent, the common message is decoded at a single user rate. The minimization is due to a bottleneck argument. Also, Theorem~\ref{TH3} states that even with a 1-bit CSI feedback sent at the beginning of each fading block, and as long as event $\mathcal{A}$ is not a zero-probability event, a positive secrecy rate can still be achieved.

At the difference of the perfect CSIT case~\cite{Liang_J1}, the power cannot be instantaneously adapted to the channel realizations and will only depend on the received 1-bit CSI feedback according to a deterministic mapping. It is worth mentioning that $p_{01}$ and $p_{02}$ in Theorem~\ref{TH3} correspond to the power allocated to common message transmissions in $\mathcal{A}$ and $\mathcal{A}^c$, respectively, whereas $p_1$ is the power allocated to the confidential message. When the feedback link has a larger capacity, i.e., more feedback bits can be sent, one bit should be used as an indication bit to point out which channel is better, as explained in the proof of Theorem~\ref{TH3}, while the remaining bits should be used to adapt the transmission power.  The resulting secrecy capacity is given in the following corollary. 

\begin{corollary}\label{CR3}
The secrecy capacity region of the block-fading BCCM with a $b$-bit CSI feedback, sent at the beginning of each coherence block over an error-free link, is given by
\begin{align}
&\mathcal{C}_\text{s}=\underset{\underset{(p_{01_q},p_{02_q},p_{1_q})\in\mathcal{P}}{\left\lbrace\mathcal{H}_q\right\rbrace_{q=1}^Q}}{\bigcup}
\begin{cases}
(\mathcal{R}_0,\mathcal{R}_1):\\
\displaystyle{\mathcal{R}_0\leq\!\displaystyle{\min_{1\leq k\leq K{+}1}}\!\Bigg\lbrace\sum_{q=1}^Q\underset{\gamma_k}{\mathbb{E}}\!\left[\log\!\left(\!1\!+\!\frac{p_{01_q}\gamma_k}{1{+}p_{1_q}\gamma_k}\!\right)\!\Big|\underline{\gamma}{\in}\mathcal{A}\!\cap\!\mathcal{H}_q\right]\!\text{\emph{Pr}}[\underline{\gamma}{\in}\mathcal{A}\!\cap\!\mathcal{H}_q]}\\
\hspace{3.1cm}+\displaystyle{\underset{\gamma_k}{\mathbb{E}}\!\left[\log\left(1{+}p_{02_q}\gamma_k\right)\big|\underline{\gamma}{\in}\mathcal{A}^c\!\cap\!\mathcal{H}_q\right]\text{\emph{Pr}}[\underline{\gamma}{\in}\mathcal{A}^c\!\cap\!\mathcal{H}_q]}\Bigg\rbrace \\
\mathcal{R}_1\leq\displaystyle{\min_{1\leq k\leq K}\sum_{q=1}^Q\underset{\gamma_k,\gamma_\text{e}}{\mathbb{E}}\left[\log\left(\frac{1{+}p_{1_q}\gamma_k}{1{+}p_{1_q}\gamma_\text{e}}\right)\Big|\underline{\gamma}{\in}\mathcal{A}\!\cap\!\mathcal{H}_q\right]\text{\emph{Pr}}[\underline{\gamma}{\in}\mathcal{A}\!\cap\!\mathcal{H}_q],}
\end{cases}
\end{align}
where $\gamma_{K+1}=\gamma_\text{e}$, $Q=2^{b{-}1}$, $\underline{\gamma}=[\gamma_1 \cdots \hspace{0.1cm}\gamma_K \hspace{0.2cm}\gamma_e]$, $\left\lbrace\mathcal{H}_q\right\rbrace_{q=1}^Q$ are the partition regions representing the space of $\underline{\gamma}$, $\mathcal{A}=\left\lbrace\underline{\gamma}:\min_{1\leq k\leq K}\mathbb{E}\left[\gamma_k\right]>\gamma_\text{e}\right\rbrace$ and 
$$\mathcal{P}=\Big\lbrace(p_{01_q},p_{02_q},p_{1_q})\!:\sum_{q=1}^Q(p_{01_q}{+}p_{1_q})~\!\text{\emph{Pr}}[\underline{\gamma}{\in}\mathcal{A}\!\cap\!\mathcal{H}_q]{+}p_{02_q}~\!\text{\emph{Pr}}[\underline{\gamma}{\in}\mathcal{A}^c\!\cap\!\mathcal{H}_q]{\leq} P_\text{avg}\Big\rbrace .$$
\end{corollary}

\textit{Proof:} The proof of Corollary~\ref{CR3} follows along similar lines as the proof of Theorem~1 by using one bit of feedback to indicate which channel is better and exploiting the remaining $b{-}1$ bits to adapt the transmission power.\vspace{0.4cm} 

In this case, it is worth mentioning that the space of the channel gain vector $\underline{\gamma}$ is partitioned into $Q$ regions. During each fading block, the index of the partition region where $\underline{\gamma}$ lies is fed back to the transmitter along with the indication bit. Furthermore, each partition index $q$ corresponds to a transmission power profile $p_{01_q}$ and $p_{1_q}$ to transmit the common and the confidential messages when $\underline{\gamma}\in\mathcal{A}$ and $p_{02_q}$ to transmit the common message solely when $\underline{\gamma}\in\mathcal{A}^c$, with $p_{01_q}$, $p_{02_q}$ and $p_{1_q}$ satisfying the average power constraint. The codebooks for the partition regions and the corresponding transmission power profiles should be known to all terminals. Also, it should be emphasized that when the feedback link has an infinite capacity, i.e., $Q\rightarrow\infty$, the secrecy capacity region in Corollary~1 coincides with the perfect CSIT result in~\cite{Liang_J1}. 
%-------------------------------------------------------------------------------------------------------------
\subsubsection{Feedback Sent Over a BEC}\label{BCCM_BEC}
%-------------------------------------------------------------------------------------------------------------
\begin{corollary}\label{CR4}
An achievable secrecy rate region of the block-fading BCCM with a 1-bit CSI feedback, sent at the beginning of each coherence block over a BEC with erasure probability~$\epsilon$, is given by 
\begin{align}
&\mathcal{R}_\text{s}=\hspace{-0.4cm}\underset{(p_{01},p_{02},p_1)\in\mathcal{P}}{\bigcup}\hspace{-0.1cm}
\begin{cases}
(\mathcal{R}_0,\mathcal{R}_1):\\
\mathcal{R}_0\leq\!\displaystyle{\min_{1\leq k\leq K{+}1}}\!\left\lbrace\displaystyle{\underset{\gamma_k}{\mathbb{E}}\!\left[\log\!\left(\!1\!+\!\frac{p_{01}\gamma_k}{1{+}p_1\gamma_k}\!\right)\!\Big|\mathit{E}^c~\!\&~\!\underline{\gamma}{\in}\mathcal{A}\right]\!(1{-}\epsilon)\text{\emph{Pr}}[\underline{\gamma}{\in}\mathcal{A}]}\right.\\
\hspace{2.5cm}\left.+\underset{\gamma_k}{\mathbb{E}}\!\left[\log\left(1{+}p_{02}\gamma_k\right)\big|\mathit{E}~\!\text{\emph{or}}\left(\mathit{E}^c~\!\&~\!\underline{\gamma}{\in}\mathcal{A}^c\right)\right]\left(\epsilon{+}(1{-}\epsilon)\text{\emph{Pr}}[\underline{\gamma}{\in}\mathcal{A}^c]\right)\right\rbrace \\
\mathcal{R}_1\leq\displaystyle{\min_{1\leq k\leq K}\underset{\gamma_k,\gamma_\text{e}}{\mathbb{E}}\left[\log\left(\frac{1{+}p_1\gamma_k}{1{+}p_1\gamma_\text{e}}\right)\Big|\mathit{E}^c~\!\&~\!\underline{\gamma}{\in}\mathcal{A}\right](1{-}\epsilon)\text{\emph{Pr}}[\underline{\gamma}{\in}\mathcal{A}],}
\end{cases}
\end{align}
where $\gamma_{K+1}=\gamma_\text{e}$, $\underline{\gamma}=[\gamma_1 \cdots \hspace{0.1cm}\gamma_K \hspace{0.2cm}\gamma_e]$, $\mathcal{A}=\left\lbrace\underline{\gamma}:\min_{1\leq k\leq K}\mathbb{E}\left[\gamma_k\right]>\gamma_\text{e}\right\rbrace$, $\mathit{E}$ represents the erasure event, and 
$\mathcal{P}=\left\lbrace(p_{01},p_{02},p_1):(p_{01}{+}p_1)(1{-}\epsilon)~\!\text{\emph{Pr}}[\underline{\gamma}{\in}\mathcal{A}]{+}p_{02}\left(\epsilon{+}(1{-}\epsilon)\text{\emph{Pr}}[\underline{\gamma}{\in}\mathcal{A}^c]\right){\leq} P_\text{avg}\right\rbrace .$
\end{corollary}

\textit{Proof:} The achievability proof is provided in Appendix~\ref{App_CR4}. We note that the conditioning on events $\mathit{E}$ and $\mathit{E}^c$ is inactive when the erasures are independent of the forward channel gains. \vspace{0.4cm}

When the 1-bit feedback is sent over a BEC, the transmission of the confidential message $W_1$ is restricted to the coherence blocks where $\min_{1\leq k\leq K}\mathbb{E}\left[\gamma_k\right]>\gamma_\text{e}$ and the feedback information is not erased. The common message $W_0$ is sent over all fading blocks. It is clear that the confidential rate $\mathcal{R}_1$ reduces as the erasure probability increases and vanishes when the erasure event is a sure event, i.e., the transmitter has no knowledge about the CSI. However, as long as $\epsilon\neq1$ and event $\mathcal{A}$ is not a zero-probability event, a positive secrecy rate can still be achieved. 

In the previous subsection, we did see that when more than one bit of feedback is sent over an error-free link, one bit is used as an indication bit while the remaining extra bits are used to adapt the transmission power. Now, in the case when the feedback bits are sent over a BEC, it would be more interesting to use the extra bits as redundant indication bits. By doing so, the probability of receiving a non-erased indication bit will increase, and this will eventually increase the probability of transmitting the secret information. The secrecy rate region is given in this case in the following corollary.  
\begin{corollary}\label{CR5}
An achievable secrecy rate region of the block-fading BCCM with a b-bit CSI feedback, sent at the beginning of each coherence block over a BEC with erasure probability~$\epsilon$, is given by
\begin{align}
&\mathcal{R}_\text{s}=\hspace{-0.4cm}\underset{(p_{01},p_{02},p_1)\in\mathcal{P}}{\bigcup}\hspace{-0.1cm}
&\hspace{0.4cm}\begin{cases}
(\mathcal{R}_0,\mathcal{R}_1):\\
\mathcal{R}_0\leq\!\displaystyle{\min_{1\leq k\leq K{+}1}}\!\left\lbrace\displaystyle{\underset{\gamma_k}{\mathbb{E}}\!\left[\log\!\left(\!1\!+\!\frac{p_{01}\gamma_k}{1{+}p_1\gamma_k}\!\right)\!\Big|\mathit{E}_{b-bit}^c~\!\&~\!\underline{\gamma}{\in}\mathcal{A}\right]\!(1{-}\epsilon^b)\text{\emph{Pr}}[\underline{\gamma}{\in}\mathcal{A}]}\right.\\
\hspace{1cm}\left.+\underset{\gamma_k}{\mathbb{E}}\!\left[\log\left(1{+}p_{02}\gamma_k\right)\big|\mathit{E}_{b-bit}~\!\text{or}\left(\mathit{E}_{b-bit}^c~\!\&~\!\underline{\gamma}{\in}\mathcal{A}^c\right)\right]\left(\epsilon^b{+}(1{-}\epsilon^b)\text{\emph{Pr}}[\underline{\gamma}{\in}\mathcal{A}^c]\right)\right\rbrace \\
\mathcal{R}_1\leq\displaystyle{\min_{1\leq k\leq K}\underset{\gamma_k,\gamma_\text{e}}{\mathbb{E}}\left[\log\left(\frac{1{+}p_1\gamma_k}{1{+}p_1\gamma_\text{e}}\right)\Big|\mathit{E}_{b-bit}^c~\!\&~\!\underline{\gamma}{\in}\mathcal{A}\right](1{-}\epsilon^b)\text{\emph{Pr}}[\underline{\gamma}{\in}\mathcal{A}],}
\end{cases}
\end{align}
%\mbox{}\vspace{-0.8cm}\\
where $\gamma_{K+1}=\gamma_\text{e}$, $\underline{\gamma}=[\gamma_1 \cdots \hspace{0.1cm}\gamma_K \hspace{0.2cm}\gamma_e]$, $\mathcal{A}=\left\lbrace\underline{\gamma}:\min_{1\leq k\leq K}\mathbb{E}\left[\gamma_k\right]>\gamma_\text{e}\right\rbrace$, $\mathit{E}_{b-bit}$ represents the event when all $b$ feedback bits are erased, and%\vspace{-0.4cm} 
$$\mathcal{P}=\left\lbrace(p_{01},p_{02},p_1):(p_{01}{+}p_1)(1{-}\epsilon^b)~\!\text{\emph{Pr}}[\underline{\gamma}{\in}\mathcal{A}]{+}p_{02}\left(\epsilon^b{+}(1{-}\epsilon^b)\text{\emph{Pr}}[\underline{\gamma}{\in}\mathcal{A}^c]\right){\leq} P_\text{avg}\right\rbrace .$$
\end{corollary}

\textit{Proof:} The proof of Corollary~\ref{CR5} follows along similar lines as the proof of Corollary~\ref{CR4} by using~$b$ redundant bits of feedback to indicate which channel is better. Here again, we note that the conditioning on events $\mathit{E}_{b-\emph{bit}}$ and $\mathit{E}_{b-\emph{bit}}^c$ is inactive when the erasures are independent of the forward channel gains.\vspace{0.4cm}

From Corollary~\ref{CR5} we can see that the probability of transmitting the secret information depends on the erasure event $\mathit{E}_{b-\emph{bit}}$, and is equal to $(1{-}\epsilon^b)\text{Pr}[\underline{\gamma}{\in}\mathcal{A}]$. That is, as long as event $\mathcal{A}$ is not a zero-probability event, increasing the number of redundant indication bits $b$ increases the probability of transmitting the secret message. This is particularly interesting when the probability of erasure~$\epsilon$ is high. 
Another interesting approach to exploit the $b$ feedback bits in this case would be to use~$i$ out of~$b$ feedback bits as redundant indication bits and the remaining~$b{-}i$ bits to adapt the power. The transmission of the secret message would be conditioned, in this case, by having at least one non-erased bit out of the~$i$ redundant bits. Also, in this case, choosing the adequate region to adapt the power would require all~$b{-}i$ bits used for power adaptation to be non-erased. 

\begin{remark}\label{Rmk6}
As stated at the beginning of this subsection, the obtained results for the BCCM consider the case when a unique feedback link is shared to send the channel information. When each receiver has its own independent feedback link, the capacity in~Theorem~\ref{TH3} and~Corollary~\ref{CR3} cannot be achieved. Each receiver will use its feedback link to indicate to the transmitter in which interval its channel gain lies. The transmitter will still be able to decide whether $\min_{1\leq k\leq K}\mathbb{E}\left[\gamma_k\right]>\gamma_\text{e}$, or not, when R/E and the legitimate receiver with the weakest channel indicate different intervals. When these intervals are the same, the transmitter cannot conclude which channel is better, and only the common message will be transmitted over those fading~blocks.  
\end{remark}
\begin{remark}\label{Rmk7}
In the event when there are multiple R/E, let us say $M$ of them, message $W_0$ will be sent to all $K{+}M$ system users while message $W_1$ will only be sent to the $K$ legitimate receivers, and has to be kept secret from the R/Es. In this case, the indication bit should be used to compare $\min_{1\leq k\leq K}\mathbb{E}\left[\gamma_k\right]$ to $\gamma_\text{e}=\max_{1\leq m\leq M}\gamma_{\text{e}_m}$ for the non-colluding scenario, and to $\gamma_\text{e}=\sum_{m=1}^M\gamma_{\text{e}_m}$ for the colluding one. By adapting the distribution of $\gamma_\text{e}$, as explained in Remark~\ref{Rmk2}, the results for the BCCM can be extended to the multi-R/E case. 
\end{remark}
%-------------------------------------------------------------------------------------------------------------
\section{Asymptotic Analysis at High-SNR}\label{AA}
%-------------------------------------------------------------------------------------------------------------
In this section, we provide an asymptotic study of the obtained ergodic secrecy rates and secrecy rate regions presented in the previous section. 
%-------------------------------------------------------------------------------------------------------------
\subsection{Broadcasting a Comman Message}\label{AA-BCM}
%-------------------------------------------------------------------------------------------------------------
\begin{corollary}\label{CR6}
In the high-SNR regime, the ergodic common message secrecy capacity of the block-fading multi-user broadcast wiretap channel with an error free $b$-bit CSI feedback sent by each legitimate receiver, at the beginning of each fading block, is characterized as
\begin{align}
\mathcal{C}_{\text{H-SNR}}^-\leq\mathcal{C}_\text{s-HSNR}\leq\mathcal{C}_{\text{H-SNR}}^+,
\end{align}
where $\mathcal{C}_{\text{H-SNR}}^-$ and $\mathcal{C}_{\text{H-SNR}}^+$ are given by
\begin{align}
&\mathcal{C}_{\text{H-SNR}}^-{=}\min_{1\leq k\leq K}\max_{\{\tau_q\}_{q=1}^Q}\sum_{q=1}^Q\text{\emph{Pr}}\left[\tau_q{\leq}\gamma_k{<}\tau_{q+1}\right]\underset{\gamma_\text{e}}{\mathbb{E}}\!\left[\left\lbrace\log\!\left(\frac{\tau_q}{\gamma_\text{e}}\right)\right\rbrace^{\hspace{-0.1cm}+}\right], \\
&\mathcal{C}_{\text{H-SNR}}^+{=}\min_{1\leq k\leq K}\underset{\gamma_\text{e},\gamma_k}{\mathbb{E}}\!\left[\left\lbrace\!\log\!\left(\frac{\gamma_k}{\gamma_\text{e}}\right)\!\right\rbrace^{\hspace{-0.1cm}+}\right],
\end{align}
with $Q{=}2^b$, and $\{\tau_q~\!|~\!0{=}\tau_0{<}\tau_1{<}\cdots{<}\tau_Q\}_{q=1}^Q$ are the reconstruction points describing the support of $\gamma_k$ with $\tau_{Q+1}{=}\infty$ for convenience.
\end{corollary}

\textit{Proof:} The result in Corollary~\ref{CR6} can be deduced directly from Theorem~\ref{TH1} by taking the limits of $\mathcal{C}_{\text{s}}^-$ and $\mathcal{C}_{\text{s}}^+$ when $P_\text{avg}\rightarrow\infty$.\vspace{0.2cm}

The obtained result in Corollary~\ref{CR6} shows that the secrecy capacity is bounded at high SNR,~i.e.,~it does not depend on $P_\text{avg}$. 

%-------------------------------------------------------------------------------------------------------------
\subsection{Broadcasting Independent Messages}\label{AA-BIM}
%-------------------------------------------------------------------------------------------------------------
%-------------------------------------------------------------------------------------------------------------
\subsubsection{Lower and Upper Bounds}
%-------------------------------------------------------------------------------------------------------------
\begin{corollary}\label{CR7}
In the high-SNR regime, the ergodic secrecy sum-capacity of the block-fading multi-user broadcast wiretap channel with an error free $b$-bit CSI feedback sent by each legitimate receiver, at the beginning of each fading block, is characterized as
\begin{align}
\mathcal{C}_{\text{H-SNR}}^-\leq\mathcal{C}_\text{s-HSNR}\leq\mathcal{C}_{\text{H-SNR}}^+,
\end{align}
where $\mathcal{C}_{\text{H-SNR}}^-$ and $\mathcal{C}_{\text{H-SNR}}^+$ are given by
\begin{align}
&\mathcal{C}_{\text{H-SNR}}^-{=}\max_{\{\tau_q\}_{q=1}^Q}\sum_{q=1}^Q\text{\emph{Pr}}\left[\tau_q{\leq}\gamma_\text{max}{<}\tau_{q+1}\right]\underset{\gamma_\text{e}}{\mathbb{E}}\!\left[\left\lbrace\log\!\left(\frac{\tau_q}{\gamma_\text{e}}\right)\right\rbrace^{\hspace{-0.1cm}+}\right], \\
&\mathcal{C}_{\text{H-SNR}}^+{=}\underset{\gamma_\text{e},\gamma_\text{max}}{\mathbb{E}}\!\left[\left\lbrace\!\log\!\left(\frac{\gamma_\text{max}}{\gamma_\text{e}}\right)\!\right\rbrace^{\hspace{-0.1cm}+}\right],
\end{align}
with $\displaystyle{\gamma_\text{max}{=}\max_{1\leq k\leq K}\gamma_k}$, $Q{=}2^b$, and $\{\tau_q~\!|~\!0{=}\tau_0{<}\tau_1{<}\cdots{<}\tau_Q\}_{q=1}^Q$ are the reconstruction points describing the support of $\gamma_\text{max}$ with $\tau_{Q+1}{=}\infty$ for convenience.
\end{corollary}

\textit{Proof:} The result in Corollary~\ref{CR7} can be deduced directly from Theorem~\ref{TH2} by taking the limits of $\mathcal{C}_{\text{s}}^-$ and $\mathcal{C}_{\text{s}}^+$ when $P_\text{avg}\rightarrow\infty$.\vspace{0.2cm}

As for the common message case, we can see that the secrecy sum-capacity does not depend on $P_\text{avg}$ at the high-SNR regime. However, since the obtained expressions are in terms of $\gamma_\text{max}$, the secrecy performance scales with the number of legitimate receivers $K$. 

%-------------------------------------------------------------------------------------------------------------
\subsubsection{Scaling Law}
%-------------------------------------------------------------------------------------------------------------
\begin{corollary}\label{CR8}
The secrecy sum-capacity when broadcasting independent messages to a large number of legitimate receivers, i.e., $K\rightarrow\infty$, over Rayleigh fading channels with an infinite average power constraint, i.e., $P_\text{avg}\rightarrow\infty$, scales as $\log\log K$.
\end{corollary}

\textit{Proof:}
On one hand, we have
\begin{align}
\lim_{K\rightarrow\infty}\mathcal{C}_\text{s-HSNR}&\geq\lim_{K\rightarrow\infty}\mathcal{C}_{\text{H-SNR}}^-\\
&\geq\lim_{K\rightarrow\infty}\sum_{q=1}^Q\text{Pr}\left[\tau_q{\leq}\gamma_\text{max}{<}\tau_{q+1}\right]\underset{\gamma_\text{e}}{\mathbb{E}}\!\left[\left\lbrace\log\!\left(\frac{\tau_q}{\gamma_\text{e}}\right)\right\rbrace^{\hspace{-0.1cm}+}\right]\\
&\geq\lim_{K\rightarrow\infty}\text{Pr}\left[\tau_Q{\leq}\gamma_\text{max}\right]\underset{\gamma_\text{e}}{\mathbb{E}}\!\left[\left\lbrace\log\!\left(\frac{\tau_Q}{\gamma_\text{e}}\right)\right\rbrace^{\hspace{-0.1cm}+}\right].
\end{align}
Then, taking $\tau_Q=\log K$, we can write
\begin{align}
\lim_{K\rightarrow\infty}\mathcal{C}_\text{s-HSNR}&\geq\lim_{K\rightarrow\infty}\text{Pr}\left[\log K{\leq}\gamma_\text{max}\right]\underset{\gamma_\text{e}}{\mathbb{E}}\!\left[\left\lbrace\log\!\left(\frac{\log K}{\gamma_\text{e}}\right)\right\rbrace^{\hspace{-0.1cm}+}\right].
\end{align}
Considering Rayleigh fading channels, the distribution of the maximum $f_{\gamma_{\text{max}}}(\gamma_{\text{max}})$ converges toward $\delta(\gamma_{\text{max}}{-}\log K)$ as $K{\rightarrow}\infty$, with $\delta(.)$ being the Dirac-Delta function. It is, then, almost sure that $\gamma_{\text{max}}=\log K$, and that $\text{Pr}\left[\log K{\leq}\gamma_\text{max}\right]=1$ as $K{\rightarrow}\infty$. Hence, we have
\begin{align}
\lim_{K\rightarrow\infty}\mathcal{C}_\text{s-HSNR}&\geq\lim_{K\rightarrow\infty}\underset{\gamma_\text{e}}{\mathbb{E}}\!\left[\left\lbrace\log\!\left(\frac{\log K}{\gamma_\text{e}}\right)\right\rbrace^{\hspace{-0.1cm}+}\right].
\end{align}
Since the variable $\gamma_\text{e}$ does not depend on $K$, the term $\underset{\gamma_\text{e}}{\mathbb{E}}\left[\log\left(\gamma_\text{e}\right)\right]$ is asymptotically dominated by $\log\log K$, i.e., $\underset{\gamma_\text{e}}{\mathbb{E}}\left[\log\left(\gamma_\text{e}\right)\right]{=}o(\log\log K)$, yielding
\begin{align}
\lim_{K\rightarrow\infty}\left(\mathcal{C}_\text{s-HSNR}{-}\log\log K\right)&\geq0.
\end{align}

On the other hand, we have
\begin{align}
\lim_{K\rightarrow\infty}\mathcal{C}_\text{s-HSNR}&\leq\lim_{K\rightarrow\infty}\mathcal{C}_{\text{H-SNR}}^+\\
&=\underset{\gamma_\text{e},\gamma_\text{max}}{\mathbb{E}}\!\left[\left\lbrace\!\log\!\left(\frac{\gamma_\text{max}}{\gamma_\text{e}}\right)\!\right\rbrace^{\hspace{-0.1cm}+}\right].
\end{align}
Once again, considering the fact that $f_{\gamma_{\text{max}}}(\gamma_{\text{max}})\rightarrow\delta(\gamma_{\text{max}}{-}\log K)$ and $\underset{\gamma_\text{e}}{\mathbb{E}}\left[\log\left(\gamma_\text{e}\right)\right]{=}o(\log\log K)$ as $K{\rightarrow}\infty$, we get 
$
\lim_{K\rightarrow\infty}\left(\mathcal{C}_\text{s-HSNR}{-}\log\log K\right)\leq0.
$
$\hfill\square$
%-------------------------------------------------------------------------------------------------------------
\subsection{Broadcasting Common and Confidential Messages}\label{AA-BCCM}
%-------------------------------------------------------------------------------------------------------------
%-------------------------------------------------------------------------------------------------------------
\subsubsection{Feedback Sent Over an Error-Free Link}\label{AA-BCCM_EF}
%-------------------------------------------------------------------------------------------------------------
\begin{corollary}\label{CR9}
In the high-SNR regime, the ergodic secrecy capacity region of the block-fading BCCM with a 1-bit CSI feedback, sent at the beginning of each coherence block over an error-free link, is given by
\begin{align}
&\mathcal{C}_\text{H-SNR}=\hspace{-0.4cm}\underset{(\alpha_{01},\alpha_{02},\alpha_1)\in\Phi}{\bigcup}
\begin{cases}
(\mathcal{R}_{0-\text{HSNR}},\mathcal{R}_{1-\text{HSNR}}):\\[6pt]
\mathcal{R}_{0-\text{HSNR}}\leq\displaystyle{\log\!\left(\!1{+}\frac{\alpha_{01}}{\alpha_1}\!\right)}\text{\emph{Pr}}[\underline{\gamma}{\in}\mathcal{A}]+\left(\log P_\text{avg}\right)\text{\emph{Pr}}[\underline{\gamma}{\in}\mathcal{A}^c]\\[4pt]
\mathcal{R}_{1-\text{HSNR}}\leq\displaystyle{\min_{1\leq k\leq K}\underset{\gamma_k,\gamma_\text{e}}{\mathbb{E}}\left[\log\left(\frac{\gamma_k}{\gamma_\text{e}}\right)\Big|\underline{\gamma}{\in}\mathcal{A}\right]\text{\emph{Pr}}[\underline{\gamma}{\in}\mathcal{A}],}
\end{cases}
\end{align}
where $\mathcal{A}=\left\lbrace\underline{\gamma}:\min_{1\leq k\leq K}\mathbb{E}\left[\gamma_k\right]>\gamma_\text{e}\right\rbrace$ and $$\displaystyle{\Phi{=}\left\lbrace(\alpha_{01},\alpha_{02},\alpha_1)\!:(\alpha_{01}{+}\alpha_1)~\!\text{\emph{Pr}}[\underline{\gamma}{\in}\mathcal{A}]{+}\alpha_{02}~\!\text{\emph{Pr}}[\underline{\gamma}{\in}\mathcal{A}^c]{\leq} 1\right\rbrace .}$$
\end{corollary}

\textit{Proof:} Considering the secrecy capacity region in Theorem~\ref{TH3} with $p_{01}{=}\alpha_{01}P_\text{avg}$, $p_{02}{=}\alpha_{02}P_\text{avg}$, and $p_{1}{=}\alpha_{1}P_\text{avg}$, such that $(\alpha_{01}{+}\alpha_1)~\!\text{Pr}[\underline{\gamma}{\in}\mathcal{A}]{+}\alpha_{02}~\!\text{Pr}[\underline{\gamma}{\in}\mathcal{A}^c]{\leq} 1$, we can write
\begin{align}
&\mathcal{R}_0\leq\displaystyle{\min_{1\leq k\leq K{+}1}}\!\left\lbrace\displaystyle{\underset{\gamma_k}{\mathbb{E}}\!\left[\log\!\left(\!1{+}\frac{\alpha_{01}P_\text{avg}\gamma_k}{1{+}\alpha_{1}P_\text{avg}\gamma_k}\!\right)\Big|\underline{\gamma}{\in}\mathcal{A}\right]\text{Pr}[\underline{\gamma}{\in}\mathcal{A}]}\right.\nonumber\\
&\left.\hspace{5cm}+\underset{\gamma_k}{\mathbb{E}}\!\left[\log\left(1{+}\alpha_{02}P_\text{avg}\gamma_k\right)\big|\underline{\gamma}{\in}\mathcal{A}^c\right]\text{Pr}[\underline{\gamma}{\in}\mathcal{A}^c]\right\rbrace ,
\end{align}
and \vspace{-0.6cm}
\begin{align}
&\mathcal{R}_1\leq\displaystyle{\min_{1\leq k\leq K}\underset{\gamma_k,\gamma_\text{e}}{\mathbb{E}}\left[\log\left(1{+}\alpha_{1}P_\text{avg}\gamma_k\right){-}\log\left(1{+}\alpha_{1}P_\text{avg}\gamma_\text{e}\right)\big|\underline{\gamma}{\in}\mathcal{A}\right]\text{Pr}[\underline{\gamma}{\in}\mathcal{A}].}
\end{align}
Without loss of generality, we assume that $\alpha_{01}$, $\alpha_{02}$, and $\alpha_1$ are non-zero power splitting factors. Taking the limit of $\mathcal{R}_0$ and $\mathcal{R}_1$ when $P_\text{avg}$ goes to $\infty$, we get
\begin{align}
&\lim_{P_\text{avg}\rightarrow\infty}\mathcal{R}_0\nonumber\\
&\hspace{0.2cm}\leq\lim_{P_\text{avg}\rightarrow\infty}\mathbb{E}\!\left[\log\!\left(\!1{+}\frac{\alpha_{01}}{\alpha_{1}}\!\right)\Big|\underline{\gamma}{\in}\mathcal{A}\right]\!\text{Pr}[\underline{\gamma}{\in}\mathcal{A}]\nonumber\\
&\hspace{5cm}{+}\left(\!\log P_\text{avg}{+}\min_{1\leq k\leq K{+}1}\underset{\gamma_k}{\mathbb{E}}\!\left[\log\left(\alpha_{02}\gamma_k\right)\big|\underline{\gamma}{\in}\mathcal{A}^c\right]\!\right)\!\text{Pr}[\underline{\gamma}{\in}\mathcal{A}^c]\\
&\hspace{0.2cm}=\lim_{P_\text{avg}\rightarrow\infty}\log\!\left(\!1{+}\frac{\alpha_{01}}{\alpha_{1}}\!\right)\text{Pr}[\underline{\gamma}{\in}\mathcal{A}]{+}\log P_\text{avg}~\!\text{Pr}[\underline{\gamma}{\in}\mathcal{A}^c],
\end{align}
and
\begin{align}
\lim_{P_\text{avg}\rightarrow\infty}\mathcal{R}_1&\leq\lim_{P_\text{avg}\rightarrow\infty}\min_{1\leq k\leq K}\underset{\gamma_k,\gamma_\text{e}}{\mathbb{E}}\left[\log\left(\frac{1{+}\alpha_{1}P_\text{avg}\gamma_k}{1{+}\alpha_{1}P_\text{avg}\gamma_\text{e}}\right)\big|\underline{\gamma}{\in}\mathcal{A}\right]\text{Pr}[\underline{\gamma}{\in}\mathcal{A}]\\
&=\lim_{P_\text{avg}\rightarrow\infty}\min_{1\leq k\leq K}\underset{\gamma_k,\gamma_\text{e}}{\mathbb{E}}\left[\log\left(\frac{\gamma_k}{\gamma_\text{e}}\right)\big|\underline{\gamma}{\in}\mathcal{A}\right]\text{Pr}[\underline{\gamma}{\in}\mathcal{A}].
\end{align}
The particular cases when $\alpha_{01}$, $\alpha_{02}$, or $\alpha_1$ are equal to zero, can be deduced similarly, and are included in the asymptotic capacity region in Corollary~\ref{CR9}. $\hfill\square$
%-------------------------------------------------------------------------------------------------------------
\subsubsection{Feedback Sent Over a BEC}\label{AA_BCCM_BEC}
%-------------------------------------------------------------------------------------------------------------
\begin{corollary}\label{CR10}
In the high-SNR regime, the achievable secrecy rate region of the block-fading BCCM with a 1-bit CSI feedback, sent at the beginning of each coherence block over a BEC with erasure probability~$\epsilon$, is given by 
\begin{align}
&\mathcal{R}_\text{H-SNR}=\underset{(\alpha_{01},\alpha_{02},\alpha_1)\in\Phi}{\bigcup}
\begin{cases}
(\mathcal{R}_{0-\text{HSNR}},\mathcal{R}_{1-\text{HSNR}}):\\[6pt]
\mathcal{R}_{0-\text{HSNR}}\leq\displaystyle{\log\!\left(\!1{+}\frac{\alpha_{01}}{\alpha_1}\!\right)}(1{-}\epsilon)\text{\emph{Pr}}[\underline{\gamma}{\in}\mathcal{A}]+\left(\log P_\text{avg}\right)\left(\epsilon{+}(1{-}\epsilon)\text{\emph{Pr}}[\underline{\gamma}{\in}\mathcal{A}^c]\right)\\[4pt]
\mathcal{R}_{1-\text{HSNR}}\leq\displaystyle{\min_{1\leq k\leq K}\underset{\gamma_k,\gamma_\text{e}}{\mathbb{E}}\left[\log\left(\frac{\gamma_k}{\gamma_\text{e}}\right)\Big|\mathit{E}^c~\!\&~\!\underline{\gamma}{\in}\mathcal{A}\right](1{-}\epsilon)\text{\emph{Pr}}[\underline{\gamma}{\in}\mathcal{A}],}
\end{cases}
\end{align}
where $\mathit{E}$ represents the erasure event, $\mathcal{A}=\left\lbrace\underline{\gamma}:\min_{1\leq k\leq K}\mathbb{E}\left[\gamma_k\right]>\gamma_\text{e}\right\rbrace$ and 
$$\Phi{=}\left\lbrace(\alpha_{01},\alpha_{02},\alpha_1):(\alpha_{01}{+}\alpha_1)(1{-}\epsilon)~\!\text{\emph{Pr}}[\underline{\gamma}{\in}\mathcal{A}]{+}\alpha_{02}\left(\epsilon{+}(1{-}\epsilon)\text{\emph{Pr}}[\underline{\gamma}{\in}\mathcal{A}^c]\right){\leq} 1\right\rbrace .$$
\end{corollary}

\textit{Proof:} The proof follows similar lines as the proof of Corollary~\ref{CR9}.\vspace{0.2cm} 

We note that in the case when $b$ bits of feedback are sent over an error-free link, the asymptotic behavior is the same as in Corollary~\ref{CR9}. This is because only one bit is used as an indication bit while the remaining bits are used to adapt the power. Power adaptation is useless at high-SNR and in this regime, we only need the 1-bit indication feedback. In the case of a BEC with $b$ feedback bits, the asymptotic secrecy rate region can be deduced from Corollary~\ref{CR10} by substituting the erasure probability~$\epsilon$ by~$\epsilon^b$. This is obtained by using all feedback bits as redundant indication~bits. 
%-------------------------------------------------------------------------------------------------------------
\section{Numerical Examples}\label{NE}
%-------------------------------------------------------------------------------------------------------------
In this section, we provide selected simulation results for the case of independent and identically distributed Rayleigh fading channels. We consider that the system's variables, the main channel gains $h_k$, $k\in\{1,\cdots ,K\}$, and the eavesdropper's channel gain $h_\text{e}$, are all drawn from the zero-mean, unit-variance complex Gaussian distribution.

\begin{figure}[t]
%\psfrag{}[l][l][1]{}
\psfrag{Secrecy Rate}[l][l][1.2]{\hspace{-1cm}Secrecy Rate (bps/Hz)}
\psfrag{Pavg}[l][l][1.2]{\hspace{-0.5cm}$P_\text{avg}$ (dB)}
\psfrag{PCSI capacity}[l][l][1]{Secrecy Capacity with Perfect Main CSI}
\psfrag{Secrecy Rate with 1 bit feedback}[l][l][1]{Secrecy Rate with a 1-bit CSI Feedback}
\psfrag{Secrecy Rate with 2 bit feedback}[l][l][1]{Secrecy Rate with a 2-bit CSI Feedback}
\psfrag{Secrecy Rate with 4 bit feedback}[l][l][1]{Secrecy Rate with a 4-bit CSI Feedback}
\psfrag{Asymptotic expressions--------------------------------------}[l][l][1]{Asymptotic expressions in Corollary~\ref{CR6}}
\vspace{-0.5cm}
\begin{center}%
%\hspace{-0.4cm}
\scalebox{0.69}{\includegraphics{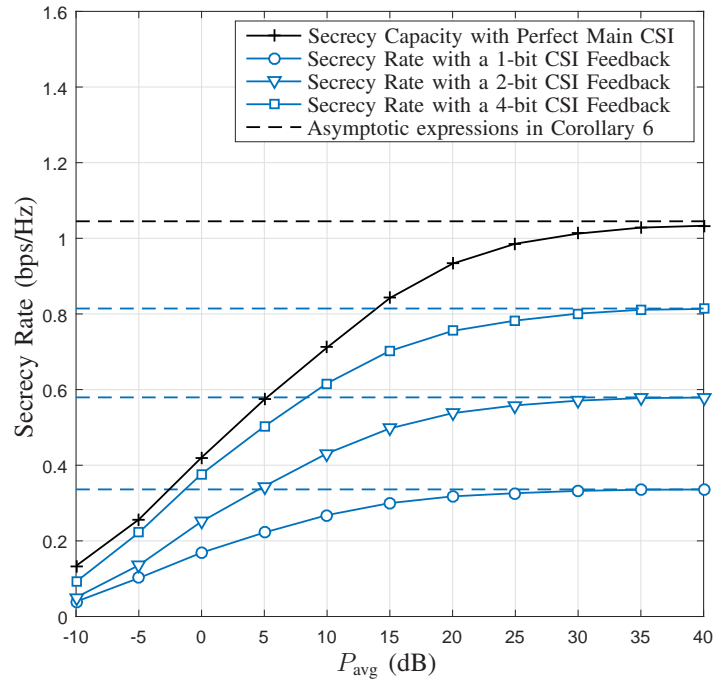}}
\end{center}
\vspace{-0.8cm}
\caption{Common message secrecy rate in Theorem~\ref{TH1} for Rayleigh fading channels with $K{=}3$.}
\label{fig:fig1}
\end{figure}

\begin{figure}[h!]
%\psfrag{}[l][l][1]{}
\psfrag{Secrecy Rate}[l][l][1.2]{\hspace{-1cm}Secrecy Sum-Rate (bps/Hz)}
\psfrag{Pavg}[l][l][1.2]{\hspace{-0.5cm}$P_\text{avg}$ (dB)}
\psfrag{PCSI capacity}[l][l][1]{Secrecy Sum-Capacity with Perfect Main CSI}
\psfrag{Secrecy Rate with 4 bit feedback}[l][l][1]{Secrecy Sum-Rate with a 4-bit CSI Feedback}
\psfrag{Asymptotic expressions---------------------------------------------}[l][l][1]{Asymptotic expressions in Corollary~\ref{CR7}}
\psfrag{K=3}[l][l][1]{$K{=}3$}
\psfrag{K=10}[l][l][1]{$K{=}10$}
\vspace{-0.4cm}
\begin{center}%
%\hspace{-0.4cm}
\scalebox{0.69}{\includegraphics{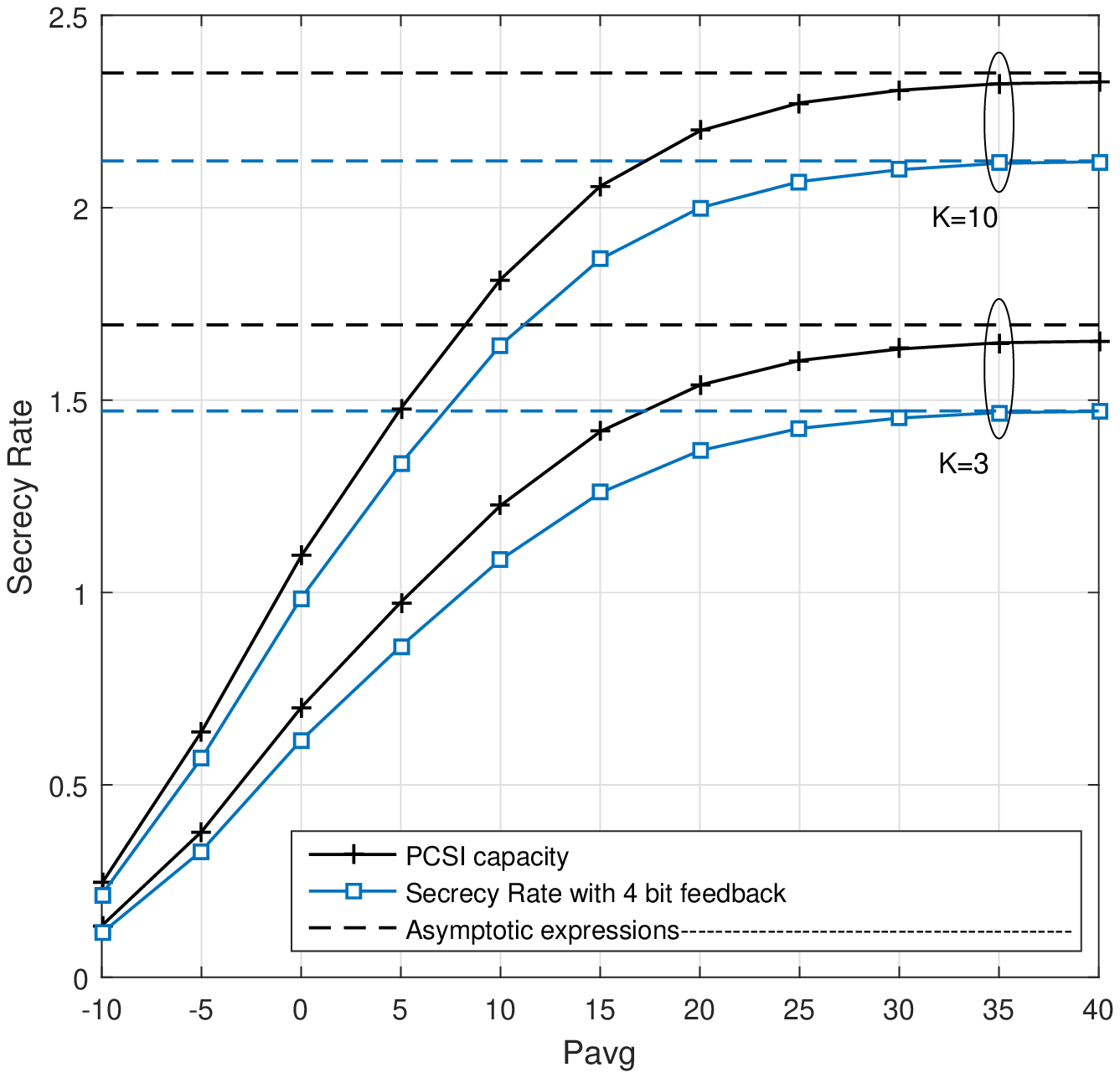}}
\end{center}
\vspace{-0.8cm}
\caption{Independent messages secrecy sum-rate in Theorem~\ref{TH2} for Rayleigh fading channels with $b{=}4$.}
\label{fig:fig2}
\end{figure}

\begin{figure}[ht]
%\psfrag{}[l][l][1]{}
\psfrag{Secrecy Rate}[l][l][1.2]{\hspace{-1cm}Secrecy Sum-Rate (bps/Hz)}
\psfrag{K}[l][l][1.2]{\hspace{-0.5cm}$K$ }
\psfrag{PCSI capacity}[l][l][1]{Secrecy Sum-Capacity with Perfect Main CSI}
\psfrag{Secrecy Rate with 1 bit feedback----------------------------------}[l][l][1]{Secrecy Sum-Rate with a 1-bit CSI Feedback}
\psfrag{Secrecy Rate with 2 bit feedback}[l][l][1]{Secrecy Sum-Rate with a 2-bit CSI Feedback}
\psfrag{Secrecy Rate with 3 bit feedback}[l][l][1]{Secrecy Sum-Rate with a 3-bit CSI Feedback}
\psfrag{Secrecy Rate with 4 bit feedback}[l][l][1]{Secrecy Sum-Rate with a 4-bit CSI Feedback}
\psfrag{K=3}[l][l][1]{$K{=}3$}
\psfrag{K=10}[l][l][1]{$K{=}10$}
\vspace{-0.5cm}
\begin{center}%
%\hspace{-0.4cm}
\scalebox{0.69}{\includegraphics{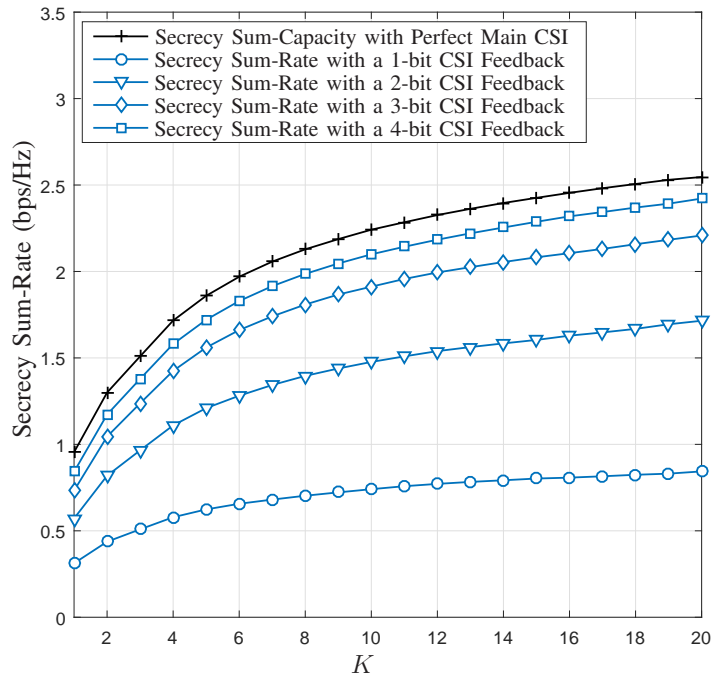}}
\end{center}
\vspace{-0.8cm}
\caption{Independent messages secrecy sum-rate in Theorem~\ref{TH2} for Rayleigh fading channels with $P_\text{avg}{=}20$ dB.}
\label{fig:fig3}
\end{figure}

\begin{figure}[h!]
%\psfrag{}[l][l][1]{}
\psfrag{Secrecy Rate}[l][l][1.2]{\hspace{-1cm}Secrecy Sum-Rate (bps/Hz)}
\psfrag{K}[l][l][1.2]{\hspace{-0.5cm}$K$ }
\psfrag{PCSI capacity}[l][l][1]{Secrecy Sum-Capacity with Perfect Main CSI}
\psfrag{Secrecy Rate with 3 bit feedback----------------------------------}[l][l][1]{Secrecy Sum-Rate with a 3-bit CSI Feedback}
\psfrag{Pavg=0dB}[l][l][1]{\hspace{-0.3cm}$P_\text{avg}{=}0$ dB}
\psfrag{Pavg=5dB}[l][l][1]{\hspace{-0.3cm}$P_\text{avg}{=}5$ dB}
\psfrag{Pavg=20dB}[l][l][1]{\hspace{-0.3cm}$P_\text{avg}{=}20$ dB}
\psfrag{K=3}[l][l][1]{$K{=}3$}
\psfrag{K=10}[l][l][1]{$K{=}10$}
\vspace{-0.4cm}
\begin{center}%
%\hspace{-0.4cm}
\scalebox{0.69}{\includegraphics{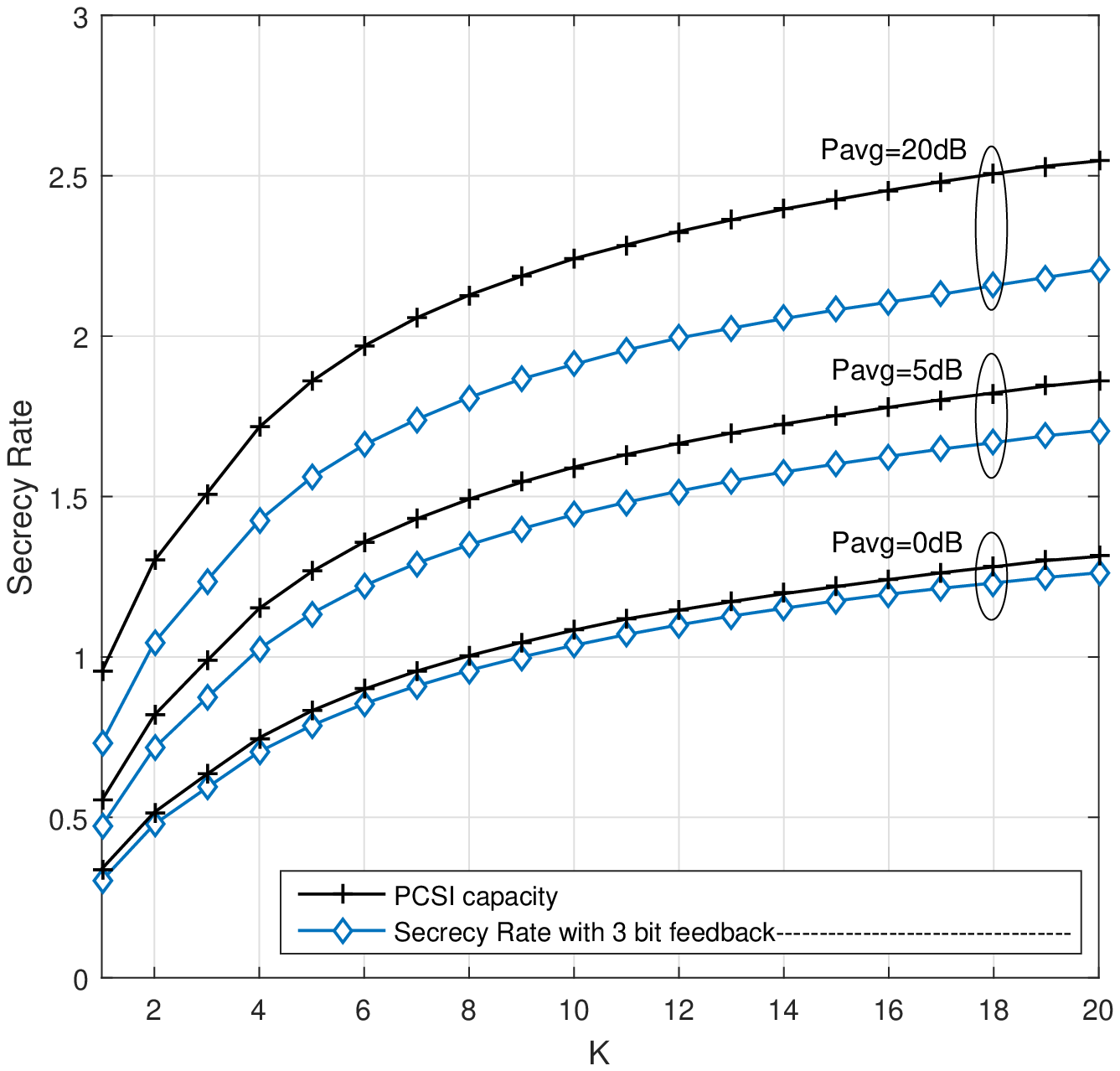}}
\end{center}
\vspace{-0.8cm}
\caption{Independent messages secrecy sum-rate in Theorem~\ref{TH2} for Rayleigh fading channels with $b{=}3$.}
\label{fig:fig4}
\end{figure}

Figure~\ref{fig:fig1} illustrates the common message achievable secrecy rate $\mathcal{C}_\text{s}^-$, presented in Theorem~\ref{TH1}, with $K{=}3$ and various $b$-bit feedback, $b{=}1,2,4.$ The secrecy capacity $\mathcal{C}_\text{s}$, from Corollary~\ref{CR1}, is also presented as a benchmark. It represents the secrecy capacity with full main CSI at the transmitter. We can see that as the capacity of the feedback link grows, i.e., the number of bits~$b$ increases, the achievable rate grows toward the secrecy capacity $\mathcal{C}_\text{s}$. The asymptotic expressions in Corollary~\ref{CR6} are also illustrated and show the boundedness of the secrecy throughput at high-SNR. The same observations can be made for the independent messages case; illustrated in figure~\ref{fig:fig2}. Two scenarios are considered here; the transmission of three independent messages to three legitimate receivers, $K{=}3$, and the transmission of ten independent messages with $K{=}10$. Both the achievable secrecy sum-rate in Theorem~\ref{TH2} and the secrecy sum-capacity in Corollary~\ref{CR2} are depicted. The impact of changing the number of legitimate receivers $K$ on the secrecy sum-rate is illustrated in Figures~\ref{fig:fig3}~and~\ref{fig:fig4} for different values of the average power constraint~$P_\text{avg}$ and of the number of feedback bits~$b$. We can see from these two figures that the secrecy throughput of the system, when broadcasting multiple messages, increases with~$K$.

%%%%%%%%%%%%%%%%%%%%%%%%%%%%%%%%%%%%%%%%%%%%%%%%%%%%%%%%%%%%%%%%%
%%%%%%%%%%%%%%%%%%%%%%%%%%%%%%%%%%%%%%%%%%%%%%%%%%%%%%%%%%%%%%%%%

%%%%%%%%%%%%%%%%%%%%%%%%%%%%%%%%%%%%%%%%%%%%%%%%%%%%%%%%%%%%%%%%%
%%%%%%%%%%%%%%%%%%%%%%%%%%%%%%%%%%%%%%%%%%%%%%%%%%%%%%%%%%%%%%%%%

%%%%%%%%%%%%%%%%%%%%%%%%%%%%%%%%%%%%%%%%%%%%%%%%%%%%%%%%%%%%%%%%%
%%%%%%%%%%%%%%%%%%%%%%%%%%%%%%%%%%%%%%%%%%%%%%%%%%%%%%%%%%%%%%%%%

%%%%%%%%%%%%%%%%%%%%%%%%%%%%%%%%%%%%%%%%%%%%%%%%%%%%%%%%%%%%%%%%%
%%%%%%%%%%%%%%%%%%%%%%%%%%%%%%%%%%%%%%%%%%%%%%%%%%%%%%%%%%%%%%%%%

%%%%%%%%%%%%%%%%%%%%%%%%%%%%%%%%%%%%%%%%%%%%%%%%%%%%%%%%%%%%%
\begin{figure}[t!]
%\psfrag{}[l][l][1]{}
\psfrag{R0}[l][l][1.2]{\hspace{-0.6cm}$\mathcal{R}_0$ (bps/Hz)}
\psfrag{R1}[l][l][1.2]{\hspace{-0.6cm}$\mathcal{R}_1$ (bps/Hz)}
\psfrag{PCSI with s2=1}[l][l][1]{Perfect CSIT case with $\sigma_\text{e}{=}1$}
\psfrag{PCSI with s2=0.5}[l][l][1]{Perfect CSIT case with \hspace{0.1cm}$\sigma_\text{e}{=}\sqrt{0.5}$}
\psfrag{1 bit feedback with s2=1}[l][l][1]{1-bit error-free feedback with \hspace{0.1cm}$\sigma_\text{e}{=}1$}
\psfrag{1 bit feedback with s2=0.5--------------------------------------}[l][l][1]{1-bit error-free feedback with $\sigma_\text{e}{=}\sqrt{0.5}$}
\begin{center}\vspace{-1cm}
\scalebox{0.69}{\includegraphics{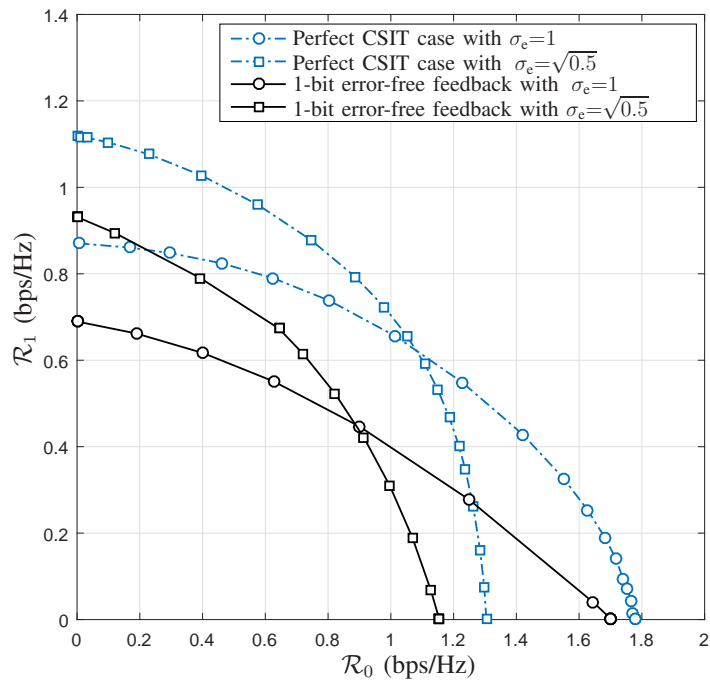}}
\end{center}\vspace{-1cm}
\caption{Secrecy capacity regions in Theorem~\ref{TH3} for Rayleigh BCCM with an error-free CSI feedback.}
\label{fig:Fig5}\vspace{-0.5cm}
\end{figure}
%%%%%%%%%%%%%%%%%%%%%%%%%%%%%%%%%%%%%%%%%%%%%%%%%%%%%%%%%%%%%

Figure~\ref{fig:Fig5} illustrates the secrecy capacity region for the BCCM with a 1-bit CSI feedback sent over an error-free link, presented in Theorem~\ref{TH3}. In this figure, we take $K{=}1$, $h_1{\sim}\mathcal{CN}(0,1)$, $h_\text{e}{\sim}\mathcal{CN}(0,\sigma_\text{e}^2)$, and $P_\text{avg}{=}5~\!\text{dB}$. The boundary of the secrecy capacity region when perfect CSI is available at the transmitter is also presented as a benchmark. We can see that even with a 1-bit CSI feedback, a positive secrecy rate is achieved, and that when the channel to receiver $\text{R}_1$ is better than the channel to receiver R/E, i.e., when $\sigma_\text{e}^2{=}0.5$, the confidential rate $\mathcal{R}_1$ improves while the common rate $\mathcal{R}_0$ decreases. 

\begin{figure}[t!]
%\psfrag{}[l][l][1]{}
\psfrag{R0}[l][l][1.2]{\hspace{-0.6cm}$\mathcal{R}_0$ (bps/Hz)}
\psfrag{R1}[l][l][1.2]{\hspace{-0.6cm}$\mathcal{R}_1$ (bps/Hz)}
\psfrag{PCSIT}[l][l][1]{Perfect CSIT case}
\psfrag{error free feedback---------------------------------------------------}[l][l][1]{1-bit feedback sent over an error-free link}
\psfrag{BEC with e=0.2}[l][l][1]{1-bit feedback sent over a BEC with $\epsilon{=}0.2$}
\psfrag{BEC with e=0.5}[l][l][1]{1-bit feedback sent over a BEC with $\epsilon{=}0.5$}
\psfrag{BEC with e=0.8}[l][l][1]{1-bit feedback sent over a BEC with $\epsilon{=}0.8$}
\begin{center}\vspace{-0.5cm}
\scalebox{0.69}{\includegraphics{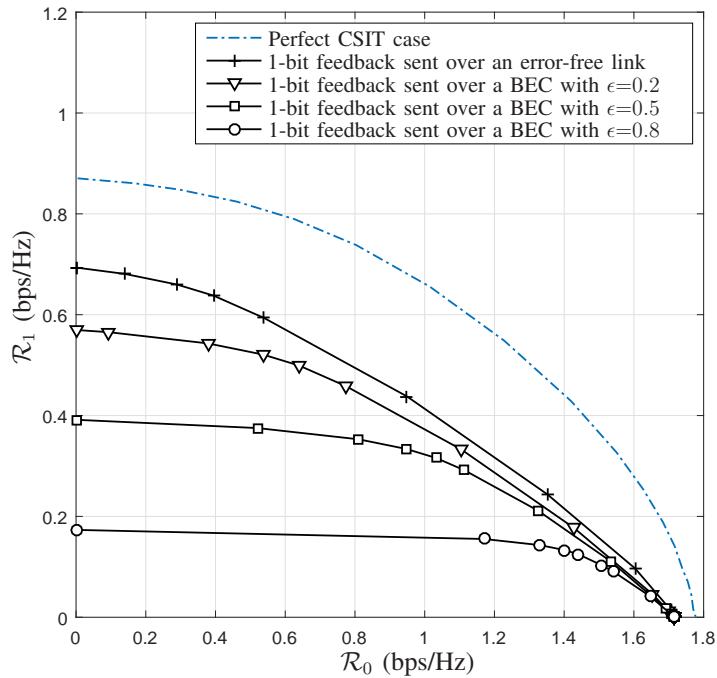}}
\end{center}\vspace{-0.8cm}
\caption{Secrecy capacity regions in Corollary~\ref{CR4} for Rayleigh BCCM with a binary erasure feedback link.}
\label{fig:Fig6}
\end{figure}

\begin{figure}[h!]
%\psfrag{}[l][l][1]{}
\psfrag{R0}[l][l][1.2]{\hspace{-0.5cm}$\mathcal{R}_0$ (bps/Hz)}
\psfrag{R1}[l][l][1.2]{\hspace{-0.5cm}$\mathcal{R}_1$ (bps/Hz)}
\psfrag{Pavg=10dB}[l][l][0.9]{\hspace{-0.1cm}$P_\text{avg}{=}10$ dB}
\psfrag{Pavg=20dB}[l][l][0.9]{\hspace{-0.1cm}$P_\text{avg}{=}20$ dB}
\psfrag{Pavg=5dB}[l][l][0.9]{\hspace{-0.1cm}$P_\text{avg}{=}5$ dB}
\psfrag{error free feedback----------------------------------------------}[l][l][1]{1-bit feedback sent over an error-free link}
\psfrag{BEC with e=0.2}[l][l][1]{1-bit feedback sent over a BEC with $\epsilon{=}0.2$}
\psfrag{BEC with e=0.8}[l][l][1]{1-bit feedback sent over a BEC with $\epsilon{=}0.8$}
\begin{center}\vspace{-0.5cm}
\scalebox{0.69}{\includegraphics{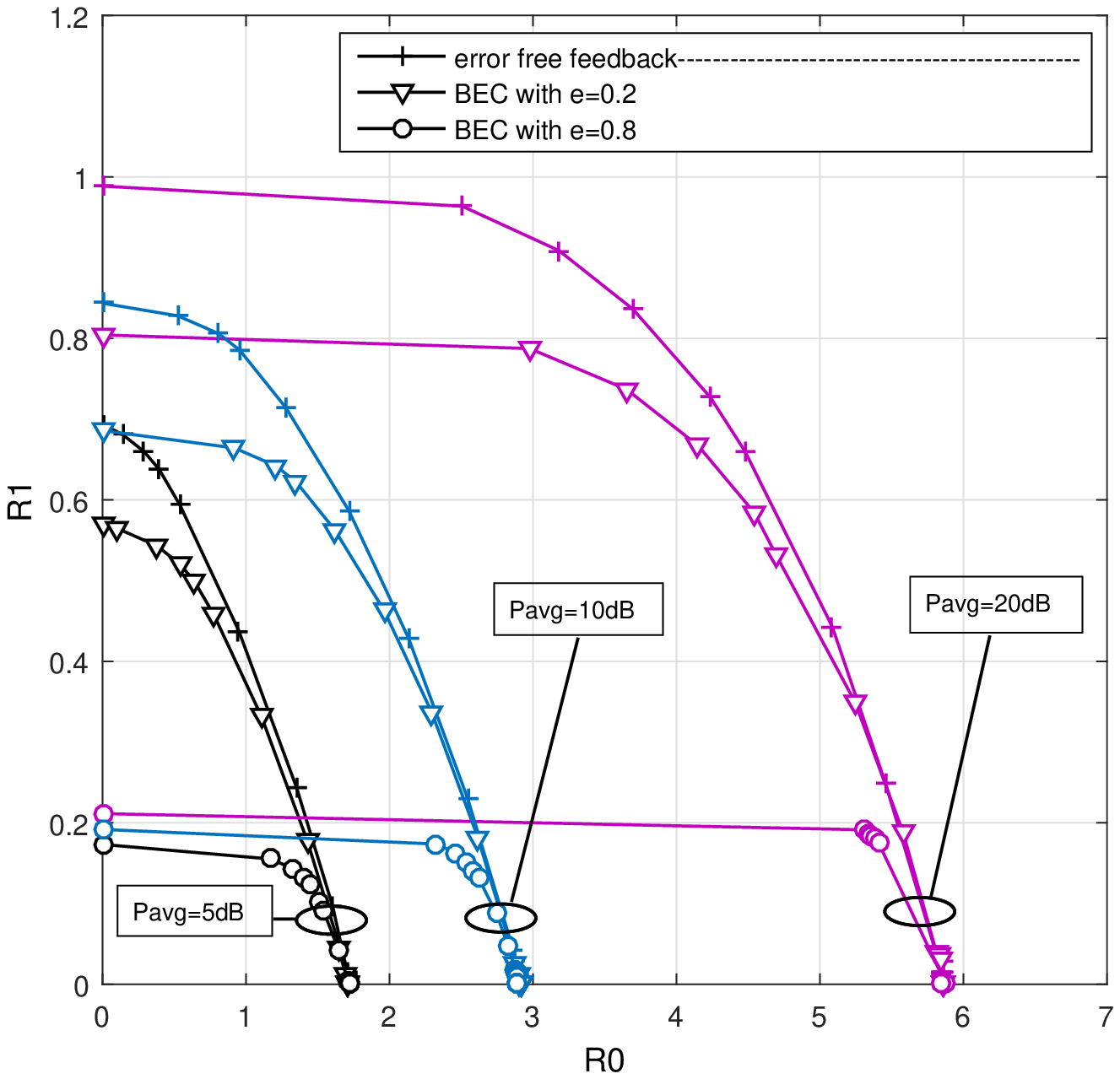}}
\end{center}\vspace{-0.8cm}
\caption{Secrecy capacity regions in Corollary~\ref{CR4} for Rayleigh BCCM with a 1-bit CSI feedback.}
\label{fig:Fig7}
\end{figure}

\begin{figure}[h!]
%\psfrag{}[l][l][1.5]{}
\psfrag{e=0.8 - b= 1,2,3 - P=10,20}[l][l][1]{}
\psfrag{R0}[l][l][1.2]{\hspace{-0.6cm}$\mathcal{R}_0$ (bps/Hz)}
\psfrag{R1}[l][l][1.2]{\hspace{-0.6cm}$\mathcal{R}_1$ (bps/Hz)}
\psfrag{Pavg=10dB}[l][l][0.9]{$P_\text{avg}{=}10$ dB}
\psfrag{Pavg=20dB}[l][l][0.9]{$P_\text{avg}{=}20$ dB}
\psfrag{error free feedback--------------------}[l][l][1]{Error-free feedback link}
\psfrag{BEC with e=0.8 and b=1}[l][l][1]{BEC with $\epsilon{=}0.8$ and $b{=}1$}
\psfrag{BEC with e=0.8 and b=2}[l][l][1]{BEC with $\epsilon{=}0.8$ and $b{=}2$}
\psfrag{BEC with e=0.8 and b=3}[l][l][1]{BEC with $\epsilon{=}0.8$ and $b{=}3$}
\begin{center}\vspace{-1cm}
\scalebox{0.69}{\includegraphics{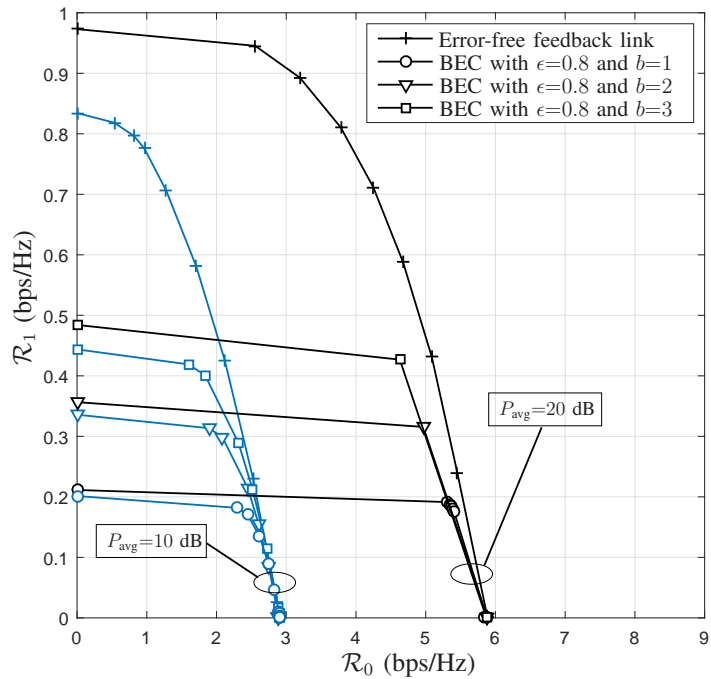}}
\end{center}\vspace{-0.8cm}
\caption{Secrecy capacity regions in Corollary~\ref{CR5} for Rayleigh BCCM with a $b$-bit CSI feedback.}\vspace{-0.3cm}
\label{fig:Fig8}
\end{figure}

The impact of having a binary erasure feedback link, on the achievable secrecy rate region, is  illustrated in Figure~\ref{fig:Fig6}, along with the boundaries on the secrecy capacity regions for the error-free feedback case and the perfect CSIT case, with $K{=}1$, $h_1{\sim}\mathcal{CN}(0,1)$, $h_\text{e}{\sim}\mathcal{CN}(0,1)$, $P_\text{avg}{=}5~\!\text{dB}$, and different values of the erasure probability $\epsilon{=}0.2,0.5, \text{and}~\!0.8$. As expected, we can see that the confidential rate $\mathcal{R}_1$ decreases as the probability of erasure increases since the transmission of the confidential message will be restricted, not only by the channel quality but also by the reception of a not erased feedback. The transmission of the common message solely is not affected by the erasure of the feedback information. Besides, we can see from Figure~\ref{fig:Fig7} that when the erasure probability is high, i.e., $\epsilon{=}0.8$, the confidential rate does not improve much even when we increase the average power constraint from $P_\text{avg}{=}5$ dB to $P_\text{avg}{=}20$ dB. Moreover, from Figure~\ref{fig:Fig8}, we can see that the secrecy rate can be significantly improved by using more redundant feedback bits.

%%%%%%%%%%%%%%%%%%%%%%%%%%%%%%%%%%%%%%%%%%%%%%%%%%%%%%%%%%%%%
%%%%%%%%%%%%%%%%%%%%%%%%%%%%%%%%%%%%%%%%%%%%%%%%%%%%%%%%%%%%%

%%%%%%%%%%%%%%%%%%%%%%%%%%%%%%%%%%%%%%%%%%%%%%%%%%%%%%%%%%%%%
%%%%%%%%%%%%%%%%%%%%%%%%%%%%%%%%%%%%%%%%%%%%%%%%%%%%%%%%%%%%%

%%%%%%%%%%%%%%%%%%%%%%%%%%%%%%%%%%%%%%%%%%%%%%%%%%%%%%%%%%%%%
%%%%%%%%%%%%%%%%%%%%%%%%%%%%%%%%%%%%%%%%%%%%%%%%%%%%%%%%%%%%%

%-------------------------------------------------------------------------------------------------------------
\section{Conclusion}\label{conclusion}
%-------------------------------------------------------------------------------------------------------------
Knowing the CSIT in a wiretap communication system is critical since the core idea behind information theoretic security is to exploit the randomness of the channel variations to achieve secrecy at the physical layer. In this work, we studied the impact of having an uncertain CSIT, obtained through a finite-rate feedback link, on the secrecy throughput of multi-user broadcast wiretap channels. We considered the cases of common message transmission, independent messages transmission, and the broadcast channel with confidential messages. The obtained results, in all three cases, show that even with a 1-bit CSI feedback, a positive secrecy rate can still be achieved. It goes without saying that the more the transmitter knows, the better the secrecy throughput is. 

Among the open challenges that could be considered as an extension of this work, one can examine the case when no assumption on the adversary's CSI is imposed, i.e., when even the statistics of the wiretapper's channel are not available at the transmitter. An interesting approach, in this case, could be found in the framework of arbitrary varying eavesdropper channel. Designing practical codes for fading wiretap channels is another open issue facing information theoretic security either with perfect or uncertain CSIT.

%------------------------------------------
\appendices
%------------------------------------------
%-------------------------------------------------------------------------------------------------------------
\section{Proof of Theorem \ref{TH1}}\label{App_TH1} 
%-------------------------------------------------------------------------------------------------------------
In this appendix, we provide a detailed proof of Theorem~\ref{TH1}. We start with the proof of achievability for the ergodic common message secrecy rate~\eqref{th1_a} followed by the proof of the upper bound~\eqref{th1_b}. We recall that, in this scenario, each legitimate receiver uses a $b$-bit feedback link to inform the transmitter about its CSI, and that all the legitimate receivers are interested in the same common secure information. 
%------------------------------------------
\subsection{Proof of Achievability}\label{App_TH1_LB} 
%------------------------------------------
Since the transmission is controlled by the feedback information, we consider that, during each fading block, if the main channel gain of the receiver with the weakest average channel gain falls within the interval $[\tau_q,\tau_{q+1})$, $q\in\{1,{\cdots},Q\}$, the transmitter conveys the codewords at rate $\mathcal{R}_q=\log\left(1{+}\tau_qP_q\right).$ Rate $\mathcal{R}_q$ changes only periodically and is held constant over the duration interval of a fading block. %This setup guarantees that when $\gamma_\text{e}{>}\tau_q$, the mutual information between the transmitter and the eavesdropper is upper bounded by $\mathcal{R}_q$. Otherwise, this mutual information will be $\log\left(1{+}\gamma_\text{e}P_q\right)$.
Besides, we adopt a probabilistic transmission model where the communication is constrained by the quality of the legitimate channels. Given the reconstruction points, $\tau_1{<}\cdots{<}\tau_Q{<}\tau_{Q+1}{=}\infty$, describing the support of each channel gain $\gamma_k, k\in\{1,\cdots ,K\}$, and since the channel gains of the $K$ receivers are independent, there are $M{=}Q^K$ different states for the received feedback information. Each of these states, $\mathcal{J}_m, m\in\{1,\cdots ,M\},$ represents one subchannel. The transmission scheme consists on transmitting an independent codeword, on each of the $M$ subchannels, with a fixed rate. We define the following rates
\begin{align}
\mathcal{R}_\text{s}^-=\min_{1\leq k\leq K}\sum_{m=1}^M\text{Pr}\left[\mathcal{J}_m\right]\underset{\gamma_\text{e}}{\mathbb{E}}\left[\left\lbrace\log\left(\frac{1{+}\tau_{k,m}P_m}{1{+}\gamma_\text{e}P_m}\right)\right\rbrace^+\right],
\end{align}
and $\displaystyle{\mathcal{R}_{\text{e},m}=\underset{\gamma_\text{e}}{\mathbb{E}}\left[\log\left(1{+}\gamma_\text{e}P_m\right)\right],}$
where $\tau_{k,m}$ is the quantized channel gain corresponding to the receiver with the weakest average SNR in state $\mathcal{J}_m$ and $P_m$ is the associated power policy satisfying the average power constraint. 

\textit{Codebook Generation:}
We construct $M$ independent codebooks $\mathcal{C}_1$, $\cdots$, $\mathcal{C}_M$, one for each subchannel, constructed similarly to the standard wiretap codes. Each codebook $\mathcal{C}_m$ is a $(n,2^{n\mathcal{R}_\text{s}^-})$ code with $2^{n(\mathcal{R}_\text{s}^-{+}\mathcal{R}_{\text{e},m})}$ codewords randomly partitioned into~$2^{n\mathcal{R}_\text{s}^-}$ bins.

\textit{Encoding and Decoding:}
Given a particular common message $w{\in}\{1,2,{\cdots},2^{n\mathcal{R}_\text{s}^-}\}$, to be transmitted, the encoder selects $M$ codewords, one for each subchannel. More specifically, if the message to be sent is $w$, then for each subchannel $m$, the encoder randomly selects one of the codewords $U_m^n$ from the $w$th bin in $\mathcal{C}_m$. During each fading block, of length $\kappa$, the transmitter experiences one of the events $\mathcal{J}_m$. Depending on the encountered channel state, the transmitter broadcasts $\kappa\mathcal{R}_q$ information bits of $U_m^{n}$ using a Gaussian codebook. By the weak law of large numbers, when the total number of fading blocks $L$ is large, the entire binary sequences are transmitted with high probability. To decode, each legitimate receiver considers the observations corresponding to all $M$ subchannels. And since the transmission is adapted with regard to the receiver with the weakest average SNR, all legitimate receivers can recover the transmitted codewords, with high probability, and hence recover message~$w$. Details on the error probability evaluation are similar to the parallel channels case~\cite{Ashish_J3}.
Since $\tau_{k,m}\in\{\tau_1,{\cdots},\tau_Q\}$, by rewriting the summation over the states of each legitimate receiver, the expression of $\mathcal{R}_\text{s}^-$ can then be reformulated as\vspace{0.2cm}
\begin{align}
\mathcal{R}_\text{s}^-&=\hspace{-0.2cm}\min_{1\leq k\leq K}\sum_{m=1}^M\text{Pr}\left[\mathcal{J}_m\right]\underset{\gamma_\text{e}}{\mathbb{E}}\left[\left\lbrace\log\left(\frac{1{+}\tau_{k,m}P_m}{1{+}\gamma_\text{e}P_m}\right)\right\rbrace^+\right]\label{stp1}\\
&=\hspace{-0.2cm}\min_{1\leq k\leq K}\sum_{m=1}^M\sum_{q=1}^Q\text{Pr}\left[\mathcal{J}_m,\tau_{k,m}{=}\tau_q\right]\underset{\gamma_\text{e}}{\mathbb{E}}\left[\left\lbrace\log\left(\frac{1{+}\tau_qP_q}{1{+}\gamma_\text{e}P_q}\!\right)\!\right\rbrace^{\hspace{-0.1cm}+}\right]\label{stp2}\\
&=\hspace{-0.2cm}\min_{1\leq k\leq K}\sum_{q=1}^Q\text{Pr}\left[\tau_q{\leq}\gamma_k{<}\tau_{q+1}\right]\underset{\gamma_\text{e}}{\mathbb{E}}\!\left[\!\left\lbrace\!\log\!\left(\!\frac{1{+}\tau_qP_q}{1{+}\gamma_\text{e}P_q}\!\right)\!\right\rbrace^{\hspace{-0.1cm}+}\right],\label{stp3}
\end{align}
where \eqref{stp2} is obtained by noting that $\tau_{k,m}\in\{\tau_1,{\cdots},\tau_Q\}$ and applying the total probability theorem, and \eqref{stp3} comes from the fact that $$\sum_{m=1}^M\text{Pr}\left[\mathcal{J}_m,\tau_{k,m}{=}\tau_q\right]=\text{Pr}\left[\tau_q{\leq}\gamma_k{<}\tau_{q+1}\right].$$

Since each user gets to know the feedback information of the other legitimate receivers, our proof is also valid when the reconstruction points $\{\tau_q\}_{q=1}^Q$ and the transmission strategies $\{P_q\}_{q=1}^Q$, associated with each legitimate receiver, are different. That is, we can choose these quantization parameters to satisfy~\eqref{th1_a}. 

\allowdisplaybreaks
\textit{Secrecy Analysis:} 
We need to prove that the equivocation rate satisfies $\mathcal{R}_\text{e}\geq\mathcal{R}_\text{s}^--\epsilon$. Let $\Gamma^L{=}\left\{\gamma_1^L,\gamma_2^L,{\cdots},\gamma_K^L\right\}$
and $F^L{=}\left\{F_1^L,F_2^L,{\cdots},F_K^L\right\}$, with $F_k(l)\in\{\tau_1,{\cdots},\tau_Q\}$ being the feedback information sent by receiver $k$ in the $l$-th fading block. We have
\begin{align}
n\mathcal{R}_\text{e}&=H(W|Y_\text{e}^n,\gamma_\text{e}^L,\Gamma^L,F^L)\\
&\geq I(W;X^n|Y_\text{e}^n,\gamma_\text{e}^L,\Gamma^L,F^L)\\
&=H(X^n|Y_\text{e}^n\!,\!\gamma_\text{e}^L\!,\!\Gamma^L\!,\!F^L){-}H(X^n|Y_\text{e}^n\!,\!\gamma_\text{e}^L\!,\!\Gamma^L\!,\!F^L\!,\!W).\label{lb1_step0}
\end{align}  

%------------------------

On one hand, we can write
\begin{align}
&H(X^n|Y_\text{e}^n,\gamma_\text{e}^L,\Gamma^L,F^L)=\sum_{l=1}^LH(X^\kappa(l)|Y_\text{e}^\kappa(l)\!,\!\gamma_\text{e}(l)\!,\!\gamma_1(l)\!,\!{\cdots}\!,\!\gamma_K(l)\!,\!F_1^L\!,\!{\cdots}\!,\!F_K^L)\label{lb1_step1}\\
&=\sum_{l=1}^L\sum_{j=1}^\kappa H(X(l,j)|Y_\text{e}^\kappa(l)\!,\!\gamma_\text{e}(l)\!,\!\gamma_1(l)\!,\!{\cdots}\!,\!\gamma_K(l)\!,\!F_1^L\!,\!{\cdots}\!,\!F_K^L)\\
&=\sum_{l=1}^L\sum_{j=1}^\kappa H(X(l,j)|\gamma_\text{e}(l)\!,\!\gamma_1(l)\!,\!{\cdots}\!,\!\gamma_K(l)\!,\!F_1^L\!,\!{\cdots}\!,\!F_K^L)\nonumber\\
&\hspace{3cm}-I(X(l,j);Y_\text{e}^\kappa(l)|\gamma_\text{e}(l)\!,\!\gamma_1(l)\!,\!{\cdots}\!,\!\gamma_K(l)\!,\!F_1^L\!,\!{\cdots}\!,\!F_K^L)\\
&=\sum_{l=1}^L\sum_{j=1}^\kappa H(X(l,j)|F_1(l)\!,\!{\cdots}\!,\!F_K(l)){-}I(X(l,j);Y_\text{e}^\kappa(l)|\gamma_\text{e}(l)\!,\!F_1(l)\!,\!{\cdots}\!,\!F_K(l))\label{lb1_step1_2}\\
&\geq\!\sum_{l\in\mathcal{D}_L}\sum_{j=1}^\kappa H(X(l,j)|F_1(l)\!,\!{\cdots}\!,\!F_K(l)){-}I(X(l,j);Y_\text{e}^\kappa(l)|\gamma_\text{e}(l)\!,\!F_1(l)\!,\!{\cdots}\!,\!F_K(l)),\label{lb1_step2}
\end{align}
where \eqref{lb1_step1} results from the memoryless property of the channel and the independence of the $X^\kappa(l)$'s, \eqref{lb1_step1_2} comes from the adopted transmission scheme, where the signal transmitted during a given fading block only depends on the feedback information sent during that block, and~\eqref{lb1_step2} is obtained by removing all the terms corresponding to the fading blocks $l\not\in\mathcal{D}_L$, with $\mathcal{D}_L=\cup_{k\in\{1,{\cdots},K\}}\left\lbrace l\in\{1,{\cdots},L\}:F_k(l)>h_\text{e}(l)\right\rbrace$.

Now, taking into account the adopted coding and transmission schemes, presented in details at the beginning of this subsection, we have
\begin{align}\label{lb1_step2_1}
H(X(l,j)|F_1(l)\!,\!{\cdots}\!,\!F_K(l))&=\sum_{q=1}^Q\text{Pr}\left[\tau_q{\leq}\gamma_{k^*}(l){<}\tau_{q+1}\right]H(X(l,j)|F_{k^*}(l)=q)\\
&=\sum_{q=1}^Q\text{Pr}\left[\tau_q{\leq}\gamma_{k^*}(l){<}\tau_{q+1}\right]\mathcal{R}_q,
\end{align}
with $k^*$ being the index of the legitimate receiver with the worst average channel gain, and
\begin{align}\label{lb1_step2_2}
&I(X(l,j);Y_\text{e}^\kappa(l)|\gamma_\text{e}(l)\!,\!F_1(l)\!,\!{\cdots}\!,\!F_K(l))\nonumber\\
&\hspace{2cm}=\sum_{q=1}^Q\text{Pr}\left[\tau_q{\leq}\gamma_{k^*}(l){<}\tau_{q+1}\right]I(X(l,j);Y_\text{e}^\kappa(l)|\gamma_\text{e}(l)\!,\!F_{k^*}(l)=q)\\
&\hspace{2cm}=\sum_{q=1}^Q\text{Pr}\left[\tau_q{\leq}\gamma_{k^*}(l){<}\tau_{q+1}\right]\log\left(1{+}\gamma_\text{e}(l)P_q\right)
\end{align}
Substituting \eqref{lb1_step2_1} and \eqref{lb1_step2_2} in \eqref{lb1_step2}, we get
\begin{align}
&H(X^n|Y_\text{e}^n,\gamma_\text{e}^L,\Gamma^L,F^L)\geq\!\sum_{l\in\mathcal{D}_L}\sum_{j=1}^\kappa\left(\sum_{q=1}^Q\text{Pr}\left[\tau_q{\leq}\gamma_{k^*}(l){<}\tau_{q+1}\right]\left(\mathcal{R}_q{-}\log\left(1{+}\gamma_\text{e}(l)P_q\right)\right){-}\epsilon'\!\right)\\
&\hspace{2.5cm}= \!\sum_{l\in\mathcal{D}_L}\kappa\!\left(\sum_{q=1}^Q\text{Pr}\left[\tau_q{\leq}\gamma_{k^*}(l){<}\tau_{q+1}\right]\left(\mathcal{R}_q{-}\log\left(1{+}\gamma_\text{e}(l)P_q\right)\right){-}\epsilon'\!\right)\\
&\hspace{2.5cm}=\!\kappa\sum_{l=1}^L\left(\sum_{q=1}^Q\text{Pr}\left[\tau_q{\leq}\gamma_{k^*}(l){<}\tau_{q+1}\right]\!\left\lbrace \!\mathcal{R}_q\!-\!\log\!\left(1{+}\gamma_\text{e}(l)P_q\right)\!\right\rbrace^{+}\hspace{-0.1cm}\!-\!\epsilon'\!\right)\\
&\hspace{2.5cm}=n\sum_{q=1}^Q\text{Pr}\left[\tau_q{\leq}\gamma_{k^*}(l){<}\tau_{q+1}\right]\underset{\gamma_\text{e}}{\mathbb{E}}\left[\left\lbrace \mathcal{R}_q{-}\log\left(1{+}\gamma_\text{e}P_q\right)\right\rbrace^+\right]-n\epsilon'\label{lb1_step3}\\
&\hspace{2.5cm}=n\mathcal{R}_\text{s}^- -n\epsilon',\label{lb1_step3p1}
\end{align}
where \eqref{lb1_step3} follows from the ergodicity of the channel as $L\rightarrow\infty$. 

On the other hand, using a list decoding argument at the eavesdropper side and applying Fano's inequality~\cite{Gopala_J1}, $\frac{1}{n}H(X^n|Y_\text{e}^n,\gamma_\text{e}^L,\Gamma^L,F^L,W)$ vanishes as $n\rightarrow\infty$ and we can write
\begin{equation}\label{lb1_step4}
H(X^n|Y_\text{e}^n,\gamma_\text{e}^L,\Gamma^L,F^L,W)\leq n\epsilon''.
\end{equation}
Substituting \eqref{lb1_step3p1} and \eqref{lb1_step4} in \eqref{lb1_step0}, we get $\mathcal{R}_\text{e}\geq \mathcal{R}_\text{s}^- -\epsilon$, with $\epsilon=\epsilon'+\epsilon''$, and $\epsilon'$ and $\epsilon''$ are selected to be arbitrarily small. 
This concludes the proof. $\hfill \square$
%------------------------------------------
\subsection{Proof of the Upper Bound}\label{App_TH1_UB} 
%------------------------------------------
To establish the upper bound on the common message secrecy capacity in~(\ref{th1_b}), we start by supposing that the transmitter sends message $w$ to only one legitimate receiver $\text{R}_k$. Using the result in \cite{Rezki_J1}, for single user transmission with limited CSI feedback, the secrecy capacity of our system can be upper bounded as 
\begin{equation}\label{UB_CM_stp1}
\mathcal{C}_{\text{s}}\!\leq\!\hspace{-0.2cm}\max_{\{\tau_q;P_q\}_{q=1}^Q}\sum_{q=0}^Q\text{Pr}\left[\tau_q{\leq}\gamma_k{<}\tau_{q+1}\right]\underset{\gamma_\text{e},\gamma_k}{\mathbb{E}}\!\left[\!\left\lbrace\!\log\!\left(\!\frac{1{+}\gamma_kP_q}{1{+}\gamma_\text{e}P_q}\!\right)\!\right\rbrace^{\hspace{-0.1cm}+}\!\Bigg|\tau_q\!\leq\!\gamma_k\!<\!\tau_{q+1}\!\right].
\end{equation}

\noindent Since the choice of the receiver to transmit to is arbitrary, we tighten this upper bound by choosing the legitimate receiver $\text{R}_k$ that minimizes this quantity, yielding 
\begin{align}\label{UB_CM_stp2}
\mathcal{C}_{\text{s}}^+&=\min_{1\leq k\leq K}\max_{\{\tau_q;P_q\}_{q=1}^Q}\sum_{q=0}^Q\text{Pr}\left[\tau_q{\leq}\gamma_k{<}\tau_{q+1}\right]\underset{\gamma_\text{e},\gamma_k}{\mathbb{E}}\!\left[\!\left\lbrace\!\log\!\left(\!\frac{1{+}\gamma_kP_q}{1{+}\gamma_\text{e}P_q}\!\right)\!\right\rbrace^{\hspace{-0.1cm}+}\!\Bigg|\tau_q\!\leq\!\gamma_k\!<\!\tau_{q+1}\!\right].
\end{align}
This concludes the proof. $\hfill \square$

%-------------------------------------------------------------------------------------------------------------
\section{Proof of Theorem \ref{TH2}}\label{App_TH2} 
%-------------------------------------------------------------------------------------------------------------
Here we provide a detailed proof of Theorem~\ref{TH2}. We start with the proof of achievability for the ergodic secrecy sum-rate~\eqref{th2_a} followed by the proof of the upper bound~\eqref{th2_b}. In this case, each legitimate receiver is interested in a different independent message, and all messages need to be kept secret from the eavesdropper. 
%------------------------------------------
\subsection{Proof of Achievability}\label{App_TH2_LB} 
%------------------------------------------
The lower bound on the secrecy sum-capacity, presented in \eqref{th2_a}, is achieved using a time division multiplexing scheme that selects periodically one receiver to transmit to. More specifically, we consider that, during each fading block, the source only transmits to the legitimate receiver with the highest $\tau_q$, and if there are more than one, we choose one of them randomly. Since we are transmitting to only one legitimate receiver at a time, the achieving coding scheme consists on using independent standard single user Gaussian wiretap codebooks.  

During each fading block, the transmitter receives~$K$ feedback information about the CSI of the legitimate receivers. Since the channel gains of the $K$ receivers are independent, there are $M{=}Q^K$ different states for the received feedback information, as discussed in the proof of achievability of Theorem 1. Each of these states, $\mathcal{J}_m; m{\in}\{1,\cdots ,M\},$ represents one subchannel. On each subchannel, the rate is fixed and the transmission is intended for the receiver with the highest $\tau_q$. The average is, then, taken over all possible subchannels. Let $\tau_m^\text{max}$ be the maximum received feedback information on channel $m$. The achievable secrecy sum-rate can be written as
\begin{align}
\mathcal{R}_\text{s}^-&=\sum_{m=1}^M\text{Pr}[\mathcal{J}_m]~\!\underset{\gamma_\text{e}}{\mathbb{E}}\left[\left\lbrace\log\left(\frac{1{+}\tau_m^\text{max}P(\tau_m^\text{max})}{1{+}\gamma_\text{e}P(\tau_m^\text{max})}\right)\right\rbrace^+\right]\label{Rs_IM_stp1}\\
&=\sum_{q=1}^Q\text{Pr}[\tau_q{\leq}\gamma
_\text{max}{<}\tau_{q+1}]~\!\underset{\gamma_\text{e}}{\mathbb{E}}\left[\left\lbrace\log\left(\frac{1{+}\tau_qP_q}{1{+}\gamma_\text{e}P_q}\right)\right\rbrace^+\right],\label{Rs_IM_stp2}
\end{align} 
where \eqref{Rs_IM_stp1} is obtained by using a Gaussian codebook with power $P(\tau_m^\text{max})$, satisfying the average power constraint, on each subchannel $m$ \cite{Gopala_J1}, and \eqref{Rs_IM_stp2} follows by using the fact that $\tau_m^\text{max}~\!{\in}~\!\{\tau_1,\cdots ,\tau_Q\}$ and rewriting the summation over these indices. Also, we note that the probability of adapting the transmission with $\tau_q$ corresponds to the probability of having $\tau_q{\leq}\gamma_\text{max}{<}\tau_{q+1}$, with $\gamma_\text{max}{=}\max_{1\leq k\leq K}\gamma_k$.
Maximizing over the main channel gain reconstruction points $\tau_q$ and the associated power transmission strategies $P_q$, concludes the proof.$\hfill \square$
%------------------------------------------
\subsection{Proof of the Upper Bound}\label{App_TH2_UB}
%------------------------------------------
To prove that~$\mathcal{C}_{\text{s}}^+$ is an upper bound on the secrecy sum-capacity, we consider a new genie-aided channel whose capacity upper bounds the capacity of the $K$-receivers channel with limited CSI feedback. The new channel has only one receiver that observes the output of the strongest main channel. The output signal of the genie-aided receiver is given by $Y_\text{max}(t){=}h_\text{max}(t)X(t)+v(t)$, at time instant $t$, with $h_\text{max}$ being the channel gain of the best legitimate channel, i.e., $|h_\text{max}|^2{=}\gamma_\text{max}$ and $\gamma_\text{max}{=}\max_{1\leq k\leq K}\gamma_k$. The new channel can then be modeled as
\begin{equation}
\begin{aligned}
&Y_\text{max}(t)=h_\text{max}(t)X(t)+v(t)\\
&Y_\text{e}(t)\hspace{0.33cm}=h_\text{e}(t)X(t)\hspace{0.1cm}+\hspace{0.1cm}w_\text{e}(t) 
\end{aligned}\hspace{0.8cm},\hspace{0.3cm} t=1,\cdots,n.
\end{equation}

Let $\tau_q$ be the feedback information sent by the new receiver to the transmitter about its channel gain, i.e., $\tau_q$ is fed back when $\tau_q{\leq}\gamma_\text{max}{<}\tau_{q+1}$. First, we need to prove that the secrecy capacity of this new channel upper bounds the secrecy sum-capacity of the $K$-receivers channel with limited CSI. To this end, it is sufficient to show that if a secrecy rate point $(\mathcal{R}_1,\mathcal{R}_2,{\cdots},\mathcal{R}_K)$ is achievable on the $K$-receivers channel with limited CSI feedback, then, a secrecy sum-rate $\sum\mathcal{R}_k$ is achievable on the new channel.

Let $(W_1,\!W_2,\!{\cdots},\!W_K)$ be the independent transmitted messages corresponding to the rates $(\mathcal{R}_1,\!\mathcal{R}_2,\!{\cdots},\!\mathcal{R}_K)$, and $(\hat{W}_1,\hat{W}_2,{\cdots},\hat{W}_K)$ the decoded messages. Thus, for any $\epsilon{>}0$ and $n$ large enough, there exists a code of length $n$ such that $\text{Pr}[\hat{W}_k\!\neq\!W_k]\!\leq\!\epsilon$ at each of the $K$ receivers, and
\begin{equation}\label{sc23}
\frac{1}{n}H\!(W_k|W_1,\cdots,W_{k\!-\!1},W_{k\!+\!1},\cdots,W_K,Y_\text{e}^n,\gamma_\text{e}^L,F^L)\geq\mathcal{R}_k\!-\!\epsilon ,
\end{equation}
with $F^L{=}\{F_1^L,F_2^L,{\cdots},F_K^L\}$, and $F_k(l)~\!{\in}~\!\{\tau_1,{\cdots},\tau_Q\}$ is the feedback information sent by receiver $k$ in the $l$-th fading block. Now, we consider the transmission of message $W{=}(W_1,W_2,\cdots,W_K)$ to the genie-aided receiver~using the same encoding scheme as for the $K$-receivers case. Adopting a decoding scheme similar to the one used at each~of the $K$ legitimate receivers, it is clear that the genie-aided~receiver can decode message $W$ with a negligible probability of error, i.e., $\text{Pr}(\hat{W}\!\neq\!W)\!\leq\!\epsilon$. For the secrecy condition, we have
\begin{align}
\frac{1}{n}H\!(W|Y_\text{e}^n\!,\!\gamma_\text{e}^L\!,\!\gamma_\text{max}^L,\!F_\text{max}^L)&=\frac{1}{n}H\!(W_1\!,\!W_2\!,\!\cdots\!,\!W_K|Y_\text{e}^n\!,\!\gamma_\text{e}^L\!,\!\gamma_\text{max}^L,\!F_\text{max}^L)\\
&\geq\sum_{k=1}^K\!\frac{1}{n}H\!(W_k|W_1,\!\cdots\!,W_{k\!-\!1}\!,\!W_{k\!+\!1}\!,\!\cdots\!,\!W_K\!,\!Y_\text{e}^n\!,\!\gamma_\text{e}^L\!,\!\gamma_\text{max}^L,\!F_\text{max}^L)\\
&\geq\sum_{k=1}^K\!\frac{1}{n}H\!(W_k|W_1,\!\cdots\!,W_{k\!-\!1}\!,\!W_{k\!+\!1}\!,\!\cdots\!,\!W_K\!,\!Y_\text{e}^n\!,\!\gamma_\text{e}^L\!,\!\gamma_\text{max}^L,\!F^L)\label{UB_IM_v0}\\
&\geq\sum_{k=1}^K\mathcal{R}_k{-}K\epsilon ,\label{UB_IM_v01}
\end{align}
where $F_\text{max}^L{=}\{F_\text{max}(1),{\cdots},F_\text{max}(L)\}$ and $F_\text{max}(l)$ is the feedback information sent by the genie-aided receiver in the $l$-th fading block, \eqref{UB_IM_v0} follows from the fact that $F_\text{max}{\in}\{F_1,{\cdots},F_K\}$ and that conditioning reduces the entropy, and where \eqref{UB_IM_v01} follows from the secrecy constraint \eqref{sc23}. 

Now, we need to prove that $\mathcal{C}_\text{s}^+$ upper bounds the secrecy capacity of the genie-aided channel. Let $\mathcal{R}_\text{e}$ be the equivocation rate of the new channel. We have
\allowdisplaybreaks
\begin{align}
\hspace{-0.2cm}n\mathcal{R}_\text{e}&=H(W|Y_\text{e}^n,\gamma_\text{e}^L,\gamma_\text{max}^L,F_\text{max}^L)\label{Re_step1}\\
&=I(W\!;\!Y_\text{max}^n|Y_\text{e}^n\!,\!\gamma_\text{e}^L\!,\!\gamma_\text{max}^L\!,\!F_\text{max}^L){+}H(W|Y_\text{max}^n\!,\!Y_\text{e}^n\!,\!\gamma_\text{e}^L\!,\!\gamma_\text{max}^L\!,\!F_\text{max}^L)\\
&\leq I(W;Y_\text{max}^n|Y_\text{e}^n,\gamma_\text{e}^L,\gamma_\text{max}^L,F_\text{max}^L){+} n\epsilon \label{Re_step2}\\
&=\sum_{l=1}^L\sum_{k=1}^\kappa\hspace{-0.1cm}I(W\!;\!Y_\text{max}(l,k)|Y_\text{e}^n\!,\!\gamma_\text{e}^L\!,\!\gamma_\text{max}^L,F_\text{max}^L,Y_\text{max}^{\kappa(l\!-\!1)\!+\!(k\!-\!1)}){+}n\epsilon \\
&=\sum_{l=1}^L\sum_{k=1}^\kappa H(Y_\text{max}(l,k)|Y_\text{e}^n,\gamma_\text{e}^L,\gamma_\text{max}^L,F_\text{max}^L,Y_\text{max}^{\kappa(l\!-\!1)\!+\!(k\!-\!1)})\nonumber\\
&\hspace{2cm}-H(Y_\text{max}(l,k)|W\!,\!Y_\text{e}^n\!,\!\gamma_\text{e}^L\!,\!\gamma_\text{max}^L\!,\!F_\text{max}^L\!,\!Y_\text{max}^{\kappa(l\!-\!1)\!+\!(k\!-\!1)}){+} n\epsilon \\
&\leq \sum_{l=1}^L\sum_{k=1}^\kappa H(Y_\text{max}(l,k)|Y_\text{e}(l,k),\gamma_\text{e}(l),\gamma_\text{max}(l),F_\text{max}^l)\\
&\hspace{2cm}-H(Y_\text{max}(l,k)|W\!,\!X(l,k)\!,\!Y_\text{e}^n,\gamma_\text{e}^L\!,\!\gamma_\text{max}^L\!,\!F_\text{max}^L\!,\!Y_\text{max}^{\kappa(l\!-\!1)\!+\!(k\!-\!1)}){+} n\epsilon\nonumber\\
&=\sum_{l=1}^L\sum_{k=1}^\kappa H(Y_\text{max}(l,k)|Y_\text{e}(l,k),\gamma_\text{e}(l),\gamma_\text{max}(l),F_\text{max}^l)\\
&\hspace{2cm}-H(Y_\text{max}(l,k)|X(l,k),Y_\text{e}(l,k),\gamma_\text{e}(l),\gamma_\text{max}(l),F_\text{max}^l){+} n\epsilon\nonumber\\
&= \sum_{l=1}^L\sum_{k=1}^\kappa\hspace{-0.1cm}I(X(l\!,\!k);\!Y_\text{max}(l\!,\!k)|Y_\text{e}(l,k)\!,\!\gamma_\text{e}(l)\!,\!\gamma_\text{max}(l)\!,\!F_\text{max}^l)\!+\!n\epsilon\\
&\leq \sum_{l=1}^L\sum_{k=1}^\kappa \hspace{-0.1cm}\left\lbrace I(X(l,k);Y_\text{max}(l,k)|\gamma_\text{max}(l),F_\text{max}^l)-I(X(l,k);Y_\text{e}(l,k)|\gamma_\text{e}(l),F_\text{max}^l)\right\rbrace^+ {+} n\epsilon \label{Re_step3}\\
&= \sum_{l=1}^L\kappa \left\lbrace I(X(l);Y_\text{max}(l)|\gamma_\text{max}(l),F_\text{max}^l)-I(X(l);Y_\text{e}(l)|\gamma_\text{e}(l),F_\text{max}^l)\right\rbrace^+ {+} n\epsilon ,\label{Re_step3p1}
\end{align}
where inequality \eqref{Re_step2} follows from the fact that $$H(W|Y_\text{max}^n,Y_\text{e}^n,\gamma_\text{e}^L,\gamma_\text{max}^L,F_\text{max}^L){\leq}H(W|Y_\text{max}^n,\gamma_\text{max}^L,F_\text{max}^L),$$ and Fano's inequality $H(W|Y_\text{max}^n,\gamma_\text{max}^L,F_\text{max}^L){\leq}n\epsilon ,$ and \eqref{Re_step3} holds true by selecting the appropriate value for the noise correlation to form the Markov chain $X(l){\rightarrow}Y_\text{max}(l){\rightarrow}Y_\text{e}(l)$ if $\gamma_\text{max}(l){>}\gamma_\text{e}(l)$ or $X(l){\rightarrow}Y_\text{e}(l){\rightarrow}Y_\text{max}(l)$ if $\gamma_\text{max}(l){\leq}\gamma_\text{e}(l)$, as explained in \cite{Liang_J1}.

The right-hand side of \eqref{Re_step3p1} is maximized by a Gaussian input. That is, taking $X(l)\sim\mathcal{CN}\left(0,\omega_l^{1/2}(F_\text{max}^l)\right)$, with the power policy $\omega_l(F_\text{max}^l)$ satisfying the average power constraint, we can write
\begin{align}
n\mathcal{R}_\text{e}&\leq\kappa\sum_{l=1}^L\mathbb{E}\!\left[\!\left\lbrace\log\left(\frac{1{+}\gamma_\text{max}(l)\omega_l(F_\text{max}^l)}{1{+}\gamma_\text{e}(l)\omega_l(F_\text{max}^l)}\right)\right\rbrace^+\right]{+}n\epsilon \\
&=\kappa\sum_{l=1}^L\mathbb{E}\!\left[\!\mathbb{E}\!\left[\!\left\lbrace\!\log\!\left(\!\frac{1{+}\gamma_\text{max}(l)\omega_l(F_\text{max}^l)}{1{+}\gamma_\text{e}(l)\omega_l(F_\text{max}^l)}\!\right)\!\right\rbrace^{\hspace{-0.15cm}+}\!\bigg|F_\text{max}(l),\gamma_\text{max}(l),\gamma_\text{e}(l)\right]\right]\hspace{-0.15cm}{+}n\epsilon \\
&\leq\kappa\sum_{l=1}^L\mathbb{E}\!\left[\displaystyle{\!\left\lbrace\log\!\left(\!\frac{1{+}\gamma_\text{max}(l)\mathbb{E}\!\left[\omega_l\scriptstyle{(F_\text{max}^l)|F_\text{max}(l),\gamma_\text{max}(l),\gamma_\text{e}(l)}\right]}{1{+}\gamma_\text{e}(l)\mathbb{E}\!\left[\omega_l\scriptstyle{(F_\text{max}^l)|F_\text{max}(l),\gamma_\text{max}(l),\gamma_\text{e}(l)}\right]}\right)\!\right\rbrace^{\hspace{-0.15cm}+}}\right]\hspace{-0.15cm}{+}n\epsilon \label{Re_step4} \\
&=\kappa\sum_{l=1}^L\mathbb{E}\!\left[\left\lbrace\log\left(\frac{1{+}\gamma_\text{max}(l)\Omega_l(F_\text{max}(l))}{1{+}\gamma_\text{e}(l)\Omega_l(F_\text{max}(l))}\right)\!\right\rbrace^{\hspace{-0.15cm}+}\right]\hspace{-0.1cm}{+}n\epsilon \label{Re_step5} \\
&=\kappa\sum_{l=1}^L\mathbb{E}\left[\left\lbrace\log\left(\frac{1{+}\gamma_\text{max}\Omega_l(F_\text{max})}{1{+}\gamma_\text{e}\Omega_l(F_\text{max})}\right)\right\rbrace^+\right]\!+\!n\epsilon ,\label{Re_step6}
\end{align}
where \eqref{Re_step4} is obtained using Jensen's inequality, $\Omega_l(F_\text{max}(l))$ in \eqref{Re_step5} is defined as $$\displaystyle{\Omega_l(F_\text{max}(l)){=}\mathbb{E}\left[\omega_l(F_\text{max}^l)|F_\text{max}(l),\gamma_\text{max}(l),\gamma_\text{e}(l)\right],}$$ and where \eqref{Re_step6} follows from the ergodicity and the stationarity of the channel gains, i.e., the expectation in \eqref{Re_step5} does not depend on the block fading index. Thus, we have
\begin{align}
\mathcal{R}_\text{e}&\leq\frac{1}{L}\sum_{l=1}^L\mathbb{E}\!\left[\left\lbrace\log\left(\frac{1{+}\gamma_\text{max}\Omega_l(F_\text{max})}{1{+}\gamma_\text{e}\Omega_l(F_\text{max})}\right)\right\rbrace^+\right]\!+\!\epsilon\\
&\leq\mathbb{E}\left[\left\lbrace\log\left(\frac{1{+}\gamma_\text{max}\Omega(F_\text{max})}{1{+}\gamma_\text{e}\Omega(F_\text{max})}\right)\right\rbrace^+\right]+\epsilon ,\label{Re_step7}
\end{align}
where \eqref{Re_step7} comes from applying Jensen's inequality once again, with $\displaystyle{\Omega(F_\text{max}){=}\frac{1}{L}\sum_{l=1}^L\Omega_l(F_\text{max}).}$
Maximizing over the main channel gain reconstruction points $\tau_q$ and the associated power transmission strategies $P_q$, for each $q\in\{1,\cdots ,Q\}$, concludes the proof.$\hfill \square$

%-------------------------------------------------------------------------------------------------------------
\section{Proof of Theorem \ref{TH3}}\label{App_TH3} 
%-------------------------------------------------------------------------------------------------------------
We provide the proof of achievability and the converse for the ergodic secrecy capacity region of the multi-users BCCM with a 1-bit CSI feedback, presented in Theorem~\ref{TH3}. In this scenario, only one bit of CSI feedback is sent to the transmitter by an entity that knows the CSI of all system users, e.g., a central controller or one of the receivers when they know each others' CSI.  
%------------------------------------------
\subsection{Proof of Achievability}\label{App_TH3_LB} 
%------------------------------------------
Since the transmission is controlled by the feedback information, we consider that, during each fading block, the 1-bit feedback indicates to the transmitter whether the channel to R/E is better than those to the legitimate receivers, i.e., the feedback is equal to one when $\min_{1\leq k\leq K}\mathbb{E}\left[\gamma_k\right]>\gamma_\text{e}$ and equal to zero otherwise. The achievability follows from \cite[Corollary~1]{Csiszar_J1} by choosing the following input distributions:
\begin{itemize}
\item For $\underline{\gamma}{\in}\mathcal{A}$, $U{\sim}\mathcal{CN}(0,\sqrt{p_{01}})$, $X^\prime{\sim}\mathcal{CN}(0,\sqrt{p_{1}})$, with $X^\prime$ independent of $U$ and $V{=}X{=}U{+}X^\prime$;
\item For $\underline{\gamma}{\in}\mathcal{A}^c$, $U{=}V{=}X{\sim}\mathcal{CN}(0,\sqrt{p_{02}})$, 
\end{itemize}
where $\mathcal{A}=\left\lbrace\underline{\gamma}:\min_{1\leq k\leq K}\mathbb{E}\left[\gamma_k\right]>\gamma_\text{e}\right\rbrace$, $U$ and $V$ are the auxiliary random variables defined in \cite{Csiszar_J1}, and the transmission powers $p_{01}, p_{02}, p_1$ are chosen to satisfy $$\text{Pr}[\underline{\gamma}{\in}\mathcal{A}]~\!(p_{01}{+}p_1){+}\text{Pr}[\underline{\gamma}{\in}\mathcal{A}^c]~\!p_{02}{\leq} P_\text{avg}.$$
Note that since we are considering a multi-users scenario in this work, we need to ensure that the common message is decodable at all system users, including R/E, and the confidential message is only decodable at the legitimate receivers. Hence, the minimum in the expressions of the secrecy rates.\footnote{Even though~\cite{Csiszar_J1} considers a two-users case, the achievability proof can be easily adapted to our CCMs system by considering the transmission to R/E and to the legitimate receiver with the worst channel (on average).} 
%------------------------------------------
\subsection{Proof of the Converse}\label{App_TH3_UB}
%------------------------------------------
$-$ \textit{Bound on the Common Rate} $\mathcal{R}_0$:
Let $F^L{=}\{F(1),F(2),\cdots ,F(L)\}$, with $F(l)\in\{0,1\}$ being the feedback information sent in the $l$-th fading block, $l\in\{1,\cdots ,L\}$, and let $\gamma_{K+1}{=}\gamma_\text{e}$ and $Y_{K+1}{=}Y_\text{e}$ in this part of the proof. Considering the transmission to the $k^\text{th}$ system user, $k\in\{1,\cdots,K\}$, we have
\allowdisplaybreaks
\begin{align}\label{ub_th1_R01}
n\mathcal{R}_0&=H(W_0|F^L)\\
&=I(W_0;Y_k^n|F^L)+H(W_0|F^L,Y_k^n)\\
&\leq I(W_0;Y_k^n|F^L)+n\eta_1\label{ub_th1_R01s1}\\
&=\sum_{l=1}^L\sum_{j=1}^\kappa I(W_0;Y_k(l,j)|F^L,Y_k^{\kappa (l-1)+(j-1)})+n\eta_1\\
&\leq\sum_{l=1}^L\hspace{-0.08cm}\sum_{j=1}^\kappa\hspace{-0.1cm} I(W_0,\!Y_{K+1}^{[\kappa (l\!-\!1)\!+\!(j\!+\!1),n]};\!Y_k(l,j)|F^L\!,\!Y_k^{\kappa (l\!-\!1)\!+\!(j\!-\!1)})\!+\!n\eta_1\\
&\leq\sum_{l=1}^L\hspace{-0.08cm}\sum_{j=1}^\kappa\hspace{-0.1cm} I(W_0,\!Y_{K+1}^{[\kappa (l\!-\!1)\!+\!(j\!+\!1),n]}\!,\!Y_k^{\kappa (l\!-\!1)\!+\!(j\!-\!1)};\!Y_k(l,j)|F^L)\!+\!n\eta_1,
\end{align}
where \eqref{ub_th1_R01s1} is obtained using Fano's inequality. 
By defining the following auxiliary random variable $U(l,j)=(W_0,\!Y_{K+1}^{[\kappa (l\!-\!1)\!+\!(j\!+\!1),n]}\!,\!Y_k^{\kappa (l\!-\!1)\!+\!(j\!-\!1)}),$ we can write
\begin{align}
n\mathcal{R}_0&\leq\sum_{l=1}^L\sum_{j=1}^\kappa I(U(l,j);Y_k(l,j)|F^L)+n\eta_1\\
&=\sum_{l\in\mathcal{A}}\sum_{j=1}^\kappa I(U(l,j);Y_k(l,j)|F^L)+\sum_{l\in\mathcal{A}^c}\sum_{j=1}^\kappa I(U(l,j);Y_k(l,j)|F^L)+n\eta_1\label{ub_th1_R01s2}
\end{align}

On one hand, when $l\in\mathcal{A}^c$, we have
\begin{align}
I(U(l,j);Y_k(l,j)|F^L)&\leq I(X(l,j);Y_k(l,j)|F^L)\label{ub_th1_R01s31}\\
&= I(X(l,j);Y_k(l,j)|F^L,h_k(l))\label{ub_th1_R01s32}\\
&\leq \underset{F^l,\underline{\gamma}(l)}{\mathbb{E}}\!\left[\log\!\left(1{+}\gamma_k(l)\omega_l(F^l)\right)\big|\underline{\gamma}(l){\in}\mathcal{A}^c\right],\label{ub_th1_R01s3}
\end{align}
where \eqref{ub_th1_R01s31} follows from $U(l,j){\rightarrow}X(l,j){\rightarrow}Y_k(l,j)$ is a Markov chain, \eqref{ub_th1_R01s32} holds since given $F^L$, $X(l,j)$ is independent of $h_1(l)$, and \eqref{ub_th1_R01s3} results since a Gaussian $X$ maximizes the right hand side of \eqref{ub_th1_R01s32}, with $\omega_l(F^l)=\mathbb{E}\left[|X(l,j)|^2\big|F^l\right]$. 

On the other hand, when $l\in\mathcal{A}$, we have
\begin{align}
I(U(l,j);Y_k(l,j)|F^L)&=I(U(l,j);Y_k(l,j)|F^L,h_k(l))\label{ub_th1_R01s33}\\
&=H(Y_k(l,j)|F^L\!,\!h_k(l)){-}H(Y_k(l,j)\big|U(l,j)\!,\!F^L\!,\!h_k(l)),
\end{align}
where \eqref{ub_th1_R01s33} follows since given $F^L$, $U(l,j)$ is independent of $h_k(l)$, with
\begin{align}
H(Y_k(l,j)\big|U(l,j),F^L,h_k(l))&\leq H(Y_k(l,j)\big|F^L,h_k(l))\\
&=\underset{F^l,\underline{\gamma}(l)}{\mathbb{E}}\!\left[\log\!\left(\pi e\!\left(1{+}\gamma_k(l)\delta_l(F^l)\right)\right)\big|\underline{\gamma}(l){\in}\mathcal{A}\right],\label{ub_th1_R01s4}
\end{align}
where \eqref{ub_th1_R01s4} follows by taking $X(l,j){\sim}\mathcal{CN}\left(0,\sqrt{\delta_l(F^l)}\right)$,~and
\begin{align}
H(Y_k(l,j)\big|U(l,j),F^L,h_k(l))&\geq H(Y_k(l,j)\big|X(l,j),F^L,h_k(l))\\
&=\log \pi e.
\end{align}
Hence, there exists $0\leq\alpha_l\leq 1$ such that
\begin{align}
&H(Y_k(l,j)\big|U(l,j),F^L,h_k(l))=\underset{F^l,\underline{\gamma}(l)}{\mathbb{E}}\!\left[\log\!\left(\pi e\!\left(1{+}\gamma_k(l)\alpha_l\delta_l(F^l)\right)\right)\big|\underline{\gamma}(l){\in}\mathcal{A}\right].
\end{align}
We can then write
\begin{align}
I(U(l,j);Y_k(l,j)|F^L)&\leq \underset{F^l,\underline{\gamma}(l)}{\mathbb{E}}\!\left[\log\!\left(\pi e\!\left(1{+}\gamma_k(l)\delta_l(F^l)\right)\right)\big|\underline{\gamma}(l){\in}\mathcal{A}\right]\nonumber\\
&\hspace{2cm}-\underset{F^l,\underline{\gamma}(l)}{\mathbb{E}}\!\left[\log\!\left(\pi e\!\left(1{+}\gamma_k(l)\alpha_l\delta_l(F^l)\right)\right)\big|\underline{\gamma}(l){\in}\mathcal{A}\right]\\
&=\underset{F^l,\underline{\gamma}(l)}{\mathbb{E}}\!\left[\log\!\left(\!1{+}\frac{\gamma_k(l)(1{-}\alpha_l)\delta_l(F^l)}{1{+}\gamma_k(l)\alpha_l\delta_l(F^l)}\right)\Big|\underline{\gamma}(l){\in}\mathcal{A}\right].\label{ub_th1_R01s5}
\end{align}
Substituting \eqref{ub_th1_R01s3} and \eqref{ub_th1_R01s5} in \eqref{ub_th1_R01s2}, we get 
\begin{align}
n\mathcal{R}_0&\leq\sum_{l\in\mathcal{A}}\kappa\underset{F^l,\underline{\gamma}(l)}{\mathbb{E}}\!\left[\log\!\left(\!1{+}\frac{\gamma_k(l)(1{-}\alpha_l)\delta_l(F^l)}{1{+}\gamma_k(l)\alpha_l\delta_l(F^l)}\right)\Big|\underline{\gamma}(l){\in}\mathcal{A}\right]\nonumber\\
&\hspace{2cm}+\sum_{l\in\mathcal{A}^c}\kappa\underset{F^l,\underline{\gamma}(l)}{\mathbb{E}}\!\left[\log\!\left(1{+}\gamma_k(l)\omega_l(F^l)\right)\big|\underline{\gamma}(l){\in}\mathcal{A}^c\right]{+}n\eta_k.
\end{align}
Noting that $\underset{F^l}{\mathbb{E}}[.]=\underset{F(l)}{\mathbb{E}}\left[\underset{F^{l-1}}{\mathbb{E}}[.|F(l)]\right]$, and applying Jensen's inequality, we get
\begin{align}
\mathcal{R}_0&\leq\frac{1}{L}\sum_{l\in\mathcal{A}}\underset{\underset{\underline{\gamma}(l)}{F(l),}}{\mathbb{E}}\!\left[\log\!\left(\!1{+}\frac{\gamma_k(l)(1{-}\alpha_l)\Delta_l(F(l))}{1{+}\gamma_k(l)\alpha_l\Delta_l(F(l))}\right)\!\Big|\underline{\gamma}(l){\in}\mathcal{A}\right]\nonumber\\
&\hspace{2cm}+\frac{1}{L}\sum_{l\in\mathcal{A}^c}\underset{\underset{\underline{\gamma}(l)}{F(l),}}{\mathbb{E}}\!\left[\log\!\left(1{+}\gamma_k(l)\Omega_l(F(l))\right)\!\big|\underline{\gamma}(l){\in}\mathcal{A}^c\right]{+}\eta_k\\
&=\frac{1}{L}\sum_{l\in\mathcal{A}}\underset{F(l),\underline{\gamma}}{\mathbb{E}}\!\left[\log\!\left(\!1{+}\frac{\gamma_k(1{-}\alpha_l)\Delta_l(F(l))}{1{+}\gamma_k\alpha_l\Delta_l(F(l))}\right)\!\Big|\underline{\gamma}{\in}\mathcal{A}\right]\nonumber\\
&\hspace{2cm}+\frac{1}{L}\sum_{l\in\mathcal{A}^c}\underset{F(l),\underline{\gamma}}{\mathbb{E}}\!\left[\log\!\left(1{+}\gamma_k\Omega_l(F(l))\right)\!\big|\underline{\gamma}{\in}\mathcal{A}^c\right]{+}\eta_k,\label{ub_th1_R01s6}
\end{align}
with $\Delta_l(F(l)){=}\underset{F^{l\!-\!1}}{\mathbb{E}}\left[\delta_l(F^l)\big|F(l)\right]$, $\Omega_l(F(l)){=}\underset{F^{l\!-\!1}}{\mathbb{E}}\hspace{-0.1cm}\left[\omega_l(F^l)\big|F(l)\right]$, and \eqref{ub_th1_R01s6} follows from the ergodicity and the stationarity of the channel gain. Applying Jensen's inequality once again, we~get
\begin{align}
\mathcal{R}_0&\leq\underset{\underline{\gamma}}{\mathbb{E}}\hspace{-0.1cm}\left[\!\log\!\left(\!1{+}\frac{\displaystyle{\frac{\gamma_k}{L_\mathcal{A}}\sum_{l\in\mathcal{A}}(1\!-\!\alpha_l)\Delta_l(F(l))}}{\displaystyle{1{+}\frac{\gamma_k}{L_\mathcal{A}}\sum_{l\in\mathcal{A}}\alpha_l\Delta_l(F(l))}}\right)\!\Bigg|\underline{\gamma}{\in}\mathcal{A}\right]\!\text{Pr}\!\left[\underline{\gamma}{\in}\mathcal{A}\right]\nonumber\\
&\hspace{2cm}+\underset{\underline{\gamma}}{\mathbb{E}}\!\left[\log\!\left(1{+}\frac{\gamma_k}{L_{\mathcal{A}^c}}\sum_{l\in\mathcal{A}^c}\Omega_l(F(l))\right)\!\Big|\underline{\gamma}{\in}\mathcal{A}^c\right]\!\text{Pr}\!\left[\underline{\gamma}{\in}\mathcal{A}^c\right]{+}\eta_k,
\end{align}
where $L_\mathcal{A}=\text{Pr}\!\left[\underline{\gamma}{\in}\mathcal{A}\right]L$ and $L_{\mathcal{A}^c}=\text{Pr}\!\left[\underline{\gamma}{\in}\mathcal{A}^c\right]L$. 
Then, by taking $\Omega(F)=\displaystyle{\frac{1}{L_{\mathcal{A}^c}}\sum_{l\in\mathcal{A}^c}\Omega_l(F(l))}$, $\Delta_1(F)=\displaystyle{\frac{1}{L_\mathcal{A}}\sum_{l\in\mathcal{A}}(1-\alpha_l)\Delta_l(F(l))}$, and $\Delta_2(F)=\displaystyle{\frac{1}{L_\mathcal{A}}\sum_{l\in\mathcal{A}}\alpha_l\Delta_l(F(l))}$, we can write
\begin{align}
\mathcal{R}_0&\leq\underset{\underline{\gamma}}{\mathbb{E}}\left[\log\!\left(\!1{+}\frac{\gamma_k\Delta_1(F)}{1{+}\gamma_k\Delta_2(F)}\right)\!\Big|\underline{\gamma}{\in}\mathcal{A}\right]\!\text{Pr}\!\left[\underline{\gamma}{\in}\mathcal{A}\right]\nonumber\\
&\hspace{1.2cm}+\underset{\underline{\gamma}}{\mathbb{E}}\!\left[\log\!\left(1{+}\gamma_k\Omega(F)\right)\!\big|\underline{\gamma}{\in}\mathcal{A}^c\right]\!\text{Pr}\!\left[\underline{\gamma}{\in}\mathcal{A}^c\right]{+}\eta_k,\label{ub_th1_R01s7}
\end{align}
with $\text{Pr}[\underline{\gamma}{\in}\mathcal{A}]~\!(\Delta_1(F){+}\Delta_2(F)){+}\text{Pr}[\underline{\gamma}{\in}\mathcal{A}^c]~\!\Omega (F){\leq} P_\text{avg}.$\vspace{0.1cm}

A similar reasoning is applied for the case $k{=}K{+}1$. Then, since the choice of the receiver to transmit to is arbitrary, we tighten this upper bound by choosing the system user that minimizes this quantity. Then, using the fact that $F{=}\rho(\underline{\gamma})$, where $\rho(.)$, is a deterministic mapping, and taking $\eta_k$ arbitrary small, we get the outer boundary on the common rate $\mathcal{R}_0$, presented in~Theorem~\ref{TH3}. \vspace{0.2cm} 

%---------------------------------
$-$ \textit{Bound on the Confidential Rate} $\mathcal{R}_1$:
Let $k\in\{1,\cdots,K\}$. We have
\allowdisplaybreaks
\begin{align}
n\mathcal{R}_1&\leq n\mathcal{R}_\text{e}\\
&=H(W_1|F^L,\underline{\gamma}^L,Y_\text{e}^n)\\
&=I(W_1;W_0|F^L,\underline{\gamma}^L,Y_\text{e}^n){+}H(W_1|F^L,\underline{\gamma}^L,W_0,Y_\text{e}^n)\\
&\leq I(W_1;Y_k^n|F^L,\underline{\gamma}^L,W_0){-}I(W_1;Y_\text{e}^n|F^L,\underline{\gamma}^L,W_0)\nonumber\\
&\hspace{2cm}+H(W_1|F^L,\underline{\gamma}^L,W_0,Y_k^n){+}H(W_0|F^L,\underline{\gamma}^L,Y_\text{e}^n)\\
&\leq I(W_1;Y_k^n|F^L\!,\underline{\gamma}^L\!,\!W_0){-}I(W_1;Y_\text{e}^n|F^L\!,\underline{\gamma}^L\!,\!W_0){+}n(\eta_k{+}\eta')\label{th1_R1s1}\\
&=\sum_{l=1}^L\sum_{j=1}^\kappa\left\lbrace I(W_1;Y_k(l,j)|F^L,\underline{\gamma}(l),W_0,Y_k^{\kappa (l{-}1){+}(j{-}1)})\right.\nonumber\\
&\hspace{2cm}-\left. I(W_1;Y_\text{e}(l,j)|F^L,\underline{\gamma}(l),W_0,Y_\text{e}^{[\kappa (l{-}1){+}(j{+}1),n]})\right\rbrace {+}n\eta_k'\\
&=\sum_{l,j}\left\lbrace I(W_1,Y_\text{e}^{[\kappa (l{-}1){+}(j{+}1),n]};Y_k(l,j)|F^L,\underline{\gamma}(l),W_0,Y_k^{\kappa (l{-}1){+}(j{-}1)})\right.\nonumber\\
&\hspace{1.7cm}-I(Y_\text{e}^{[\kappa (l{-}1){+}(j{+}1),n]};Y_k(l,j)|F^L,\underline{\gamma}(l),W_1,W_0,Y_k^{\kappa (l{-}1){+}(j{-}1)})\nonumber\\
&\hspace{2cm}-I(W_1,Y_k^{\kappa (l{-}1){+}(j{-}1)};Y_\text{e}(l,j)|F^L,\underline{\gamma}(l),W_0,Y_\text{e}^{[\kappa (l{-}1){+}(j{+}1),n]})\nonumber\\
&\hspace{2.3cm}+\left.I(Y_k^{\kappa (l{-}1){+}(j{-}1)};Y_\text{e}(l,j)|F^L,\underline{\gamma}(l),W_1,W_0,Y_\text{e}^{[\kappa (l{-}1){+}(j{+}1),n]})\right\rbrace{+}n\eta_k'\\
&=\sum_{l,j}\left\lbrace I(W_1,Y_\text{e}^{[\kappa (l{-}1){+}(j{+}1),n]};Y_k(l,j)|F^L,\underline{\gamma}(l),W_0,Y_k^{\kappa (l{-}1){+}(j{-}1)})\right.\nonumber\\
&\hspace{2cm}-\left. I(W_1,Y_k^{\kappa (l{-}1){+}(j{-}1)};Y_\text{e}(l,j)|F^L,\underline{\gamma}(l),W_0,Y_\text{e}^{[\kappa (l{-}1){+}(j{+}1),n]})\right\rbrace{+}n\eta_k'\label{th1_R1s2}\\
&=\sum_{l,j}\left\lbrace I(W_1;Y_k(l,j)|F^L,\underline{\gamma}(l),W_0,Y_k^{\kappa (l{-}1){+}(j{-}1)},Y_\text{e}^{[\kappa (l{-}1){+}(j{+}1),n]})\right.\nonumber\\
&\hspace{2cm}-\left. I(W_1;Y_\text{e}(l,j)|F^L,\underline{\gamma}(l),W_0,Y_k^{\kappa (l{-}1){+}(j{-}1)},Y_\text{e}^{[\kappa (l{-}1){+}(j{+}1),n]})\right\rbrace{+}n\eta_k',\label{th1_R1s3}
\end{align}
where \eqref{th1_R1s1} is obtained using Fano's inequality, \eqref{th1_R1s2} and \eqref{th1_R1s3} follow from Lemma 7 in \cite{Csiszar_J1}, and $\eta_k'{=}\eta_k{+}\eta'$. By defining $\displaystyle{U(l,j){=}(W_0,\!Y_\text{e}^{[\kappa (l\!-\!1)\!+\!(j\!+\!1),n]}\!,\!Y_k^{\kappa (l\!-\!1)\!+\!(j\!-\!1)}), V(l,j){=}(W_1,U(l,j)),}$ such that $U(l,j){\rightarrow}V(l,j){\rightarrow}X(l,j){\rightarrow}(Y_k(l,j),Y_\text{e}(l,j))$ is a Markov chain, we can write%\vspace{0.2cm}
\begin{align}
n\mathcal{R}_1&\leq\sum_{l=1}^L\sum_{j=1}^\kappa\left\lbrace I(V(l,j);Y_k(l,j)|F^L,\underline{\gamma}(l),U(l,j))\right.\nonumber\\
&\hspace{4cm}\left.-I(V(l,j);Y_\text{e}(l,j)|F^L,\underline{\gamma}(l),U(l,j))\right\rbrace{+}n\eta_k'.
\end{align}
When $l\in\mathcal{A}^c$, we have\vspace{0.2cm}
\begin{align}
&I(V(l,j);Y_k(l,j)|F^L,\underline{\gamma}(l),U(l,j))-I(V(l,j);Y_\text{e}(l,j)|F^L,\underline{\gamma}(l),U(l,j))\nonumber\\
&\hspace{2cm}\leq I(V(l,j);Y_k(l,j),Y_\text{e}(l,j)|F^L,\underline{\gamma}(l),U(l,j))\nonumber\\
&\hspace{4cm}-I(V(l,j);Y_\text{e}(l,j)|F^L,\underline{\gamma}(l),U(l,j))\\
&\hspace{2cm}=I(V(l,j);Y_\text{e}(l,j)|F^L,\underline{\gamma}(l),U(l,j))\nonumber\\
&\hspace{3cm}+I(V(l,j);Y_k(l,j)|F^L,\underline{\gamma}(l),Y_\text{e}(l,j),U(l,j))\nonumber\\
&\hspace{4cm}-I(V(l,j);Y_\text{e}(l,j)|F^L,\underline{\gamma}(l),U(l,j))\\
&\hspace{2cm}=0,\label{th1_R1s4}
\end{align}
where \eqref{th1_R1s4} results since $\displaystyle{X(l,j){\rightarrow}Y_k(l,j){\rightarrow}Y_\text{e}(l,j)}$ is a Markov chain when $l{\in}\mathcal{A}^c$. Hence, we have
\begin{align}
\mathcal{R}_1&\leq\frac{1}{n}\sum_{l\in\mathcal{A}}\sum_{j=1}^\kappa\left\lbrace I(V(l,j);Y_k(l,j)|F^L,\underline{\gamma}(l),U(l,j))\right.\nonumber\\
&\hspace{3.5cm}\left.-I(V(l,j);Y_\text{e}(l,j)|F^L,\underline{\gamma}(l),U(l,j))\right\rbrace{+}\eta_k' \\
&\leq\frac{1}{n}\sum_{l\in\mathcal{A}}\sum_{j=1}^\kappa\left\lbrace I(X(l,j);Y_k(l,j)|F^L,\underline{\gamma}(l),U(l,j))\right.\nonumber\\
&\hspace{3.5cm}\left.-I(X(l,j);Y_\text{e}(l,j)|F^L,\underline{\gamma}(l),U(l,j))\right\rbrace{+}\eta_k' \\
&\leq\underset{\underline{\gamma}}{\mathbb{E}}\left[\log\!\left(1{+}\gamma_k\Delta_2(F)\right){-}\log\!\left(1{+}\gamma_\text{e}\Delta_2(F)\right)\!\Big|\underline{\gamma}{\in}\mathcal{A}\right]\!\text{Pr}\!\left[\underline{\gamma}{\in}\mathcal{A}\right]{+}\eta_k' ,\label{th1_R1s5}
\end{align}
where \eqref{th1_R1s5} follows along similar lines as for the common rate case. We, then, tighten this upper bound by choosing the legitimate user that minimizes this quantity. $\hfill \square$

%-------------------------------------------------------------------------------------------------------------
\section{Proof of Corollary \ref{CR4}}\label{App_CR4} 
%-------------------------------------------------------------------------------------------------------------
The achievability follows by considering the same feedback scheme as in Theorem \ref{TH3}, i.e., during each fading block, the 1-bit feedback indicates to the transmitter which channel is better (the feedback is equal to one when $\min_{1\leq k\leq K}\mathbb{E}\left[\gamma_k\right]>\gamma_\text{e}$ and equal to zero otherwise). Besides, we consider that the confidential message is only transmitted over the coherence blocks where the feedback bit is not erased and is equal to one. The input distributions are chosen, in this case, as follows
\begin{itemize}
\item When $\underline{\gamma}{\in}\mathcal{A}$ and no erasure occurs, $U{\sim}\mathcal{CN}(0,\sqrt{p_{01}})$, $X^\prime{\sim}\mathcal{CN}(0,\sqrt{p_{1}})$, with $X^\prime$ independent of $U$ and $V{=}X{=}U{+}X^\prime$;
\item When $\underline{\gamma}{\in}\mathcal{A}^c$ or when an erasure occurs, $U=V=X{\sim}\mathcal{CN}(0,\sqrt{p_{02}})$, 
\end{itemize}
where $\mathcal{A}=\left\lbrace\underline{\gamma}:\min_{1\leq k\leq K}\mathbb{E}\left[\gamma_k\right]>\gamma_\text{e}\right\rbrace$, $U$ and $V$ are the auxiliary random variables defined in \cite{Csiszar_J1}, and the transmission powers $p_{01}, p_{02}, p_1$ are chosen to satisfy 
\begin{align}
&(p_{01}{+}p_1)~\!\text{Pr}[\underline{\gamma}{\in}\mathcal{A}~\!\text{and no erasure occurs}]+p_{02}~\!\text{Pr}[\underline{\gamma}{\in}\mathcal{A}^c~\!\text{or an erasure occurs}]{\leq} P_\text{avg},
\end{align}
which reduces to $(p_{01}{+}p_1)(1{-}\epsilon)~\!\text{Pr}[\underline{\gamma}{\in}\mathcal{A}]{+}p_{02}\left(\epsilon{+}(1{-}\epsilon)\text{Pr}[\underline{\gamma}{\in}\mathcal{A}^c]\right){\leq} P_\text{avg}.$ $\hfill \square$

%%-------------------------------------------------------------------------------------------------------------
%
\bibliographystyle{IEEEtran}
\bibliography{References}

\begin{thebibliography}{10}
\providecommand{\url}[1]{#1}
\csname url@samestyle\endcsname
\providecommand{\newblock}{\relax}
\providecommand{\bibinfo}[2]{#2}
\providecommand{\BIBentrySTDinterwordspacing}{\spaceskip=0pt\relax}
\providecommand{\BIBentryALTinterwordstretchfactor}{4}
\providecommand{\BIBentryALTinterwordspacing}{\spaceskip=\fontdimen2\font plus
\BIBentryALTinterwordstretchfactor\fontdimen3\font minus
  \fontdimen4\font\relax}
\providecommand{\BIBforeignlanguage}[2]{{%
\expandafter\ifx\csname l@#1\endcsname\relax
\typeout{** WARNING: IEEEtran.bst: No hyphenation pattern has been}%
\typeout{** loaded for the language `#1'. Using the pattern for}%
\typeout{** the default language instead.}%
\else
\language=\csname l@#1\endcsname
\fi
#2}}
\providecommand{\BIBdecl}{\relax}
\BIBdecl

\bibitem{Shannon_J1}
C.~E. Shannon, ``Communication theory of secrecy systems,'' \emph{Bell Systems
  Technical Journal}, vol.~28, pp. 656--719, Oct. 1949.

\bibitem{Wyner_J1}
A.~D. Wyner, ``The wiretap channel,'' \emph{Bell System Technical Journal},
  vol.~54, no.~8, pp. 1355--1387, 1975.

\bibitem{Csiszar_J1}
I.~Csisz\'{a}r and J.~K\"{o}rner, ``Broadcast channels with confidential
  m­essages,'' \emph{IEEE Transactions on Information Theory}, vol.~24, no.~3,
  pp. 339--348, 1978.

\bibitem{Leung_J1}
S.~Leung-Yan-Cheong and M.~Hellman, ``The {G}aussian wiretap channel,''
  \emph{IEEE Transactions on Information Theory}, vol.~24, no.~4, pp. 451--456,
  Jul. 1978.

\bibitem{Gopala_J1}
P.~Gopala, L.~Lai, and H.~E. Gamal, ``On the secrecy capacity of fading
  channels,'' \emph{IEEE Transactions on Information Theory}, vol.~54, no.~10,
  pp. 4687--4698, Oct. 2008.

\bibitem{Liang_J1}
Y.~Liang, H.~Poor, and S.~Shamai, ``Secure communication over fading
  channels,'' \emph{IEEE Transactions on Information Theory}, vol.~54, no.~6,
  pp. 2470--2492, Jun. 2008.

\bibitem{Ashish_J3}
A.~Khisti, A.~Tchamkerten, and G.~W. Wornell, ``Secure broadcasting over fading
  channels,'' \emph{IEEE Transactions on Information Theory}, vol.~54, no.~6,
  pp. 2453--2469, Jun. 2008.

\bibitem{Ekrem_J2}
E.~Ekrem and S.~Ulukus, ``Secrecy capacity of a class of broadcast channels
  with an eavesdropper,'' \emph{EURASIP Journal on Wireless Communications and
  Networking, Special Issue on Wireless Physical Layer Security}, Article ID
  824235, 29 pages, 2009.

\bibitem{Hyadi_J1}
A.~Hyadi, Z.~Rezki, A.~Khisti, and M.-S. Alouini, ``Secure broadcasting with
  imperfect channel state information at the transmitter,'' \emph{IEEE
  Transactions on Wireless Communications}, vol.~15, no.~3, pp. 2215--2230,
  Mar. 2016.

\bibitem{Tekin_C1}
E.~Tekin and A.~Yener, ``The {G}aussian multiple access wire-tap channel
  with collective secrecy constraints,'' in \emph{Proc. International
  Symposiumon on Information Theory (ISIT'2006)}, Seattle, US, Jul. 2006, pp.
  1164--1168.

\bibitem{Ekrem_C1}
E.~Ekrem and S.~Ulukus, ``On the secrecy of multiple access wiretap channel,''
  in \emph{Proc. 46th Allerton conference on Communication Control and
  Computing}, Urbana-Champaign, IL,US, Sep. 2008, pp. 1014--1021.

\bibitem{Ashish_C1}
A.~Khisti, G.~Wornell, A.~Wiesel, and Y.~Eldar, ``On the {G}aussian {MIMO}
  wiretap channel,'' in \emph{Proc. International Symposiumon on Information
  Theory (ISIT'2007)}, Nice, France, Jun. 2007, pp. 2471--2475.

\bibitem{LiJ_J2}
J.~Li and A.~P. Petropulu, ``On ergodic secrecy rate for gaussian {MISO}
  wiretap channels,'' \emph{IEEE Transactions on Signal Processing}, vol.~10,
  no.~4, pp. 1176--1187, Apr. 2011.

\bibitem{Ashish_J1}
A.~Khisti and G.~Womell, ``Secure transmission with multiple antennas {P}art
  {I}: The {MISOME} wiretap channel,'' \emph{IEEE Transactions on Information
  Theory}, vol.~56, no.~7, pp. 3088--3104, Jul. 2010.

\bibitem{Oggier_J1}
F.~Oggier and B.~Hassibi, ``The secrecy capacity of the {MIMO} wiretap
  channel,'' \emph{IEEE Transactions on Information Theory}, vol.~57, no.~8,
  pp. 4961--4972, Aug. 2011.

\bibitem{Ashish_J2}
A.~Khisti and G.~Womell, ``Secure transmission with multiple antennas {P}art
  {II}: The {MIMOME} wiretap channel,'' \emph{IEEE Transactions on Information
  Theory}, vol.~56, no.~11, pp. 5515--5532, Nov. 2010.

\bibitem{Ekrem_J1}
E.~Ekrem and S.~Ulukus, ``The secrecy capacity region of the {G}aussian {MIMO}
  multi-receiver wiretap channel,'' \emph{IEEE Transactions on Information
  Theory}, vol.~57, no.~4, pp. 2083--2114, Apr. 2011.

\bibitem{LiuR_J1}
R.~Liu and H.~V. Poor, ``Secrecy capacity region of a multiple-antenna
  {G}aussian broadcast channel with confidential messages,'' \emph{IEEE
  Transactions on Information Theory}, vol.~55, no.~3, p. 1235–1249, Mar.
  2009.

\bibitem{Xu_J1}
J.~Xu, Y.~Cao, and B.~Chen, ``Capacity bounds for broadcast channels with
  confidential messages,'' \emph{IEEE Transactions on Information Theory},
  vol.~55, no.~10, pp. 4529--4542, Oct. 2009.

\bibitem{Zou_C1}
S.~Zou, Y.~Liang, L.~Lai, and S.~Shamai, ``Rate splitting and sharing for
  degraded broadcast channel with secrecy outside a bounded range,'' in
  \emph{Proc. International Symposiumon on Information Theory (ISIT'2015)},
  Hong Kong, Jun. 2015, pp. 1357--1361.

\bibitem{Heath_C1}
R.~W. Heath and A.~Paulraj, ``A simple scheme for transmit diversity using
  partial channel feedback,'' in \emph{Proc. IEEE Asilomar Conference on in
  Signals, Systems and Computers}, Pacific Grove, CA, US, Nov. 1998, pp.
  1073--1078.

\bibitem{Blum_C1}
R.~S. Blum, ``{MIMO} with limited feedback of channel state information,'' in
  \emph{Proc. IEEE International Conference on Acoustics, Speech, and Signal
  Processing (ICASSP '2003)}, Hong Kong, China, Apr. 2003, pp. 89--92.

\bibitem{Love_J2}
D.~J. Love, R.~W. Heath, and T.~Strohmer, ``Grassmannian beamforming for
  multiple-input multiple-output wireless systems,'' \emph{IEEE Transactions on
  Information Theory}, vol.~49, no.~10, pp. 2735--2747, Oct. 2003.

\bibitem{Lau_J1}
V.~Lau, Y.~Liu, and T.-A. Chen, ``On the design of {MIMO} block-fading channels
  with feedback-link capacity constraint,'' \emph{IEEE Transactions on
  Communications}, vol.~52, no.~1, pp. 62--70, Jan. 2004.

\bibitem{Murthy_J1}
C.~R. Murthy and B.~D. Rao, ``Quantization methods for equal gain transmission
  with finite rate feedback,'' \emph{IEEE Transactions on Signal Processing},
  vol.~55, no.~1, pp. 233--245, Jan. 2007.

\bibitem{Love_J1}
D.~Love, R.~Heath, V.~Lau, D.~Gesbert, B.~Rao, and M.~Andrews, ``An overview of
  limited feedback in wireless communication systems,'' \emph{IEEE Journal on
  Selected Areas in Communications}, vol.~26, no.~8, pp. 1341--1365, Oct. 2008.

\bibitem{Rezki_J1}
Z.~Rezki, A.~Khisti, and M.-S. Alouini, ``Ergodic secret message capacity of
  the wiretap channel with finite-rate feedback,'' \emph{IEEE Transactions on
  Wireless Communications}, vol.~13, no.~6, pp. 3364--3379, Jun. 2014.

\bibitem{Hyadi_J3}
A.~Hyadi, Z.~Rezki, and M.-S. Alouini, ``Secure multiple-antenna block-fading
  wiretap channels with limited {CSI} feedback,'' \emph{IEEE Transactions on
  Wireless Communications}, vol.~16, no.~10, pp. 6618--6634, Jul. 2017.

\bibitem{LiuS_J1}
S.~Liu, Y.~Hong, and E.~Viterbo, ``Guaranteeing positive secrecy capacity for
  {MIMOME} wiretap channels with finite-rate feedback using artificial noise,''
  \emph{IEEE Transactions on Wireless Communications}, vol.~14, no.~8, pp.
  4193--4203, Aug. 2015.

\bibitem{LiN_J1}
N.~Li, X.~Tao, and J.~Xu, ``Ergodic secrecy sum-rate for downlink multiuser
  {MIMO} systems with limited {CSI} feedback,'' \emph{IEEE Communications
  Letters}, vol.~18, no.~6, pp. 969--972, Jun. 2014.

\bibitem{Zhang_J2}
X.~Zhang, M.~R. McKay, X.~Zhou, and R.~W.~H. Jr, ``Artificial-noise aided
  secure multi-antenna transmission with limited feedback,'' \emph{IEEE
  Transactions on Wireless Communications}, vol.~14, no.~5, pp. 2742--2754, May
  2015.

\bibitem{LinSC_J1}
S.-C. Lin, T.-H. Chang, Y.-L. Liang, Y.-W. Hong, and C.-Y. Chi, ``On the impact
  of quantized channel feedback in guaranteeing secrecy with artificial noise:
  The noise leakage problem,'' \emph{IEEE Transactions on Wireless
  Communications}, vol.~10, no.~3, pp. 901--915, Mar. 2011.

\bibitem{Bassi_C1}
G.~Bassi, P.~Piantanida, and S.~Shamai, ``The wiretap channel with generalized
  feedback: Secure communication and key generation,'' in \emph{IEEE
  Information Theory Workshop - Fall (ITW'2015)}, Jeju, South Korea, Oct. 2015,
  pp. 282--286.

\bibitem{Bassi_J1}
------, ``The wiretap channel with generalized feedback: Secure communication
  and key generation,'' \emph{IEEE Transactions on Information Theory},
  vol.~65, no.~4, pp. 2213--2233, Apr. 2019.

\bibitem{Bassi_C2}
------, ``Secret key generation over noisy channels with common randomness,''
  in \emph{IEEE International Symposium on Information Theory (ISIT'2016)},
  Barcelona, Spain, Jul. 2016, pp. 510--514.

\bibitem{ZhengTX_J1}
T.-X. Zheng and H.-M. Wang, ``Optimal power allocation for artificial noise
  under imperfect {CSI} against spatially random eavesdroppers,'' \emph{IEEE
  Transactions on Vehicular Technology}, vol.~65, no.~10, pp. 8812--8817, Oct.
  2016.

\bibitem{Rezki_J2}
Z.~Rezki, A.~Khisti, and M.-S. Alouini, ``On the secrecy capacity of the
  wiretap channel under imperfect main channel estimation,'' \emph{IEEE
  Transactions on Communications}, vol.~62, no.~10, pp. 3652--3664, Sep. 2014.

\bibitem{ChuZ_J1}
Z.~Chu, H.~Xing, M.~Johnston, and S.~L. Goff, ``Secrecy rate optimizations for
  a {MISO} secrecy channel with multiple multiantenna eavesdroppers,''
  \emph{IEEE Transactions on Wireless Communications}, vol.~15, no.~1, pp.
  283--297, Jan. 2016.

\bibitem{Zhou_J1}
X.~Zhou and M.~McKay, ``Secure transmission with artificial noise over fading
  channels:achievable rate and optimal power allocation,'' \emph{IEEE
  Transactions on Vehicular Technology}, vol.~59, no.~8, pp. 3831--3842, Oct.
  2010.

\bibitem{Ferdinand_J1}
N.~S. Ferdinand, D.~B. da~Costa, and M.~Latva-aho, ``Effects of outdated {CSI}
  on the secrecy performance of {MISO} wiretap channels with transmit antenna
  selection,'' \emph{IEEE Communications Letters}, vol.~17, no.~5, pp.
  864--867, May 2013.

\bibitem{HuangY_J1}
Y.~Huang, F.~S. Al-Qahtani, T.~Q. Duong, and J.~Wang, ``Secure transmission in
  {MIMO} wiretap channels using general-order transmit antenna selection with
  outdated {CSI},'' \emph{IEEE Transactions on Communications}, vol.~63, no.~8,
  pp. 2959--2971, Aug. 2015.

\bibitem{Hu_J1}
J.~Hu, Y.~Cai, N.~Yang, and W.~Yang, ``A new secure transmission scheme with
  outdated antenna selection,'' \emph{IEEE Transactions on Forensics and
  Security}, vol.~10, no.~11, pp. 2435--2446, Nov. 2015.

\bibitem{LiuTY_J2}
T.-Y. Liu, P.-H. Lin, Y.-W.~P. Hong, and E.~Jorswieck, ``To avoid or not to
  avoid {CSI} leakage in physical layer secret communication systems,''
  \emph{IEEE Communications Magazine}, vol.~53, no.~12, pp. 19--25, Dec. 2015.

\bibitem{Hyadi_J2}
A.~Hyadi, Z.~Rezki, and M.-S. Alouini, ``An overview of physical layer security
  in wireless communication systems with {CSIT} uncertainty,'' \emph{IEEE
  Access}, vol.~4, pp. 6121--6132, Sep. 2016.

\bibitem{HeX_J1}
X.~He and A.~Yener, ``{MIMO} wiretap channels with unknown and varying
  eavesdropper channel states,'' \emph{IEEE Transactions on Information
  Theory}, vol.~60, no.~11, pp. 6844--6869, Nov. 2014.

\bibitem{HeX_C1}
------, ``The interference wiretap channel with an arbitrarily varying
  eavesdropper: Aligning interference with artificial noise,'' in \emph{Proc.
  of the 50th Annual Allerton Conference on Communication, Control, and
  Computing (Allerton'12)}, Monticello, IL, Oct. 2012, pp. 204--211.

\bibitem{HeX_J5}
X.~He, A.~Khisti, and A.~Yener, ``{MIMO} multiple access channel with an
  arbitrarily varying eavesdropper: Secrecy degrees of freedom,'' \emph{IEEE
  Transactions on Information Theory}, vol.~59, no.~8, pp. 4733--4745, Aug.
  2013.

\bibitem{HeX_J6}
------, ``{MIMO} broadcast channel with an unknown eavesdropper: Secrecy
  degrees of freedom,'' \emph{IEEE Transactions on Communications}, vol.~62,
  no.~1, pp. 246--255, Jan. 2014.

\end{thebibliography}

%%\balance

\end{document}